\documentclass[a4paper,fleqn,usenatbib,useAMS]{mnras}
\usepackage[T1]{fontenc}
\usepackage{ae,aecompl}
\usepackage{graphicx}
\usepackage{natbib}
\usepackage{amssymb}
\usepackage{amsmath}
\usepackage{pdflscape}
\usepackage{xcolor}
\usepackage{newtxtext,newtxmath}

\usepackage{array}
\newcolumntype{H}{>{\setbox0=\hbox\bgroup}c<{\egroup}@{}}

\def\lesssim{\lower.5ex\hbox{$\; \buildrel < \over \sim \;$}}
\def\gtrsim{\lower.5ex\hbox{$\; \buildrel > \over \sim \;$}}

\title[Galactic winds in FIRE-2]{Characterizing mass, momentum, energy and metal outflow rates of multi-phase galactic winds in the FIRE-2 cosmological simulations}

\author[V. Pandya et al.] {Viraj Pandya$^{1,2}$\thanks{email: viraj.pandya@ucsc.edu}, Drummond B. Fielding$^2$, Daniel Angl\'es-Alc\'azar$^{3,2}$, Rachel S. Somerville$^2$,
\newauthor Greg L. Bryan$^{4,2}$, Christopher C. Hayward$^2$, Jonathan Stern$^5$, Chang-Goo Kim$^{6,2}$,
\newauthor Eliot Quataert$^6$, John C. Forbes$^2$, Claude-Andr\'{e} Faucher-Gigu\`{e}re$^5$, Robert Feldmann$^7$,
\newauthor Zachary Hafen$^8$, Philip F. Hopkins$^{9}$, Du\v{s}an Kere\v{s}$^{10}$, Norman Murray$^{11}$, Andrew Wetzel$^{12}$\\
$^1$UCO/Lick Observatory, Department of Astronomy and Astrophysics, University of California, Santa Cruz, CA 95064, USA\\
$^2$Center for Computational Astrophysics, Flatiron Institute, New York, NY 10011, USA\\
$^3$Department of Physics, University of Connecticut, 196 Auditorium Road, U-3046, Storrs, CT 06269, USA\\
$^4$Department of Astronomy, Columbia University, 550 West 120th Street, New York, NY 10027, USA\\
$^5$Department of Physics \& Astronomy and CIERA, Northwestern University, 1800 Sherman Ave, Evanston, IL 60201, USA\\
$^6$Department of Astrophysical Sciences, Princeton University, Princeton, NJ 08544, USA\\
$^7$Institute for Computational Science, University of Zurich, Winterthurerstrasse 190, CH-8057 Zurich, Switzerland\\
$^8$Department of Physics and Astronomy, University of California, Irvine, CA 92697, USA\\
$^{9}$TAPIR, Mailcode 350-17, California Institute of Technology, Pasadena, CA 91125, USA\\
$^{10}$Department of Physics, Center for Astrophysics and Space Sciences, University of California at San Diego, La Jolla, CA 92093, USA\\
$^{11}$Canadian Institute for Theoretical Astrophysics, University of Toronto, 60 St. George Street, Toronto, ON M5S 3H8, Canada\\
$^{12}$Department of Physics and Astronomy, University of California, Davis, CA 95616, USA\\
}

\date{Accepted ???. Received ??? in original form ???}

\begin{document}
\label{firstpage}
\pagerange{\pageref{firstpage}--\pageref{lastpage}}
\maketitle

\begin{abstract}
We characterize mass, momentum, energy and metal outflow rates of multi-phase galactic winds in a suite of FIRE-2 cosmological ``zoom-in'' simulations from the Feedback in Realistic Environments (FIRE) project. We analyze simulations of low-mass dwarfs, intermediate-mass dwarfs, Milky Way-mass halos, and high-redshift massive halos. Consistent with previous work, we find that dwarfs eject about 100 times more gas from their interstellar medium (ISM) than they form in stars, while this mass ``loading factor'' drops below one in massive galaxies. Most of the mass is carried by the hot phase ($>10^5$ K) in massive halos and the warm phase  ($10^3-10^5$ K) in dwarfs; cold outflows ($<10^3$ K) are negligible except in high-redshift dwarfs. Energy, momentum and metal loading factors from the ISM are of order unity in dwarfs and significantly lower in more massive halos. Hot outflows have $2-5\times$ higher specific energy than needed to escape from the gravitational potential of dwarf halos; indeed, in dwarfs, the mass, momentum, and metal outflow rates increase with radius whereas energy is roughly conserved, indicating swept up halo gas. Burst-averaged mass loading factors tend to be larger during more powerful star formation episodes and when the inner halo is not virialized, but we see effectively no trend with the dense ISM gas fraction. We discuss how our results can guide future controlled numerical experiments that aim to elucidate the key parameters governing galactic winds and the resulting associated preventative feedback.
\end{abstract}


\section{Introduction}
Supernova (SN) driven winds play a fundamental role in modern models of galaxy formation by helping to regulate star formation. Without SN-driven winds, models would predict an overabundance of dwarf galaxies compared to observations \citep[e.g.,][]{whitefrenk91,benson03,keres09}, overestimate the average stellar masses formed within dwarf halos \citep[e.g.,][]{dekel86,springel03}, and fail to match the redshift evolution of several observed scaling relations \citep[e.g.,][]{somerville01}. In addition to regulating star formation, galactic winds are thought to affect the thermodynamic state and metal content of the circumgalactic medium \citep[CGM; e.g., see the review by][]{tumlinson17} as well as chemically enrich the intergalactic medium \citep[IGM; e.g.,][]{oppenheimer06}. Winds may also fuel a significant fraction of late-time star formation in more massive halos by recycling back into the interstellar medium \citep[ISM; e.g.,][]{oppenheimer10,henriques13,white15,anglesalcazar17}. In lower mass halos, SN-driven winds may more easily escape and heat the CGM/IGM, causing preventative feedback effects by suppressing gas accretion in the first place \citep[e.g.,][]{vandevoort11,dave12,lu15,pandya20} and decreasing the metal and dust content of dwarfs \citep[e.g.,][]{dave11,feldmann15}.

Despite their central importance, a complete characterization of galactic winds in a cosmological context and their implications for galaxy evolution has remained elusive. In the current landscape of models, genuinely emergent wind properties have been predicted by ``resolved'' ISM simulations but these only represent a relatively small sub-galactic region and generally assume $z=0$ Milky Way-like global conditions \citep[e.g.,][]{walch15,martizzi16,li17,kim18,kim20}. Extending SN-driven wind predictions to global galaxy scales has been challenging, but much progress has been made using idealized high-resolution simulations of dwarfs and more massive galaxies \citep[e.g.,][]{hopkins12,fielding17a,smith18,hu19,li20b}. On cosmological scales, all large-volume models are effectively phenomenological: they must implement wind scalings ``by hand'' and rely on subgrid approaches that require tunable free parameters such as hydrodynamically decoupled wind particles or temporary shut off of cooling \citep[e.g.,][]{springel03,stinson06,dave16}. In between these approaches sit a relatively new generation of cosmological ``zoom-in'' simulations such as the Feedback In Realistic Environments project\footnote{\url{http://fire.northwestern.edu}} \citep[FIRE;][]{hopkins14,hopkins18}, where in some cases SN remnants can be resolved. When SN remnants are unresolved, the subgrid approach is to deposit the additional momentum expected from the unresolved energy-conserving phase of SN remnants using even higher resolution simulations for calibration instead of observational tuning \citep{hopkins18b}. In addition, a variety of physical processes are accounted for in such zoom-in simulations that may not otherwise be captured in small-scale simulations (e.g., self-consistent clustering of star formation, cosmological gas accretion, galaxy mergers, the large-scale propagation of winds into the CGM, etc.). It is timely to ask how the emergent wind properties from such zoom-in simulations compare to those of higher resolution subgalactic simulations\footnote{In this work, we will measure wind properties at distances from the ISM that are typically outside of the domain of small-scale simulations, hence providing crucial complementary information. A more detailed analysis of winds closer to ISM breakout is deferred to future work \citep[but see][]{gurvich20}.}, and to derive new wind scalings that can be implemented into large-volume simulations and semi-analytic models \citep[SAMs; as presented by, e.g.,][]{muratov15}.

When analyzing galactic winds, it is common practice to focus on ``mass loading factors'' and ``metal loading factors,'' which respectively describe gas mass outflow rates and metal outflow rates conveniently normalized by reference star formation rates and supernova metal injection rates. It has long been appreciated that dwarf halos preferentially have higher mass and metal loading factors \citep[e.g.,][]{dekel86,maclow99,efstathiou00}. The common interpretation of this is that dwarfs have shallower potential wells and hence SN ejecta can more easily escape. Simple arguments suggest that we should expect a power law relation between the mass loading factor and global halo circular velocity whose slope will be steeper if winds are ``energy-conserving'' and shallower if they are ``momentum-conserving'' \citep{murray05}. Much work has gone into testing this simple energy- and momentum-driven dichotomy using hydrodynamical simulations \citep[e.g.,][]{hopkins12,muratov15,christensen16}, and the language of this framework is commonly used to justify assumed wind scalings in SAMs and simulations with insufficient resolution to capture SN remnant evolution \citep[e.g.,][]{somerville08,oppenheimer10,anglesalcazar14}. While characterizing winds in this way has provided useful insights, a more detailed analysis of the thermodynamic properties of multi-phase winds (i.e., temperature and velocity distributions) provides additional clues about whether their driving energy source is kinetic or thermal, and enables more careful consistency checks between different simulations (and against observations).

In addition to characterizing the mass and metal loading of galactic winds, it is also crucial to explicitly measure their multi-phase energy and momentum loading factors: how much of the energy and momentum input by SNe also make it out of the ISM? The explicit calculation of energy and momentum loading factors can help test whether winds are energy-driven or momentum-driven in a simple way, and help to interpret any secondary heating or ``pushing'' effects on the CGM/IGM. In recent years, small-scale, high-resolution idealized simulations have made quantitative predictions for energy and momentum loadings, with a common finding that the cold phase carries most of the mass whereas the hot phase carries most of the energy \citep{kim18,fielding18,hu19,li20,kim20}. These idealized numerical experiments have also been able to correlate their loading factors against the granular conditions of the ISM in which SNe go off rather than just the global halo circular velocity \citep[e.g.,][]{creasey13,fielding17,li20,kim20}. A similarly comprehensive analysis of multi-phase galactic winds in cosmological simulations would provide major insights on how ``ejective feedback'' (quantified by mass and metal loading) and ``preventative feedback'' (quantified by energy and momentum loading) may act in concert to regulate galaxy evolution. 

In this paper, we build on the analysis of winds in the FIRE-1 zoom-in simulations \citep{hopkins14} by \citet{muratov15,muratov17,anglesalcazar17,hafen19,hafen20}. We use a suite of simulations run using the updated FIRE-2 code, which model the same stellar processes as the FIRE-1 simulations but use a new hydrodynamic solver \citep{hopkins18}. Thus, we expect many of the overall predictions to be similar between FIRE-1 and FIRE-2. Motivated by analysis procedures for small-scale idealized simulations, we implement a sophisticated method for identifying galactic winds by considering their bulk kinetic, thermal and potential energies. Instead of focusing only on the total mass and metal loading factors as is common practice, our multi-dimensional analysis focuses on the temperature dependence of all four loading factors (mass, momentum, energy and metals) and how this varies as a function of galaxy mass. With the FIRE-2 simulation suite, we will comment on the nature of SN-driven galactic winds across a wide range of halo masses (low-mass dwarfs, intermediate-mass dwarfs, MW halos and their high-redshift dwarf progenitors, and more massive halos at high redshift). We also present scaling relations for the loading factors not just with the global halo circular velocity (as is commonly done), but also with several ``quasi-local'' ISM properties as a first step toward connecting the larger-scale emergent loadings with the smaller-scale conditions of the ISM in which the winds are launched. 

This paper is organized as follows. \autoref{sec:models} describes the FIRE-2 simulations and \autoref{sec:analysis} details our analysis methods. In \autoref{sec:results_ism}, we present wind loading factors near the ISM and in \autoref{sec:results_halo} we describe results for winds leaving the halo at $R_{\rm vir}$. We discuss our results in \autoref{sec:discussion} and summarize in \autoref{sec:summary}. We assume a standard flat $\Lambda$CDM cosmology consistent with the FIRE-2 code and \citet{planck14}; i.e., $h=0.7$, $\Omega_M=0.27$, $\Omega_{\Lambda}=0.73$ and $f_b\equiv\Omega_b/\Omega_M\approx0.16$.

\section{Simulation Description}\label{sec:models}
We use a suite of cosmological ``zoom-in'' simulations run using the FIRE-2 code \citep{hopkins18}. Our analysis focuses on a ``core'' suite of 13 FIRE-2 halos: 4 low-mass dwarfs with $M_{\rm vir}\sim10^{10}M_{\odot}$ at $z=0$ (m10q, m10v, m10y, m10z), 6 intermediate-mass dwarfs with $M_{\rm vir}\sim10^{11}M_{\odot}$ by $z=0$ (m11a, m11b, m11q, m11c, m11v, m11f), and 3 MW-mass halos with $M_{\rm vir}\sim10^{12}M_{\odot}$ by $z=0$ (m12i, m12f, m12m). These halos were first presented in \citet{wetzel16,garrisonkimmel17,chan18,hopkins18}. To this core suite, we also add the four FIRE-2 massive halos (A1, A2, A4, and A8 with $M_{\rm vir}\sim10^{12.5}-10^{13}M_{\odot}$ at $z=1$) presented by \citet{anglesalcazar17b} and further analyzed in \citet{cochrane19,wellons20,stern20}. These halos are denoted as ``m13'' throughout the paper, were only run down to $z=1$, and were previously simulated with the FIRE-1 model as part of the MassiveFIRE suite \citep{feldmann16,feldmann17}. While the m10, m11 and m12 halos agree well with empirical stellar-to-halo-mass relations, the m13 halos have unrealistically high stellar masses and central densities by $z=1$ \citep[e.g.,][]{parsotan21}, and hence should not be taken as representative of the observed population (this is a regime where feedback from supermassive black holes may have an appreciable effect but this is not included in these simulations).

We refer the reader to \citet{hopkins18} for a detailed description of the simulations and methodology. Here we only briefly review the most relevant aspects, with a particular emphasis on the explicit stellar feedback model. The core FIRE-2 simulations model the same physical processes as in FIRE-1 but use a new Lagrangian ``meshless finite-mass'' hydrodynamic solver as opposed to the ``pressure--entropy'' formulation of smoothed particle hydrodynamics \citep{hopkins15}. The FIRE-2 simulations implement a broad range of physics, including deposition of mass, momentum, energy and metals due to both Type Ia and Type II SNe, stellar winds, radiation pressure, and photo-ionization and photo-electric heating. There is a spatially-uniform but redshift-dependent UV background based on \citet{fauchergiguere09}.

The relatively high resolution of the FIRE-2 simulations (Lagrangian particle masses of $\sim250M_{\odot}$ in the low-mass dwarfs, up to $\sim7100M_{\odot}$ for the MW halos and $\sim33000M_{\odot}$ for the m13 runs) allows stellar feedback to be modeled locally and explicitly. In particular, the generation and propagation of winds is not explicitly dependent on global halo properties and does not require subgrid approaches of limited predictive power (e.g., hydrodynamically-decoupled winds, shut-off of cooling, thermal bombs).
Of course, not all SN remnants will be resolved, especially in the more massive halos which have comparatively worse resolution. As detailed in \citet{hopkins18b}, this is ``corrected'' for in FIRE-2 by depositing onto nearby gas particles the additional momentum expected from the unresolved energy-conserving Sedov-Taylor phase (due to $PdV$ work). The thermal energy output by the unresolved SN remnant is also self-consistently reduced to account for radiative cooling after the energy-conserving phase. In cases where the SN remnant is resolved, the FIRE-2 subgrid model deposits the full SN kinetic and thermal energy and allows the hydrodynamic solver to explicitly calculate any heating and momentum boosting. Note that while some small-scale simulations suggest that a resolution of $\lesssim100M_{\odot}$ may be necessary to properly capture the evolution of SN remnants \citep[e.g.,][]{kim15,steinwandel20}, the combination of multiple stellar feedback effects (e.g., early radiative feedback) with self-consistent clustering of star formation in FIRE-2 may act to alleviate this resolution requirement. \citet[][Figure 9]{hopkins18b} showed that the FIRE subgrid model remains converged to the high-resolution result up to resolutions of $2000M_{\odot}$ for an m10 halo \citep[see also][who re-simulated a few FIRE-2 dwarfs with $30M_{\odot}$ resolution]{wheeler19}.

As our work builds on the analysis of FIRE-2 presented in \citet{pandya20}, here we use the same halo catalogs and merger trees generated using the Rockstar and consistent-trees codes \citep{behroozi13a,behroozi13b}. For halo masses and radii, we adopt the \citet{bryan98} virial overdensity definition. We only focus on the main central halo in each of these simulations and do not analyze winds from satellites. We also do not attempt to exclude gas associated with satellites from large-scale outflow measurements of the central halo.

\section{Analysis}\label{sec:analysis}
In this section, we describe how we select outflowing gas, define multi-phase outflows, and compute loading factors.

\subsection{Accurately defining outflows}
\subsubsection{Selecting outflowing particles}
It is common practice to define outflows in cosmological simulations using a single cut on the halo-centric radial velocity of particles (regardless of using the shell/Eulerian or particle-tracking/Lagrangian methods). The simplest cut often adopted is $v_{\rm rad}>0$ km/s which would select all particles that are traveling radially away from the halo center \citep[as done by, e.g.,][]{fauchergiguere11,muratov15}. This can confuse slow random motions with galactic outflows. The other extreme is to select only particles at a given radius whose $v_{\rm rad}>v_{\rm esc}(r)$ where $v_{\rm esc}(r)$ is the local escape velocity at that radius. This cut is often used to define the subset of fastest moving ``wind'' particles among the whole distribution of outflowing particles. There are variations on this radial velocity cut method in the literature: \citet{muratov15} use the velocity dispersion of the underlying virialized DM halo particles, \citet{mitchell20} use $0.25V_{\rm max}$ where $V_{\rm max}$ is the maximum circular velocity of the halo, and \citet{nelson19} compute the cumulative mass fraction of outflowing particles with radial velocities above sequentially increasing velocity thresholds. 

However, using a single cut on $v_{\rm rad}$ alone is sub-optimal for defining winds for the following reason.\footnote{From an observational perspective, the simplest $v_{\rm rad}>0$ km/s cut might be justified and desirable (especially if detailed kinematics, phase information and gravitational potential constraints are unavailable), though in practice a larger threshold velocity is usually adopted to avoid ISM contamination. Nevertheless, here we are interested in robustly identifying and characterizing winds from a simulation perspective.} Consider that every gas particle possesses three forms of energy: kinetic, thermal and potential energy. A single cut on $v_{\rm rad}$ alone assumes the extreme case of ``ballistic motion.'' But since we are dealing with gas, we must account for the fact that the thermal energy of gas particles can serve as a source of acceleration assuming adiabatic expansion (i.e., no external heating, cooling or interactions). This has long been realized in the literature for small-scale resolved ISM/CGM simulations \citep[e.g.,][]{martizzi16,kim20,schneider20} but has not been fully leveraged for cosmological simulations \citep[though see][]{hopkins12}.  
Here we introduce a slightly more sophisticated methodology to accurately define outflowing particles. First, we make a simple cut on $v_{\rm rad}>0$ km/s. This selects all particles that are flowing radially outwards. However, a large fraction of these particles may have relatively small radial velocities arising from underlying random velocity fluctuations. We only want to select particles that will be able to travel a significant distance. Hence, for every gas particle, we calculate the radial component of the total Bernoulli velocity $v_{\rm B,total}$, which is a measure of the total specific energy \citep[e.g.,][]{hopkins12,li17,kim18,fielding18}: 
\begin{equation}\label{eqn:vbern}
v_{\rm B,total}^2 \equiv \frac{1}{2}v_r^2 + \frac{c_s^2}{\gamma-1} - \frac{1}{2}v^2_{\rm esc}\;.
\end{equation}
The first term is the specific radial kinetic energy quantified by the halo-centric particle radial velocity squared. The second term is the specific enthalpy assuming an ideal gas whose equation of state has adiabatic index $\gamma$ and sound speed $c_s=\sqrt{\gamma\frac{kT}{\mu m_p}}$. We assume a monatomic ideal gas, hence $\gamma=\frac{5}{3}$. 

The third term is equivalent to the specific gravitational potential energy, $\Phi$. The simulation code internally keeps track of $\Phi$ for each particle to compute its gravitational acceleration, but unfortunately $\Phi$ is not one of the properties output in the particle snapshot files. Computing $\Phi$ in post-processing is tricky because the mass distribution is heterogeneous and that can disproportionately affect the potential for some particles, even if they have the same halo-centric distance. For simplicity, we assume the mass distribution can be approximated as spherically symmetric, which allows us to relate the potential to the enclosed mass profile in a simple way:
\begin{equation}
\Phi(r) = -\int_r^{r_{\rm \infty}} \frac{GM(<r)}{r^2} dr
\end{equation}
where $r_{\rm \infty}$ is an arbitrarily large radius. Given that we are working with cosmological zoom-in simulations, we adopt the following strategy. We set the zeropoint of the potential at $r_{\rm \infty}=2R_{\rm vir}$ since that is the turnaround radius for a virialized system and particles traveling beyond $2R_{\rm vir}$ are likely unbound from the halo anyway (also, our zoom regions can start to become contaminated by low-resolution DM beyond $2R_{\rm vir}$). Within $2R_{\rm vir}$, we compute the enclosed mass profile based on all star, gas and high-resolution DM particles using spherical shells of width $0.01R_{\rm vir}$.

We re-write $\Phi$ as an escape velocity using the energy conservation equation and assuming the gas particle at $r$ is already maximally cold (i.e., ignoring any changes in enthalpy):
\begin{equation}
\frac{1}{2}v_{\rm esc}^2(r) + \Phi(r) = \Phi(2R_{\rm vir})\;.
\end{equation}
In this way, we can derive the radial profile of escape velocity, which tells us how fast a particle must initially be going (at minimum) to fully climb out of the halo potential: 
\begin{equation}
v_{\rm esc}(r) = \sqrt{2(\Phi(2R_{\rm vir}) - \Phi(r))}\;.
\end{equation}

The quantity $v_{\rm B,total}^2$ represents the radial component of the total specific energy of a gas particle at its current position. Note that $v_{\rm B,total}^2$ can be negative, which means that a particle is bound (i.e., its kinetic energy plus enthalpy is less than its potential energy). By comparing this initial Bernoulli velocity to a hypothetical final Bernoulli velocity at some other larger halo-centric distance, we can assess whether a given gas particle has enough starting energy to make it to that larger distance (neglecting interactions). For a particle to be able to travel from its current radius $r_1$ to some secondary radius $r_2$, its initial Bernoulli velocity must be larger than the potential energy at that secondary radius.\footnote{This neglects the effect of heating by the UV background that prevents gas from cooling to arbitrarily low temperature. Thus, in principle for the secondary radius we should add the $\frac{c_s^2}{\gamma-1}$ term assuming the sound speed for gas in thermal equilibrium with the UV background at $\sim10^4K$, roughly $15$ km/s. In practice, this makes a negligible difference for outflow selection (most of the gas tends to be escaping in low-mass halos anyway, and for MW-mass halos this $10^4$K gas sound speed term is an order of magnitude lower than the escape velocity term).} We use this to impose an additional criterion that selects only gas with sufficiently large $v_{\rm B,total}$ relative to the escape velocity at some target distance (defined in the next section): 
\begin{equation}\label{eqn:vB_comparison}
v_{\rm B,total}^2(r_1)> -\frac{1}{2}v^2_{\rm esc}(r_2)\;.
\end{equation}
This criterion along with $v_{\rm rad}>0$ km/s is a more physically meaningful and robust way to select wind particles compared to either $v_{\rm rad}>0$ km/s or $v_{\rm rad}>v_{\rm esc}(r)$ alone. It is effectively an intermediate case that avoids the very slow moving turbulent motions while still selecting the hotter and slower components of the wind. This definition is also a natural way to quantitatively distinguish between genuinely escaping winds and outflows expected to remain bound out to some larger radius. Note that \autoref{eqn:vB_comparison} does not account for possible time-varying pressure gradients in the inner CGM, which winds would also need to overcome in addition to the (nearly static) gravitational potential. We will show later that even though we are neglecting this complication, we measure weaker winds when there is a substantial hot hydrostatic corona as in the more massive halos.

\subsubsection{Computing outflow fluxes}
We compute outflow fluxes in two characteristic spherical shells:
\begin{enumerate}
\item $0.1-0.2R_{\rm vir}$ (ISM boundary shell)
\item $1.0-1.1R_{\rm vir}$ (virial boundary shell)
\end{enumerate}
In each of these two shells, we must select particles that have enough energy to make it to some secondary radius, $r_2$, if not farther (assuming an adiabatic flow). There is inevitably a large range of arbitrary choices that could be made for $r_2$. For the ISM shell, we adopt a secondary radius of $r_2=0.5R_{\rm vir}$, which we take to represent the ``middle'' of the CGM. Choosing a smaller target radius would pick up additional cooler/slower outflows, but we note that our ISM shell is already quite far out ($0.1-0.2R_{\rm vir}$). In Appendix \ref{sec:vBalt}, we illustrate how our results would change if we used a factor of two smaller or larger target distance. For the virial shell, we adopt a secondary radius of $r_2=2.0R_{\rm vir}$. This lets us select particles at $1.0-1.1R_{\rm vir}$ that have at least enough energy to make it to $2.0R_{\rm vir}$, if not farther. Since particles can be considered unbound if they travel beyond the turnaround radius of $2R_{\rm vir}$, this is a natural way to estimate genuinely escaping outflows from the halo. Finally, since we will compare the halo outflow rate to the preceding ISM outflow rate, we also define a second more restrictive ISM outflow criterion by choosing $r_2=2R_{\rm vir}$. This lets us additionally estimate the subset of ISM outflows that have enough energy to get not just to $0.5R_{\rm vir}$ but rather escape to $2R_{\rm vir}$ or beyond.

Finally, with outflowing particles identified for each of the two shells above, we compute their total mass, momentum, energy and metal mass outflow rates as follows: 
\begin{equation}
\dot{M}_{\rm out} = \sum_i \frac{m_i v_{\rm r,i}}{\Delta L}
\end{equation}
\begin{equation}
\dot{p}_{\rm out} = \sum_i \dot{M}_{\rm out,i} v_{\rm r,i} \left(1+\frac{1}{\gamma\mathcal{M}_i^2}\right)
\end{equation}
\begin{equation}
\dot{E}_{\rm out} = \sum_i \dot{M}_{\rm out,i} v_{\rm B,i}^2
\end{equation}
\begin{equation}
\dot{M}_{\rm Z,out} = \sum_i \dot{M}_{\rm out,i} Z_i
\end{equation}
Here, the subscript $i$ runs over all the selected outflowing particles in the shell, $\Delta L=0.1R_{\rm vir}$ is the width of our ISM and virial shells, $v_r$ is the radial velocity, $\mathcal{M}\equiv v_r/c_s$ is the Mach number, and $Z$ is the metal mass fraction of the particle. Note that the second term in the momentum flux accounts for the thermal pressure component (defined as $P=\rho c_s^2 / \gamma$), which can be substantial for hot outflows or more generally when $\mathcal{M}$ is small. $v_{\rm B}$ is the Bernoulli velocity neglecting the gravitational term and including the transverse kinetic energy component (as opposed to $v_{\rm B,total}$ in \autoref{eqn:vbern}):
\begin{equation}\label{eqn:vbern2}
v_{\rm B}^2 = \frac{1}{2}v^2+\frac{3}{2}c_s^2\;,
\end{equation}
where $v$ is the magnitude of the total halo-centric particle velocity vector instead of just $v_r$. We neglect the gravitational term for $\dot{E}_{\rm out}$ because we want to quantify how much specific kinetic energy and enthalpy are being transported by outflows (these quantities, including the transverse velocity components, will be responsible for any heating and pushing of ambient gas). The gravitational term comes in earlier when we first want to identify escaping and bound outflows. 

\begin{figure*} 
\begin{center}
\includegraphics[width=0.85\hsize]{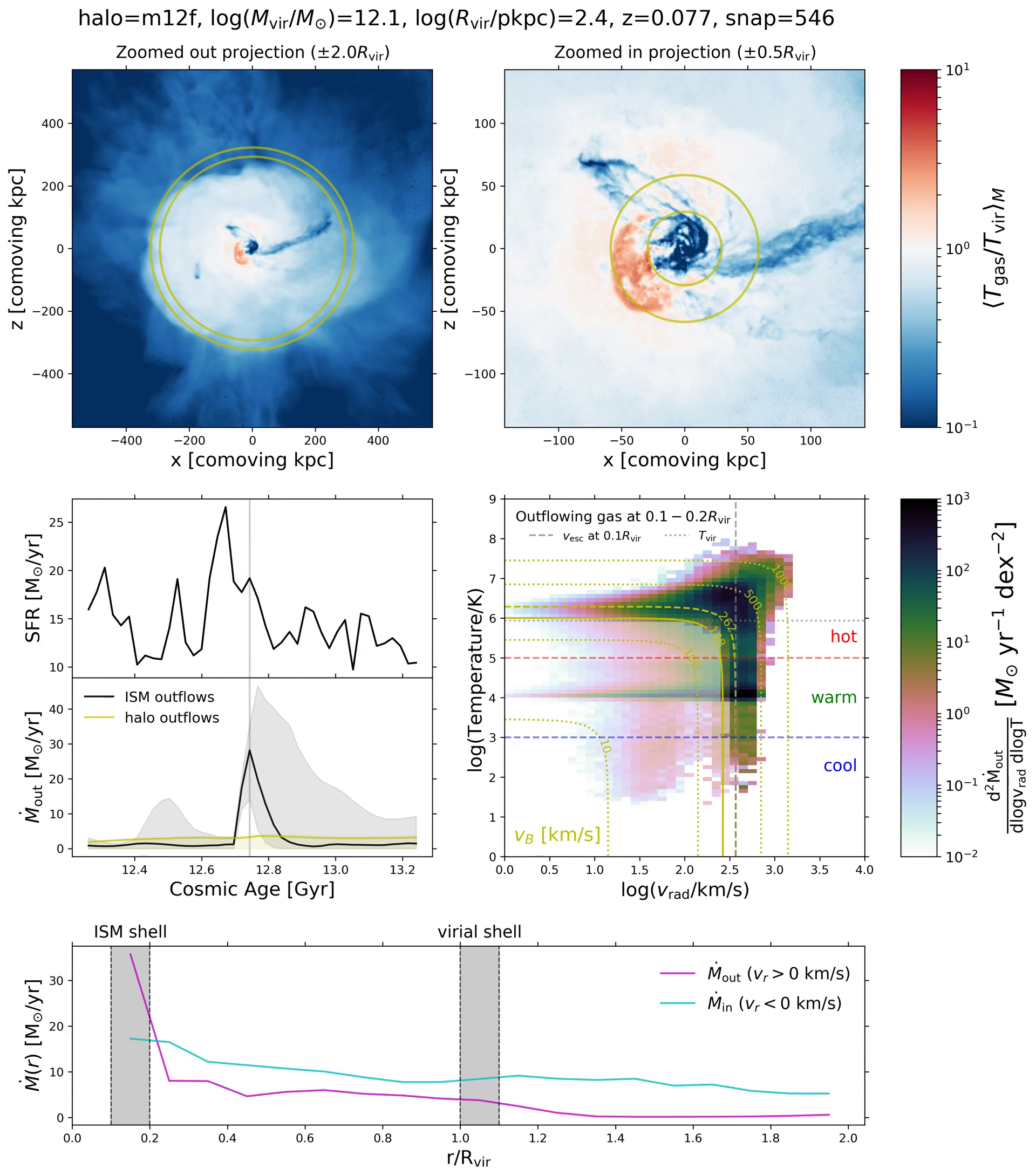}
\end{center}
\caption{Visualizing and quantifying a strong outflow at $z\sim0$ in a MW-mass halo (m12f). This is a single frame from a movie that is available for download. \textit{Top-left:} Zoomed-out projection ($\pm2R_{\rm vir}$) of the mass-weighted average gas temperature. The colorbar has been normalized by the halo virial temperature. The two white circles demarcate our virial shell ($1.0-1.1R_{\rm vir}$). \textit{Top-right:} Similar to the top-left panel but now a zoomed-in projection ($\pm0.5R_{\rm vir}$). The two white circles mark our ISM shell ($0.1-0.2R_{\rm vir}$). \textit{Middle-left:} Time series of the SFR (top panel) and the mass outflow rate in the ISM and viral shells (bottom panel). The lines in the lower part of this panel show mass outflow rate measurements based on our fiducial Bernoulli velocity wind criterion. The shaded regions show how more extreme cuts would lead to different estimates: $v_{\rm rad}>0$ km/s gives an upper bound to the mass outflow rate whereas $v_{\rm rad}>v_{\rm esc}$ picks up only the fastest material and hence leads to a lower bound. The vertical gray line marks the current snapshot time. \textit{Middle-right:} Phase diagram of temperature and radial velocity for the multi-phase ISM outflows identified using our Bernoulli velocity method. The colorbar shows the mass outflow rate in logarithmic bins of temperature and radial velocity. The horizontal red and blue lines demarcate our cool, warm and hot outflow temperature regimes. The horizontal dotted gray line indicates the halo virial temperature (computed at $R_{\rm vir}$) and the vertical gray dashed line denotes $v_{\rm esc}$ at $0.1R_{\rm vir}$. Yellow dotted contours show lines of constant Bernoulli velocity, with the potential difference between $0.1R_{\rm vir}$ and either $0.5R_{\rm vir}$ or $2R_{\rm vir}$ shown as the solid and dashed yellow contours, respectively. The transparent histogram below the solid yellow contour shows what is excluded from our $v_{\rm B}$ cut. Selecting only outflows with $v_{\rm rad}>v_{\rm esc}$ would miss the slower but still hot wind component, which our Bernoulli velocity method successfully captures. \textit{Bottom:} Radial profile of instantaneous mass flux for both outflows (magenta) and inflows (cyan) between $0.1-2R_{\rm vir}$ in spherical shells of width $0.1R_{\rm vir}$. For simplicity, we use $v_{\rm rad}=0$ km/s as the dividing point between outflows and inflows. Our ISM and virial shells are marked as the gray bands. This panel can be used to follow the radial evolution of individual outflow episodes and qualitatively infer CGM entrainment or wind mass losses. \texttt{This movie and others are available for download online at \url{https://vimeo.com/showcase/8705020}.}}
\label{fig:example}
\end{figure*}

\begin{figure*} 
\begin{center}
\includegraphics[width=0.85\hsize]{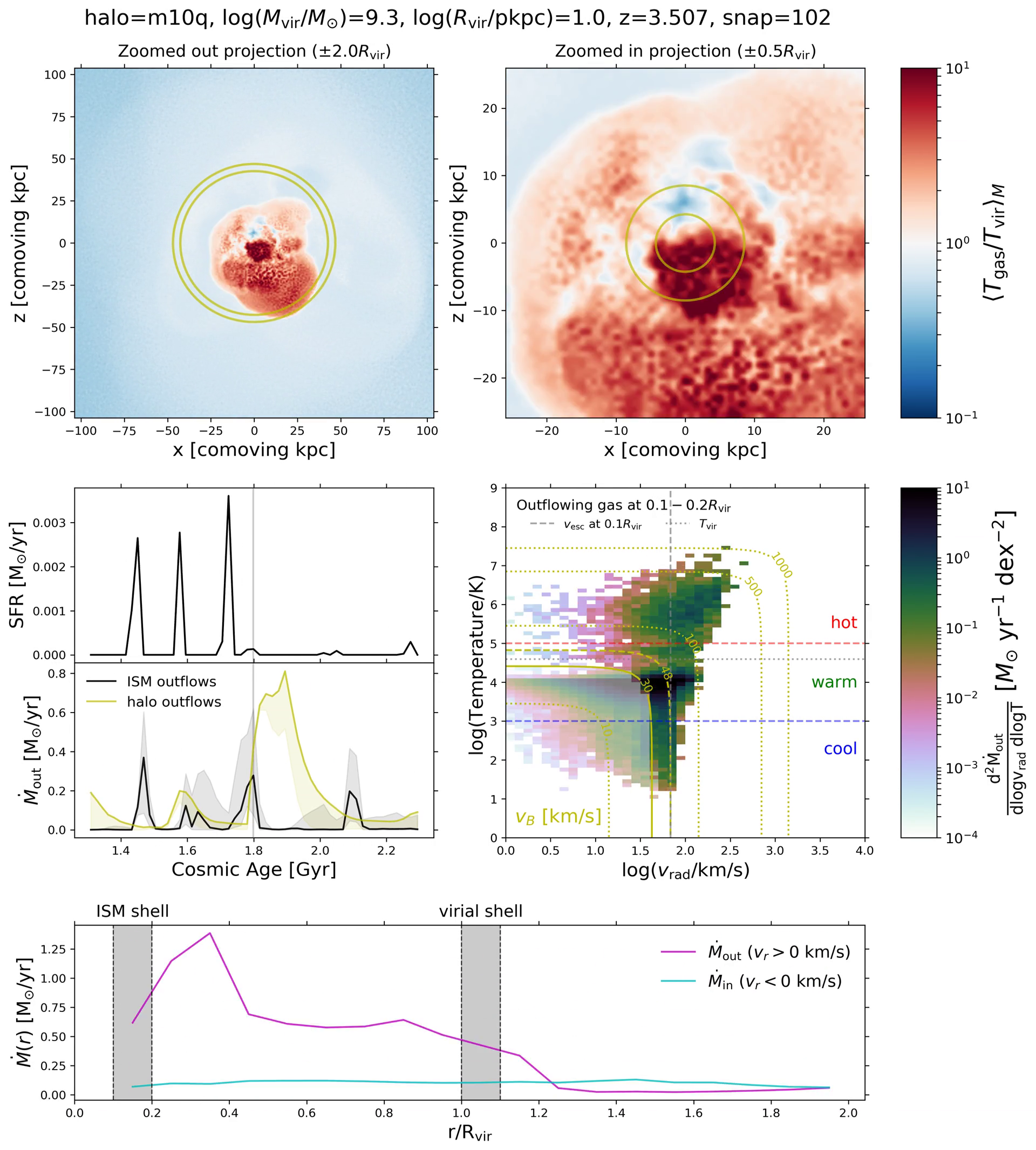}
\end{center}
\caption{Similar to \autoref{fig:example} but now for a low-mass dwarf (m10q). This movie frame is during a major outflow episode at high redshift $z\sim3.5$, a regime where dwarfs are often characterized as having mass loadings of $\sim100$ or more. If we divide the values of the individual ISM mass outflow rate peaks by their associated preceding SFR spikes (bottom-left panel), we would indeed infer instantaneous mass loadings of $\sim100$. The halo-scale mass loadings (magenta) are even larger due to entrainment of CGM gas by outflows. Note how there is a hot bubble in the projection panels created by the strong outflows. \texttt{This movie and others are available for download online at \url{https://vimeo.com/showcase/8705020}.}}
\label{fig:example2}
\end{figure*}

\autoref{fig:example} and \autoref{fig:example2} respectively show examples of strong outflows in a MW halo at $z\sim0$ and a dwarf halo at $z\sim3$. The phase diagram of temperature versus radial velocity shows that our Bernoulli velocity wind criterion successfully captures the slower but still very hot buoyant wind component, which would otherwise be missed by simply requiring $v_{\rm rad}>v_{\rm esc}$. At the same time, our method excludes cold ballistic outflows that are moving too slowly to travel a significant distance and instead are likely tracing turbulence in the inner CGM (much of this gas may rapidly recycle back into the ISM via fountain flows; \citealt{anglesalcazar17}). There is generally a time lag between peaks in the star formation history (SFH) and subsequent spikes in the mass outflow rate time series. As outflows propagate from the inner halo to the outer halo, they can either deposit or sweep up mass in the CGM. This can be inferred qualitatively from the time evolution of the radial profile of $\dot{M}_{\rm out}$ since the amplitude and width of individual outflow spikes may change as they move to larger radius.

\subsection{Multi-phase outflow selection criteria}
It is important to distinguish between outflows of different temperatures since that can clarify whether the driving energy source is kinetic or thermal. The simplest way to do this is based on atomic cooling physics. We can divide the temperature distribution into roughly three phases: 
\begin{enumerate}
\item $T<10^3 K$ (cold outflows)
\item $10^3<T<10^5 K$ (warm outflows)
\item $T>10^5 K$ (hot outflows).
\end{enumerate}
These temperature cuts correspond to physically distinct regimes.\footnote{Our warm gas is also termed cold gas in some CGM studies since it is much colder than the virial temperature of MW-mass halos.} The cut at $10^5$ K corresponds to the peak of the cooling curve, so material is expected to separate naturally about this temperature. Likewise, the cut at $10^3$ K corresponds to the unstable part of the cooling curve at the usual pressures and photoelectric heating rates found in the ISM/inner CGM, so gas is also expected to naturally separate about this temperature. Lastly, a significant amount of gas can be expected to have $T\sim10^4$ K since that is roughly the equilibrium temperature between photoionization from the UV background and recombination cooling. These cuts, therefore, mirror the delineations that are expected to arise naturally in and around galaxies. These temperature bins also trace what observers can measure: the cold phase corresponds to molecular/atomic outflows, the warm phase traces partially ionized gas that will produce H$\alpha$ emission and absorption from singly and doubly ionized metals, and the hot phase traces highly ionized gas that produces X-ray emission. 

\begin{figure*} 
\begin{center}
\includegraphics[width=\hsize]{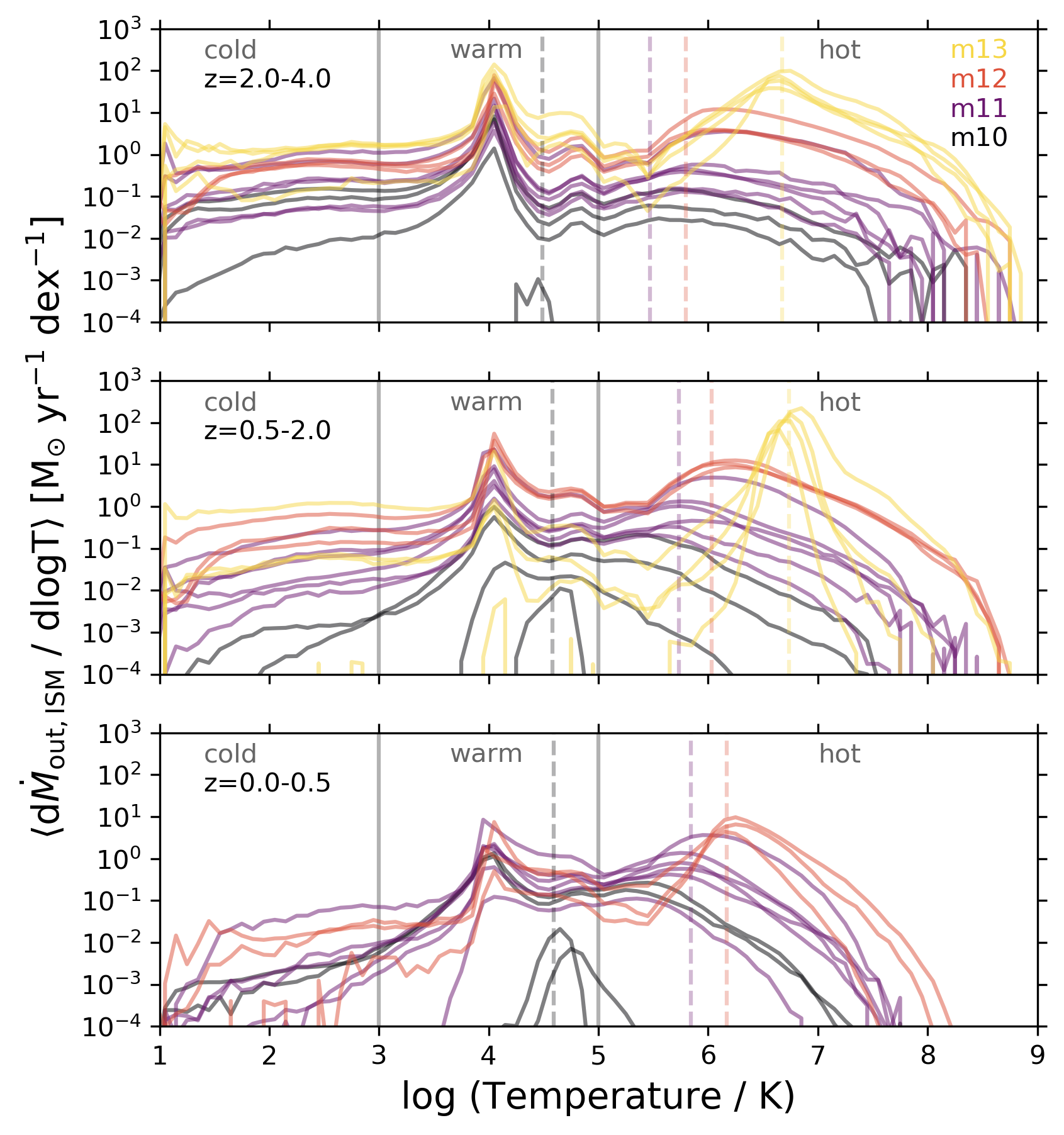}
\end{center}
\caption{Temperature distribution of ISM-scale winds. These distributions are based on $\dot{M}_{\rm out}$-weighted averages over three broad redshift bins. Solid gray vertical lines are cooling physics-based temperature cuts at $10^3$ and $10^5$ K. Dashed vertical colored lines illustrate example virial temperatures of the inner halo (at $0.1R_{\rm vir}$) for representative halos from each mass bin. The virial temperatures roughly align with the temperature distribution peak for hot (virialized) outflows. Cold outflows are more prominent at high redshift. The two lowest mass halos generally do not have multiphase outflows. Note that the m13 halos were only run down to $z=1$ and so are absent from the bottom panel.}
\label{fig:mdot_temp_hist}
\end{figure*}

In \autoref{fig:mdot_temp_hist}, we plot the average temperature distribution of the mass outflow rate of our halos through the ISM shell. The distribution is averaged over three broad redshift bins using the $\dot{M}_{\rm out,ISM}$ in each snapshot as the weight. We see that outflows in our simulations are inherently multi-phase, except in the two lowest mass halos. The cold phase is more pronounced at higher redshift. The peak in the warm regime at $\sim10^4$K likely reflects the equilibrium temperature between heating and cooling, and the broad peak in the hot regime corresponds to the virial temperature in the inner halo, computed as 
\begin{equation}
T_{\rm vir}=35.9\frac{V_{\rm circ}^2}{\textrm{km s}^{-1}} \;\textrm{K} 
\end{equation}
where $V_{\rm circ}$ is the circular velocity at $0.1R_{\rm vir}$ (as opposed to the common practice of using $V_{\rm circ}$ at the virial radius). The virial temperatures of the lowest mass halos are themselves below $10^5$K, so there is no pronounced peak in their hot outflow rates. The two lowest mass dwarfs, in particular, show a cut-off in their cold outflow rates at $\sim10^4$K. We will see later that this means the warm phase is remarkably important for outflows in dwarfs.

\subsection{Computing wind loading factors}\label{sec:loadings}
\subsubsection{Reference fluxes}
Lastly, it is useful to compare the wind fluxes to reference fluxes at the ISM scale. By dividing the two, we can estimate the loading factor $\eta$ and get a sense of how much mass, energy, momentum and metal mass is being ejected versus what was input from star formation and SNe. Computing the reference fluxes is non-trivial in cosmological simulations because of the wide range of processes that are simultaneously at play. We therefore limit ourselves to considering Type II SNe (we expect these to dominate over Type Ia SNe, radiative heating and mass loss from normal stellar evolution, and other processes, but see our discussion of caveats in \autoref{sec:systematics}). In line with \citet{kim20}, we adopt the following reference fluxes:
\begin{enumerate}
\item $\dot{M}_{\rm ref} =$ SFR
\item $\dot{p}_{\rm ref} = \dot{N}_{\rm SN} \frac{E_{\rm SN}}{v_{\rm cool}} = \frac{\rm SFR}{100 M_{\odot}} \frac{E_{\rm SN}}{v_{\rm cool}}$
\item $\dot{E}_{\rm ref} = \dot{N}_{\rm SN} E_{\rm SN} = \frac{\rm SFR}{100 M_{\odot}} E_{\rm SN}$
\item $\dot{M}_{\rm Z,ref} = \dot{N}_{\rm SN} M_{\rm ej} Z_{\rm SN} = \frac{\rm SFR}{100 M_{\odot}} M_{\rm ej} Z_{\rm SN}$
\end{enumerate}
Here, the total instantaneous galaxy SFR is computed by summing over the individual SFRs predicted by all gas particles\footnote{Alternatively, we could have summed the masses of star particles younger than, say, 20 Myr and then divided by that timescale. We do not expect our conclusions to change had we used this different SFR definition.} within $0.1R_{\rm vir}$. Then $\dot{N}_{\rm SN}=\frac{\rm SFR}{100 M_{\odot}}$ is the supernova rate; we adopt the common assumption that one SN occurs for every $100M_{\odot}$ of stars formed under reasonable assumptions for the IMF. This is consistent with the FIRE-2 assumptions of a \citet{kroupa01} IMF and the \texttt{STARBURST99} stellar population models \citep{leitherer99}; see section 2.5 of \citet{hopkins18} for more details. $E_{\rm SN}=10^{51}$ erg is the total mechanical energy assumed to be released by a single Type II SN. $v_{\rm cool}=200$ km/s is the terminal velocity of the supernova remnant after it has shocked and swept up ambient ISM material (note that this is lower than the actual injection velocity of $\approx2000$ km/s). We assume the mean SN ejecta mass is $M_{\rm ej}=10M_{\odot}$ of which $2M_{\odot}$ is metal mass (so that the mean SN ejecta metallicity is $Z_{\rm SN}=0.2$). This is equivalent to the \citet{muratov17} approach of defining $\dot{M}_{\rm Z,ref} = y\times$SFR, where they use $y=0.02$ for the chemical yield of one SN per $100M_{\odot}$ stars formed (see their Footnote 4). 

\subsubsection{Redshift-averaged loading factors}
To compute loading factors $\eta$, we cannot simply divide the wind fluxes by their corresponding reference fluxes in a given snapshot because of the time lag between generation and propagation of outflows (see again the time series in the bottom left of \autoref{fig:example} and \autoref{fig:example2}). The bursty nature of SF in dwarfs means that there will be extended periods of zero SF, which can lead to artificially high loading factor estimates (if the time delay and burst integration are not properly accounted for). On the other hand, continuous, steady-state SF in more massive halos at late times also makes it challenging to derive accurate delay times and detect local maxima in the SFH \citep{muratov15,hung19,kim20}. Given the small-scale complexity of our time series and the fact that we are analyzing outflows over $\sim10$ Gyr, we adopt the redshift-averaging approach of \citet{muratov15}. We compute time integrals of the wind and reference fluxes over sufficiently long timescales so as to encompass multiple stellar feedback episodes. This avoids dependence of loading factors on averaging timescale (for sufficiently long timescales). As an example, for mass loading factors, we integrate the total mass of stars formed (as expected from the instantaneous gas particle SFRs) and the total mass of wind blown out over a large redshift interval, and divide the latter by the former. We do a similar calculation for cumulatively summed momentum, energy and metal loading factors.

We define the same three redshift bins as \citet{muratov15}: low-redshift ($z=0.0-0.5$, 5 Gyr), intermediate-redshift ($z=0.5-2.0$, 5 Gyr), and high-redshift ($z=2.0-4.0$, 2 Gyr). Although these redshift bins are extremely long, they have the advantage of giving us a robust estimate of the average loading factor for both the ISM shell and the virial shell, effectively marginalizing over the difference in delay times for $0.1R_{\rm vir}$ and $R_{\rm vir}$. This will allow us to more confidently compare our halo loading factors to our ISM loading factors and constrain any losses/gains in mass, energy, momentum and metals as outflows transit the CGM.

In Appendix \ref{sec:tables}, we provide tabulated measurements of our redshift-averaged loading factors and the galaxy/halo properties that we correlate against in this paper.

\subsubsection{Burst-averaged loading factors}
In addition to our fiducial redshift-averaged loading factors, we compute burst-averaged ISM loading factors for individual outflow episodes.\footnote{We will not attempt to derive individual burst halo-scale loading factors in this paper because the delay time and halo outflow duration can be substantially longer. There may also be significant variation in how different outflow episodes evolve as they transit the CGM.} Any burst-averaged loading factor algorithm must involve three steps: (1) time-shifting, (2) peak detection, and (3) burst integration. We now describe our approach for each of these in turn.

Given that our time series span most of the history of the universe, we adopt the following strategy. For each halo, we first split the whole time series from $z=4$ to $z=0$ ($\sim12$ Gyr) into twelve 1 Gyr chunks. Then we cross-correlate the SFH and mass outflow rate history in each chunk to derive a single time lag for that chunk (using \texttt{numpy.correlate}). We define the time lag as the value at which the cross-correlation function peaks. Since it is unlikely for the time delay to exceed $\sim200$ Myr, we limit the cross-correlation window length to $\pm6$ snapshots, which roughly translates to a $\pm120$ Myr window (given the typical snapshot spacing of $\sim20$ Myr). Our chunking approach allows for the possibility that the time lag can systematically increase towards later times. Indeed, we find that the time lag roughly increases from $\sim20-40$ Myr at $z\sim4$ to $\sim50$ Myr at $z\sim1$, and finally to $\sim100$ Myr at $z\sim0$. \citet[][their Appendix B]{muratov15} found a similar systematic increase in the time lag towards low redshift and suggested it could be because halo radii grow with time while outflow velocities do not increase as dramatically (so outflows take longer to get to $0.1R_{\rm vir}$). Based on visual inspection, we find our simple cross-correlation algorithm to work remarkably well. In dwarfs, SFHs are bursty so the time lag is most easily constrained. In more massive halos, although the SFH is continuous, there can still be peaks (often broad) in the mass outflow rate history that help constrain the cross-correlation. As we will show below, even when the time-shifting is imperfect for individual episodes, our burst integration baselines are usually wide enough to smooth over this error.

Next, in a given chunk, we detect peaks in the shifted mass outflow rate time series using the automated \texttt{scipy.signal.find\_peaks} routine. This is a powerful algorithm that identifies local maxima based on their ``topographic prominence'' (i.e., how the amplitude of a peak compares to the amplitude of its direct neighbors). The function also estimates the peak baseline by extending a horizontal line on both sides of the peak until intersection with part of an even higher peak. For efficiency, we limit the extent of this baseline search window to a total of 8 snapshots (for a total possible burst duration of $\sim160$ Myr). Although the width of an ISM outflow spike is unlikely to exceed $\sim100$ Myr, we find that allowing for this slightly larger max baseline helps correct for any imperfect time shifts due to the single-lag cross-correlation described earlier. All other routine parameters were set to their defaults.

With the peak centers and baselines for outflow spikes in hand, we numerically integrate the references fluxes and (time-shifted) outflow fluxes within each burst window. While our adopted peak detection algorithm performs well (based on visual inspection), any time series analysis is fraught with uncertainty and some filtering criteria must be applied to remove unwanted, noisy detections. For simplicity, we only have two selection criteria for bursts.\footnote{For the m13 halos, we further choose to only include bursts at $z=2-4$ since both the SF and $\dot{M}_{\rm out}$ history are continuous at $z<2$ and it is not clear that the derived time lags are meaningful.} First, we remove outflow episodes where the corresponding burst-integrated stellar mass is zero; these scenarios likely reflect mergers and other inner halo activity. Second, adapting \citet[][their Appendix B]{muratov15}, we only keep bursts whose integrated wind mass is at least 10\% of the wind mass of the most powerful burst within their 1 Gyr time chunk. This choice is inevitably arbitrary but it is designed to pick up the clearer, well-defined and more interesting outflow episodes. While this does mean we have a floor on our burst-averaged outflow fluxes, the loading factors can still be arbitrarily low depending on the starburst strength (in practice, we can recover values as low as $\eta_{\rm M}\sim0.1$ in the low-redshift MW halos, for example). As our results will show, our burst-averaged loading factors also agree remarkably well with our fiducial redshift-averaged measurements. This serves to validate the two very different approaches while also allowing us to get a sense of the scatter in wind loading factor trends.

\begin{figure} 
\begin{center}
\includegraphics[width=\hsize]{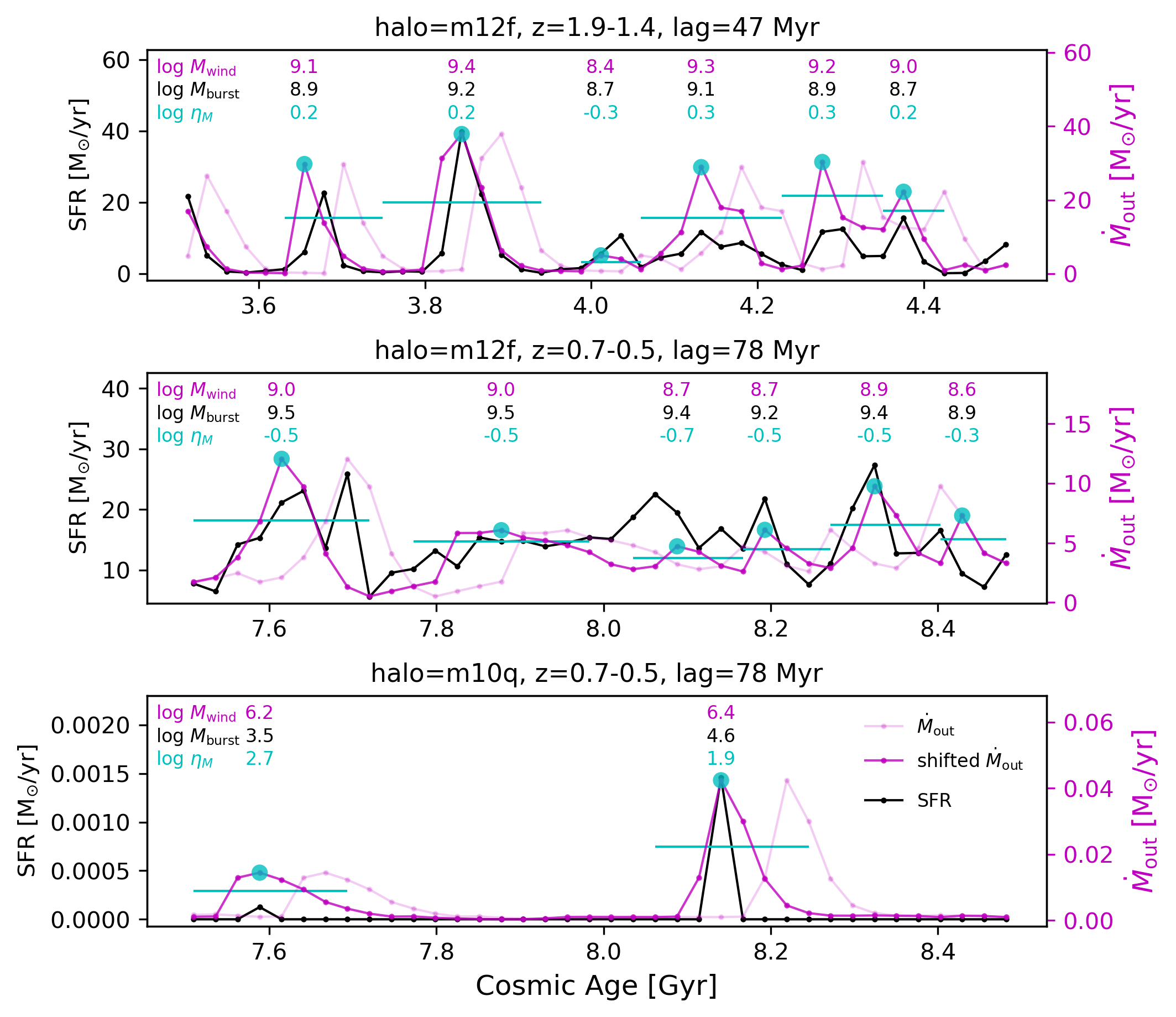}
\end{center}
\caption{Illustration of our automated algorithm for measuring burst-averaged loading factors. These are three representative 1 Gyr time chunks showing m12f at high-redshift when SF is bursty (top), m12f at lower redshift when SF is more continuous (middle), and m10q at the same lower redshift when its SF is still bursty (bottom). In each panel, the SFH is shown in black, the original unshifted mass outflow rate history as transparent magenta, and the time-shifted mass outflow rate history as opaque magenta. The cyan dots and horizontal lines identify peaks and their baselines, respectively. Above each detected peak, we write the burst-integrated wind mass, stellar mass, and burst-averaged mass loading factor. Note how the burst-averaged mass loading is $\sim5-10$ times weaker in m12f at lower redshift compared to high redshift. Note also how highly mass-loaded the two bursts in m10q are despite being at the same lower redshift, and also how far apart these two bursts are in time ($\sim600$ Myr with zero SF in between).} 
\label{fig:instant}
\end{figure}

In \autoref{fig:instant}, we illustrate our time-shifting, peak detection and burst integration results for three representative 1 Gyr time chunks using m12f and m10q as examples. Burst-averaged mass loading factors are found to be $\sim5-10$ times higher in m12f at high-redshift ($z=1.9-1.4$) than at a lower redshift ($z=0.7-0.5$). In the lower redshift chunk, SF is continuous with a non-zero baseline unlike at high-redshift for m12f. However, broad outflow peaks are still apparent and the cross-correlation result seems sensible. In the same lower redshift chunk, m10q only has two starbursts that are spaced far apart (by $\sim600$ Myr) and the outflows are highly mass-loaded with $\eta_M\sim500$ and $80$. To better characterize and understand these trends, we will later correlate all individual detected outflow episodes using their associated burst-averaged physical properties.

\section{ISM wind loading factors}\label{sec:results_ism}
Here we present our ISM loading factors as a function of a few galaxy/halo properties. It is beyond the scope of this paper to identify a ``universal'' halo property (or combination of properties) with which to unambiguously correlate the loading factors. We start with a brief comparison to previous work on mass and metal loading for FIRE-1 halos as a function of both stellar mass and halo virial velocity. We then investigate how all four multi-phase loading factors (mass, momentum, energy and metals) vary with stellar mass and redshift for the FIRE-2 halos. Finally, we correlate our burst-averaged loading factors versus burst interval-averaged gas and SFR surface densities and a few other interesting physical properties.

For completeness, we provide tabulated data in Appendix \ref{sec:tables}, which the reader can use to explore dependence on other global or quasi-local quantities. For purely illustrative purposes, we also provide fitting functions to approximate many of the trends. However, we caution that the scatter is often large and the optimal functional form is not always obvious. For simplicity, we fit (sometimes broken) power laws in log-log space. We use the \texttt{scipy.optimize.curve\_fit} implementation of the Levenberg-Marquardt damped least-squares method (without weighting). We report one standard deviation uncertainties on fitted parameters using the square root of the diagonal entries of the covariance matrix. Unless indicated otherwise, our fits are only done to the broad redshift-averaged measurements and generally include the overly massive m13 halos. In a future work, we will present scalings and quantify scatter in a form that can be implemented into SAMs.

\subsection{Comparison to FIRE-1}
\begin{figure} 
\begin{center}
\includegraphics[width=0.95\hsize]{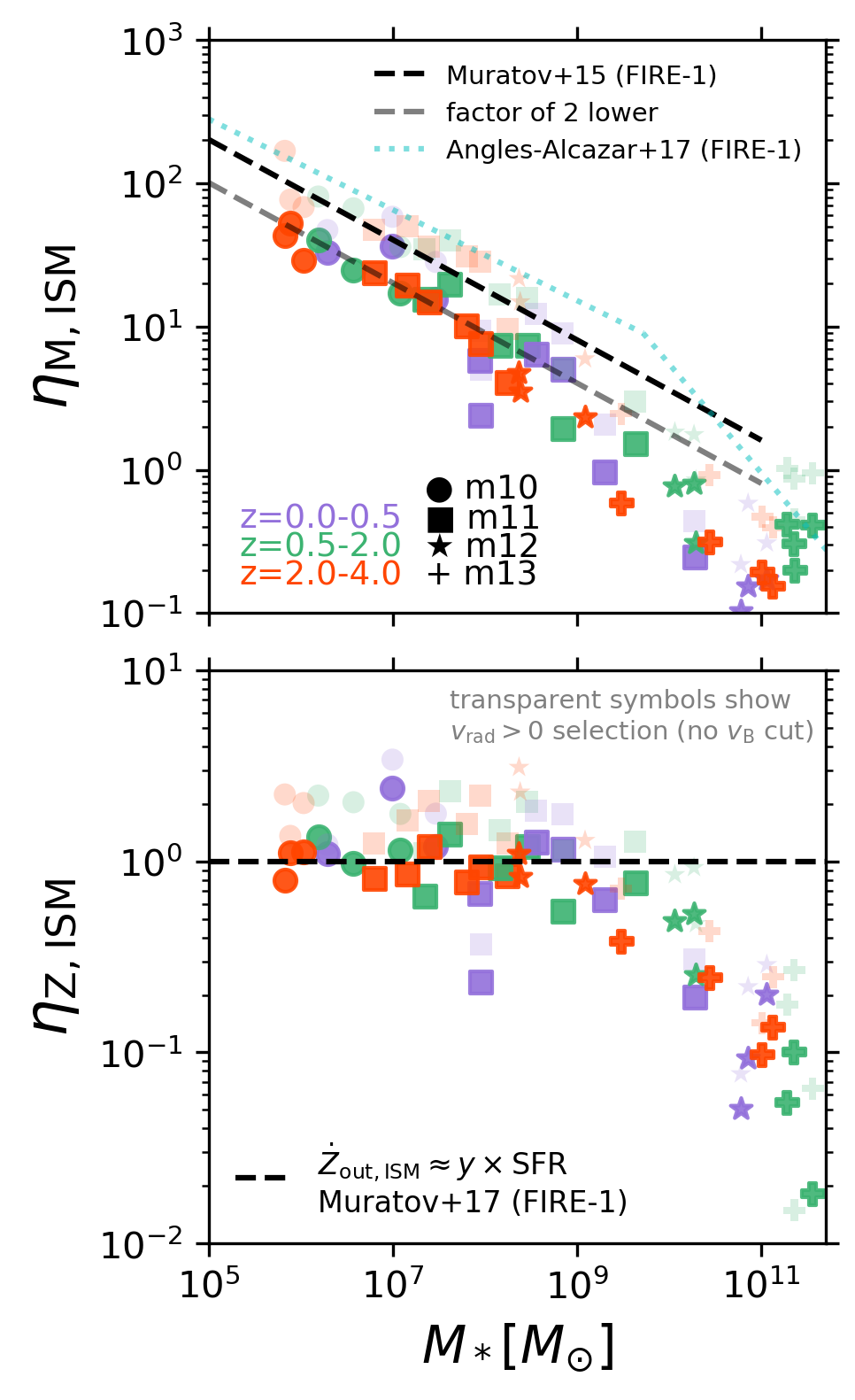}
\end{center}
\caption{Comparison of our ISM-scale mass and metal loading factors as a function of stellar mass to previous FIRE-1 work \citep{muratov15,muratov17,anglesalcazar17}. \textit{Top:} Our fiducial mass loadings are roughly $\sim2\times$ lower than \citet{muratov15}, who found $\eta_{\rm M,ISM}\propto(M_*/M_{\odot})^{-0.35}$ (excluding the low-redshift m12 halos from their fit). This is due to our Bernoulli velocity wind criterion excluding slower, turbulent flows which would otherwise lead to larger mass loadings as in \citet{muratov15}; the transparent symbols show that we would agree with FIRE-1 remarkably well if we use the same wind selection criteria. We also plot the particle tracking-based measurements of mass loadings in FIRE-1 from \citet{anglesalcazar17}, which are even higher since they track outflows directly out of the ISM (much of which recycles back). \textit{Bottom:} Despite our stricter wind selection criterion, our metal loadings agree with \citet{muratov17}, suggesting that the switch from FIRE-1 to FIRE-2 subgrid physics affected outflow metallicities. The transparent symbols show that we would predict even larger metal loadings than FIRE-1 if we use the same wind selection criteria as \citet{muratov17}. Nevertheless, our overall conclusion is similar: nearly all metals produced by Type II SNe are ejected from the ISM of dwarfs but retained within the ISM of more massive galaxies.}
\label{fig:previous}
\end{figure}

In \autoref{fig:previous}, we compare our FIRE-2 measurements of mass and metal loading factors vs. stellar mass to the FIRE-1 results of \citet{muratov15} and \citet{muratov17}, respectively. 

Our mass loading factors are roughly a factor of two lower than \citet{muratov15}, who found a redshift-independent relation with stellar mass: $\eta_{\rm M,ISM}\propto(M_*/M_{\odot})^{-0.35}$. Similar to FIRE-1, our mass loading factors drop off more steeply at $M_*\gtrsim10^{9}M_{\odot}$ \citep[note that the low-redshift m12 halos were not used to fit the FIRE-1 relation;][]{muratov15}. Our lower normalization relative to FIRE-1 is driven by our different particle selection schemes: our Bernoulli velocity wind criterion excludes slower-moving, turbulent flows whereas the simpler $v_{\rm rad}>0$ km/s selection of \citet{muratov15} includes this slow component (and hence leads to upper limits). We have verified that if we use all particles with $v_{\rm rad}>0$ km/s and place the ISM shell at $0.2-0.3R_{\rm vir}$ instead of $0.1-0.2R_{\rm vir}$ \citep[to be even more consistent with][]{muratov15}, then our mass loading factors increase and become remarkably similar to FIRE-1. We also compare to the particle tracking-based measurements of mass loadings in FIRE-1 from \citet{anglesalcazar17}, which are even higher since they tracked outflows directly out of the ISM (much of which recycles back). 

Our metal loading factors agree with FIRE-1 from \citet{muratov17} despite our more stringent wind selection criteria. Had we selected outflows with $v_{\rm rad}>0$ km/s at $0.2-0.3R_{\rm vir}$ instead of $0.1-0.2R_{\rm vir}$ \citep[like][]{muratov17}, we would predict about a factor of two higher metal loading factors than our fiducial measurements. This suggests that although the subgrid physics change from FIRE-1 to FIRE-2 did not greatly affect the overall mass loading factors, there was an appreciable effect on the metal loading and hence metallicity of winds. Nevertheless, our conclusions remain broadly similar to \citet{muratov17}: ISM metal outflows in dwarfs are comparable to the yield of type II SNe (i.e., $\eta_{\rm Z,ISM}\sim1$), with relatively lower ISM metal outflows in the more massive halos.

\subsection{Global halo circular velocity}
\begin{figure} 
\begin{center}
\includegraphics[width=0.9\hsize]{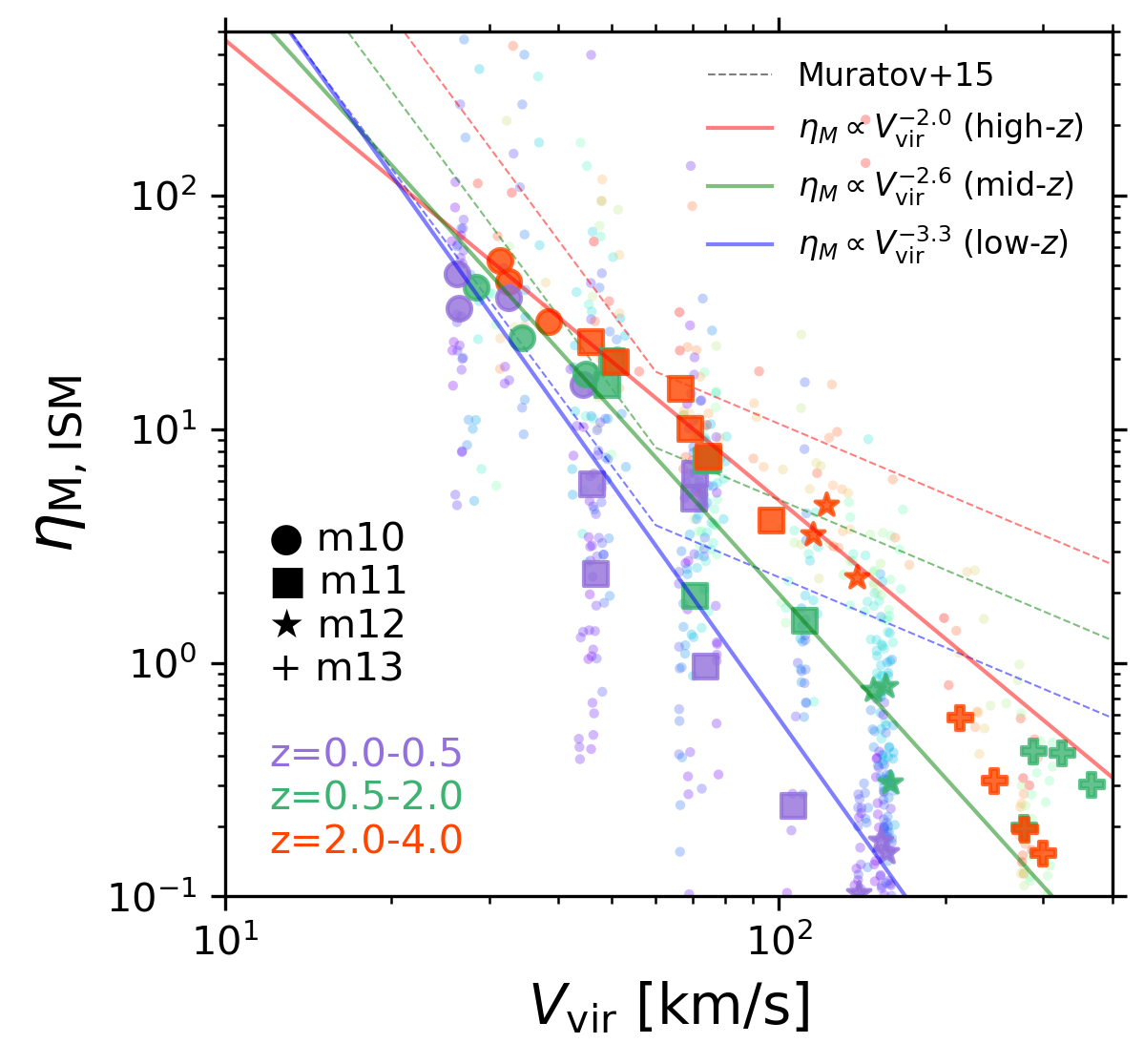}
\end{center}
\caption{ISM mass loading as a function of halo circular velocity. The big markers show our fiducial redshift-averaged measurements whereas the small dots show our individual burst-averaged mass loading measurements color-coded by redshift ($V_{\rm vir}$ for the latter is an $\dot{M}$ weighted average over individual burst intervals). \citet{muratov15} found a shallower slope of ISM mass loading with halo circular velocity for FIRE-1 halos with $V_{\rm vir}>60$ km/s and interpreted it to mean a transition from energy-driven winds in dwarfs to momentum-driven winds in higher mass halos (dotted colored lines). We do not see this flattening with our more stringent outflow selection criteria in FIRE-2 except possibly at the very massive end ($V_{\rm vir}\gtrsim300$ km s$^{-1}$) starting at intermediate redshift. Instead, our measurements are roughly consistent with a single power law that goes as $V_{\rm vir}^{-2}$ at high-redshift (red line), with a steepening at later times (green and blue lines).}
\label{fig:Vvirscaling}
\end{figure}

Next we plot the mass loading as a function of virial velocity in \autoref{fig:Vvirscaling}. We follow the common practice of plotting ISM mass loading versus global halo circular velocity\footnote{Note that we define $V_{\rm vir}$ as the circular velocity at $R_{\rm vir}$ using our own calculated enclosed mass profile accounting for stars, gas and high-resolution dark matter (see \autoref{sec:analysis}). Some studies take $V_{\rm vir}$ directly from a halo finder, but this may not account for the reduced baryon fractions of dwarfs if only the dark matter particles are used.}; we find similar scalings when plotting ISM mass loading versus circular velocity at $0.1R_{\rm vir}$ or halo-scale mass loading versus circular velocity at $R_{\rm vir}$. The theoretical motivation for comparing mass loading to circular velocity is that the power law is expected to be steeper for energy-driven winds ($\eta_M\propto V^{-2}$) and shallower for momentum-driven winds \citep[$\eta_M\propto V^{-1}$); see][]{murray05}. \citet{muratov15} found a very steep slope for the FIRE-1 dwarfs ($\propto V^{-3.2}$), and then a transition to a shallower slope ($\propto V^{-1}$) for more massive halos with $V_{\rm vir}>60$ km/s.

At high-redshift, we find that our measurements follow
\begin{equation}
\eta_{\rm M,ISM} = 10^{4.6}\left(\frac{V_{\rm vir}}{\textrm{km s$^{-1}$}}\right)^{-2.0}\textrm{ for } z=2.0-4.0
\end{equation}
with a coefficient error of $\pm0.2$ dex and power law exponent error of $\pm0.1$ (the m13 halos are excluded from the fit). This is consistent with the expectation for energy-conserving winds. We do not see the need for a broken power law with a shallower slope for more massive halos at high-redshift. If anything, the m13 halos at high-redshift fall off more steeply than expected for a $\propto V_{\rm vir}^{-2}$ scaling, perhaps suggesting they retain more of their outflows in the ISM as fuel for rapid early star formation.

At intermediate redshift, the relation steepens:  
\begin{equation}
\eta_{\rm M,ISM} = 10^{5.5}\left(\frac{V_{\rm vir}}{\textrm{km s$^{-1}$}}\right)^{-2.6}\textrm{ for } z=0.5-2.0
\end{equation}
with a coefficient error of $\pm0.5$ dex and power law exponent error of $\pm0.2$ (again excluding the m13 halos). The relation steepens even further by low redshift: 
\begin{equation}
\eta_{\rm M,ISM} = 10^{6.4}\left(\frac{V_{\rm vir}}{\textrm{km s$^{-1}$}}\right)^{-3.3}\textrm{ for } z=0.0-0.5
\end{equation}
with errors of $\pm0.6$ dex and $\pm0.3$ for the coefficient and power law exponent, respectively. There is a hint that a broken power law may be appropriate at intermediate-redshift given the elevated $\eta_{\rm M,ISM}$ of the m13 halos, but this would be at a much higher pivot point ($\gtrsim200$ km s$^{-1}$) than the 60 km s$^{-1}$ found by \citet{muratov15}. The MW halos follow our simple unbroken scalings at both intermediate- and low-redshift. As we will discuss later, the stronger redshift dependence when plotting against halo virial velocity instead of stellar mass may reflect the fact that the stellar-to-halo-mass ratio, at fixed halo mass, gets larger at later times whereas $V_{\rm vir}$ does not evolve as dramatically (this is particularly true for the massive halos).

\subsection{Multi-phase ISM mass loadings}
\begin{figure*} 
\begin{center}
\includegraphics[width=0.95\hsize]{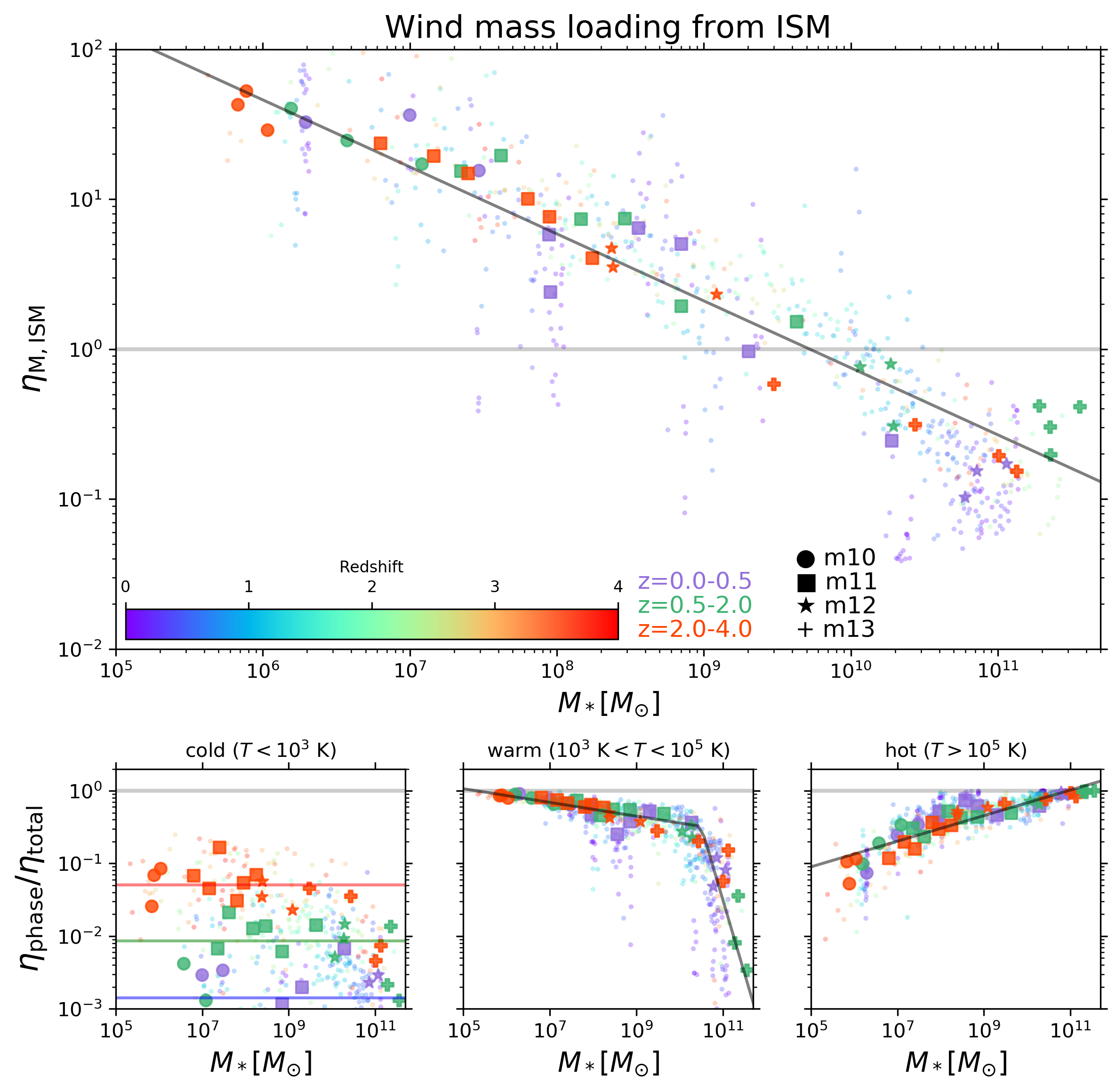}
\end{center}
\caption{Evolution of multi-phase ISM mass loading factors with stellar mass. The bigger markers show our fiducial redshift-averaged measurements and the smaller dots are our individual burst-averaged measurements color-coded by redshift ($M_*$ for the latter is an $\dot{M}$ weighted average over each individual burst interval, whereas for the former it is simply the unweighted mean $M_*$ over each redshift interval). The horizontal gray line denotes order unity and the other lines show approximate fits (see text). \textit{Top:} For all outflow phases combined, $\eta_{\rm M}$ is less than 1 for massive halos and rises to $\sim100$ for dwarfs. \textit{Bottom-left:} The fraction of mass loading in the cold phase is far less than one but correlates strongly with redshift. High redshift halos can have $\sim10\%$ of their mass loading in the cold phase, but this drops to less than $1\%$ for most halos at later times. \textit{Bottom-middle:} the warm outflow phase dominates by mass fraction in the dwarfs. Note also the much tighter correlation and lack of redshift dependence compared to the cold phase. \textit{Bottom-right:} the hot phase dominates in the massive halos but steadily drops off toward lower mass halos. Interestingly, when the total $\eta_{\rm M,ISM}<1$, the hot phase dominates, whereas in the dwarfs with $\eta_{\rm M,ISM}\gg1$ the warm phase dominates.}
\label{fig:etaM_all}
\end{figure*}

In \autoref{fig:etaM_all}, we plot multi-phase ISM mass loading factors versus stellar mass. For total loadings (all phases combined), we plot the actual loading factors whereas for the individual phases we plot fractions for clarity (i.e., $\eta_{\rm phase} / \eta_{\rm total}$). 

The topmost panel is similar to \autoref{fig:previous}: when combining all phases, dwarfs have significantly higher mass loading factors than more massive halos. The total mass loadings in low-mass dwarfs are of order $\sim100$, steadily dropping towards $\sim10$ for intermediate-mass dwarfs and becoming less than unity for the m12 and m13 halos (despite the latter being at high redshift). The total mass loadings approximately follow
\begin{equation}
\eta_{\rm M,ISM} = 10^{4.3} (M_*/M_{\odot})^{-0.45}
\end{equation}
with errors of $\pm0.2$ dex and $\pm0.02$ for the coefficient and power law exponent, respectively.\footnote{Unless indicated otherwise, all fits are based on the broad redshift-averaged measurements and do not include the individual burst-averaged points.} At low redshift, a few of the m11 halos and all three m12 halos are a factor of a few below our approximate fit \citep[see also FIRE-1;][]{muratov15,hayward17}. This may reflect the deeper potential wells at later times as well as the changing structure of the ISM and inner CGM over time, as we will discuss in \autoref{sec:discussion}. 

The cold mass loading fractions correlate strongly with redshift but are generally flat with stellar mass (at a given redshift). High redshift dwarfs (including the progenitors of the MW halos) have cold mass loading fractions approaching $\sim0.1$. At lower redshifts, the cold mass loading fractions drop to $0.01$ or less. We find that the cold mass loading fractions can be approximated simply as
\begin{equation}
f_{\rm M,cold} \equiv \frac{\eta_{\rm M,ISM,cold}}{\eta_{\rm M,ISM}} = 10^{-3.2} (1+z)^{3.1}
\end{equation}
with errors of $\pm0.1$ dex and $\pm0.3$ for the coefficient and power law exponent, respectively. We have excluded the m13 halos from the fits.

By comparison, the warm and hot phases show much less scatter. In the m13 and $z\sim0$ MW halos, the hot phase carries most of the outflowing mass. In contrast, the warm phase carries most of the outflow mass in dwarfs, including the high-redshift MW progenitor dwarfs. One possible reason for this is that winds in dwarfs sweep up significant amounts of ambient gas, and this ambient gas may roughly track the halo virial temperature which in dwarfs would fall in our warm regime ($T_{\rm vir}<10^5$ K). Hence, while the winds in dwarfs are predominantly warm in an atomic cooling sense, they are still ``dynamically hot'' in a global thermodynamics sense.\footnote{Another possible reason is that the higher resolution in the dwarfs may lead to better resolved radiative cooling, especially since their virial temperatures are close to the peak of the cooling curve. However, since the overall densities and metallicities would still be rather low in dwarfs, radiative cooling may be negligible \citep[e.g.,][]{lochhaas21}. Indeed, we can infer this from the high energy loading factors presented below.} In the high-redshift dwarf progenitors of MW halos, the cold mass loadings are comparable to the hot mass loadings, yet the warm mass loadings dominate over the other two phases by a factor of $\sim10$. There is no strong redshift dependence for the warm and hot mass loading fractions, but the warm mass loading fractions drop significantly at high stellar masses. We fit a broken power law to the warm mass loading fractions with the break fixed at $10^{10.5}M_{\odot}$:

\begin{equation}\scriptsize
f_{\rm M,warm} \equiv \frac{\eta_{\rm M,ISM,warm}}{\eta_{\rm M,ISM}} = 
\begin{cases} 
10^{-0.5} (M_*/10^{10.5}M_{\odot})^{-0.09} & \textrm{for } M_*\lesssim10^{10.5}M_{\odot}\\
10^{-0.5} (M_*/10^{10.5}M_{\odot})^{-2.0} & \textrm{for } M_*\gtrsim10^{10.5}M_{\odot}
\end{cases}
\end{equation}
with errors of $\pm0.1$ dex for the coefficient, $\pm0.05$ for the low-mass exponent and $\pm0.3$ for the high-mass exponent. For the hot mass loading fractions, we assume a single power law:  
\begin{equation}
f_{\rm M,hot} \equiv \frac{\eta_{\rm M,ISM,hot}}{\eta_{\rm M,ISM}} = 10^{-1.9} (M_*/M_{\odot})^{0.18}
\end{equation}
with errors of $\pm0.1$ dex for the coefficient and $\pm0.01$ for the power law exponent.

\subsection{Multi-phase ISM momentum loadings}
\begin{figure*} 
\begin{center}
\includegraphics[width=0.95\hsize]{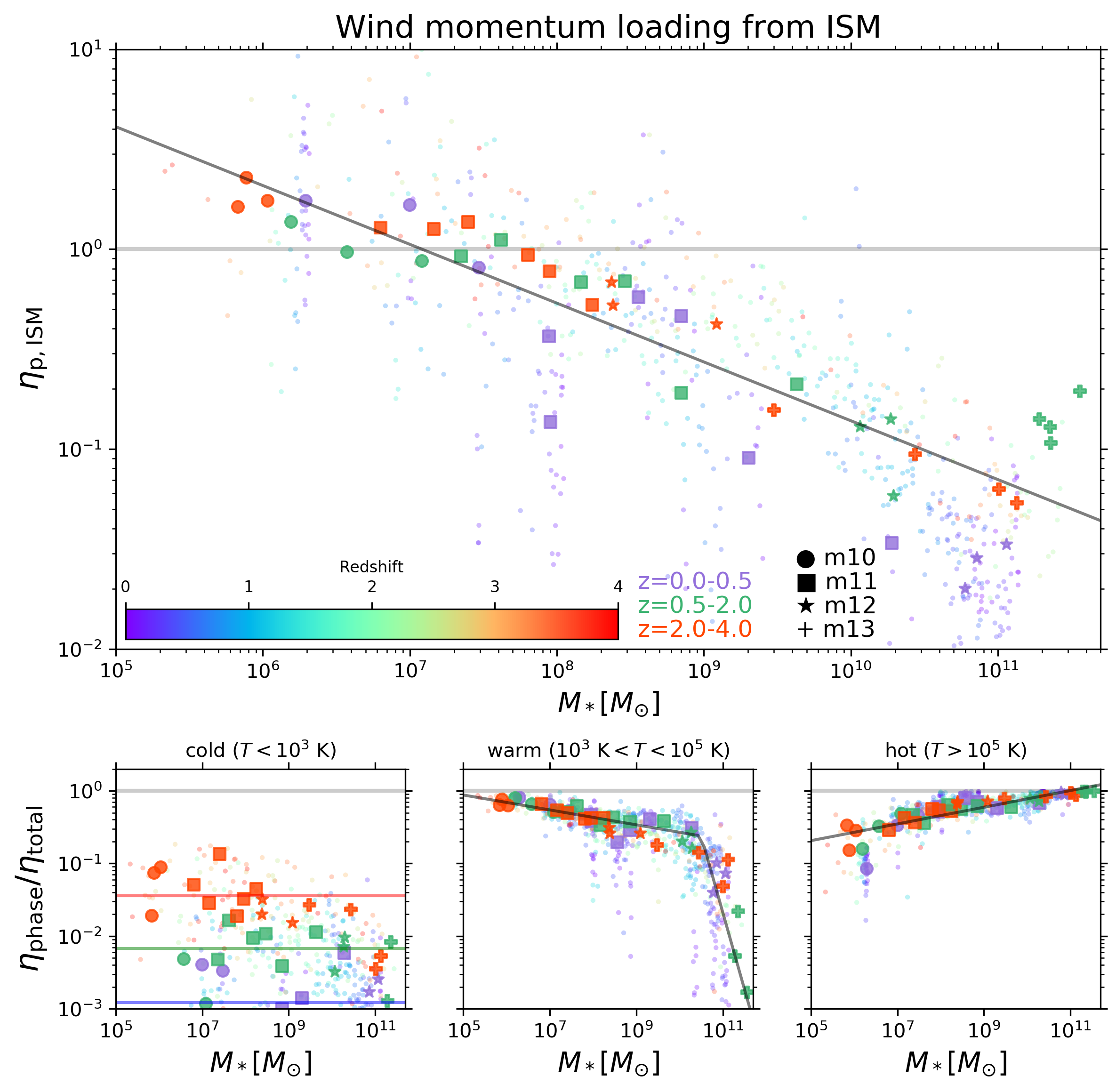}
\end{center}
\caption{Similar to \autoref{fig:etaM_all} but now for the ISM momentum loading factor. \textit{Top:} The overall momentum loading factor is of order unity for higher redshift dwarfs, dropping to $\sim0.1$ for some dwarfs at $z\sim0$. For massive halos, overall momentum loadings are generally less than 0.1. \textit{Bottom-left:} The fraction of momentum loading in the cold phase is negligible except for high redshift dwarfs where it can approach $\sim10\%$. \textit{Bottom-middle:} The warm phase carries nearly all of the momentum in the lowest mass dwarfs, gradually dropping to $\lesssim10\%$ for more massive halos. \textit{Bottom-right:} In contrast, the hot phase carries nearly all of the momentum in massive halos, gradually dropping to $\sim10\%$ for the lowest mass dwarfs.}
\label{fig:etap_all}
\end{figure*}

In \autoref{fig:etap_all}, we show the multi-phase ISM momentum loading factors versus stellar mass. With all phases combined, the total momentum loadings are $\lesssim0.1$ for the MW halos at $z\sim0$ as well as the m13 halos at high-redshift. In contrast, the momentum loadings are of order unity in the dwarfs, including the high-redshift progenitors of MW halos. For the lowest mass halos, the momentum loading exceeds one, which may reflect some SN/superbubble clustering phenomenon \citep[e.g.,][]{gentry17,fielding18,fauchergiguere18}. There is some scatter in the momentum loadings of dwarfs but averaged over long timescales their values exceed those of the more massive halos by about a factor of ten. We approximate the scaling between total momentum loading and stellar mass as
\begin{equation}
\eta_{\rm p,ISM} = 10^{2.1} (M_*/M_{\odot})^{-0.29}
\end{equation}
with errors of $\pm0.2$ dex for the coefficient and $\pm0.02$ for the power law exponent. 

Splitting by phase, the cold momentum loadings are negligible in low-redshift dwarfs, MW halos at $z\sim0$ and the m13 halos. However, cold momentum loading fractions are more substantial in high redshift dwarfs: the progenitors of MW halos have values of a few percent whereas some lower mass dwarfs exceed $10\%$. A simple redshift-dependent formula can reasonably approximate the cold momentum loading fractions (excluding the m13 halos):
\begin{equation}
f_{\rm p,cold} = 10^{-3.2} (1+z)^{2.9}
\end{equation}
with errors of $\pm0.1$ dex for the coefficient and $\pm0.3$ for the power law exponent. 

The warm and hot momentum loading fractions are significantly higher than the cold momentum loadings for all galaxies considered. For the more massive halos (including the MW halos at $z\sim0$), the hot momentum loading fractions are much larger than the warm momentum loading fractions and approach order unity. In intermediate-mass dwarfs, the hot and warm momentum loading fractions are comparable, and in the lowest mass dwarfs the warm phase carries nearly all of the momentum. The importance of warm momentum loading in low-mass dwarfs may be related to their virial temperatures being lower than $10^5$ K: their outflows may not satisfy our lower limit for hot temperatures but may still be ``dynamically hot.'' We approximate our trends for the warm phase using a broken power law with the break fixed at $10^{10.5}M_{\odot}$:
\begin{equation}\footnotesize
f_{\rm p,warm} = 
\begin{cases} 
10^{-0.6} (M_*/10^{10.5}M_{\odot})^{-0.10} & \textrm{for } M_*\lesssim10^{10.5}M_{\odot}\\
10^{-0.6} (M_*/10^{10.5}M_{\odot})^{-2.1} & \textrm{for } M_*\gtrsim10^{10.5}M_{\odot}
\end{cases}
\end{equation}
with errors of $\pm0.1$ dex for the coefficient, $\pm0.05$ for the low-mass exponent and $\pm0.3$ for the high-mass exponent. For the hot phase, we find simply
\begin{equation}
f_{\rm p,hot} = 10^{-1.3} (M_*/M_{\odot})^{0.12}
\end{equation}
with a coefficient error of $\pm0.1$ dex and power law exponent error of $\pm0.01$.

\subsection{Multi-phase ISM energy loadings}
\begin{figure*} 
\begin{center}
\includegraphics[width=0.95\hsize]{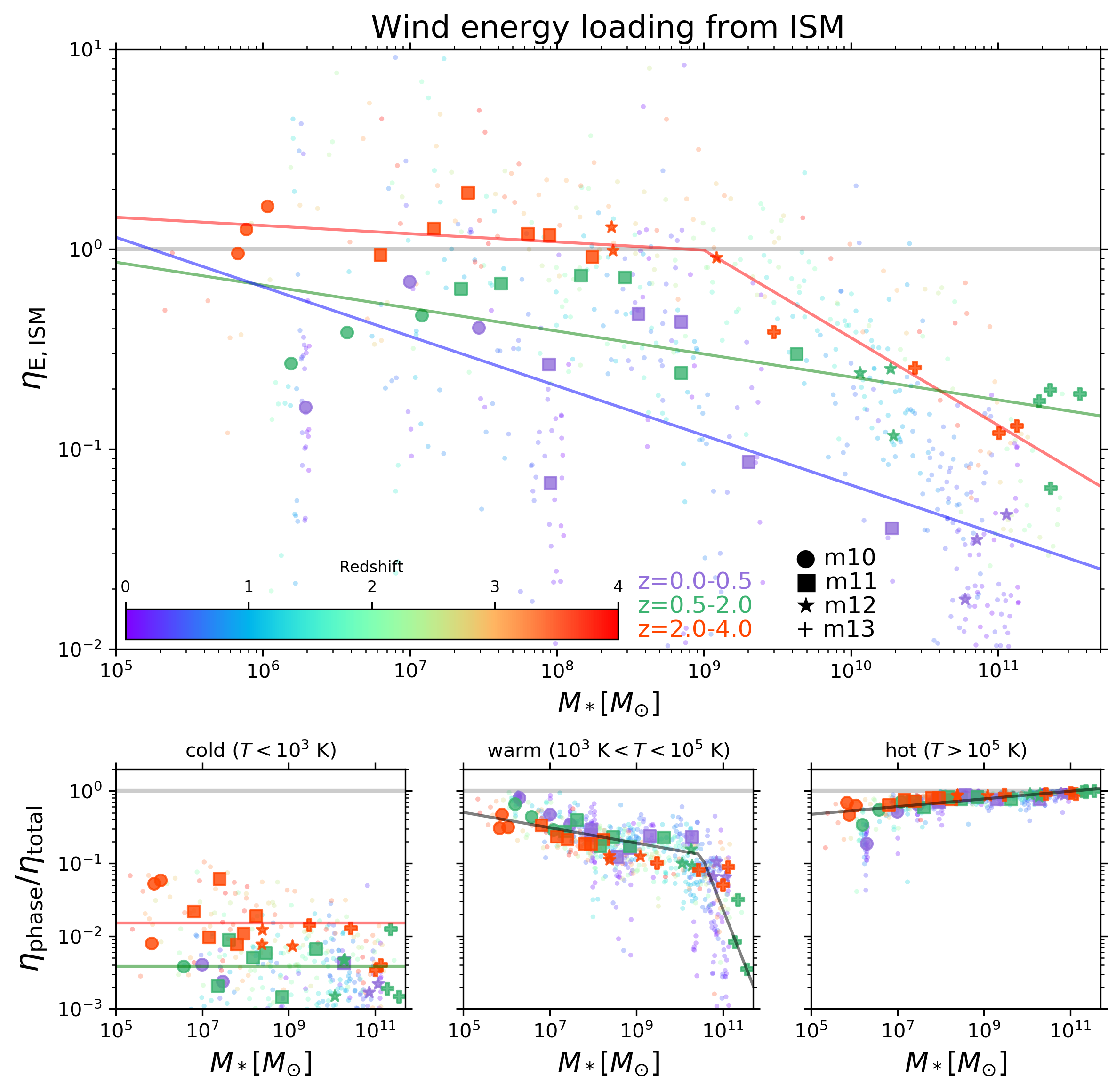}
\end{center}
\caption{Similar to \autoref{fig:etaM_all} but now for the evolution of ISM energy loading factor. \textit{Top:} Overall energy loadings are of order unity for dwarfs at high-redshift, and between $0.1$ and 1 for lower redshift dwarfs. MW and m13 halos have $\eta_E\sim0.1$ especially at lower redshift. \textit{Bottom-left:} The fraction of energy loading carried by the cold phase is negligible, except in some low-mass dwarfs at high-redshift where it is almost $\sim10\%$ (signifying large kinetic energy). \textit{Bottom-middle:} The warm phase carries only about $\sim10\%$ of the energy loadings in massive halos but becomes increasingly important for low-mass dwarfs, where its fractional contribution approaches $\sim50\%$ or higher. \textit{Bottom-right:} The hot phase carries $\sim50\%$ of the energy in the lowest mass dwarfs and effectively becomes the sole carrier of energy in massive halos ($f_{\rm hot}\sim100\%$).}
\label{fig:etaE_all}
\end{figure*}

In \autoref{fig:etaE_all}, we plot multi-phase ISM energy loading factors versus stellar mass. When we consider all phases combined, the total ISM energy loadings are less than $0.1$ in the MW halos at low-redshift. The same is true for the m13 halos at both intermediate and high redshift, which have similarly low energy loadings. In contrast, dwarfs at high redshift have energy loadings of order unity. At lower redshifts, dwarfs show more scatter in their total energy loadings, but still maintain preferentially higher energy loadings compared to the massive halos (generally $\eta_{\rm E,ISM}\gtrsim0.2$). Taking into account this complicated redshift and mass dependence, we parameterize the energy loadings as a broken power law at high-redshift (with the break point fixed at $10^9M_{\odot}$) and two distinct power laws for the other two redshift bins: 
\begin{equation}\scriptsize
\eta_{\rm E,ISM} = 
\begin{cases} 
10^{1.3} (M_*/M_{\odot})^{-0.25} & \textrm{for } z=0.0-0.5\\
10^{0.5} (M_*/M_{\odot})^{-0.11} & \textrm{for } z=0.5-2.0\\
10^{-0.005} (M_*/10^9M_{\odot})^{-0.04} & \textrm{for } z=2.0-4.0 \;\&\; M_*\lesssim10^9M_{\odot}\\
10^{-0.005} (M_*/10^9M_{\odot})^{-0.44} & \textrm{for } z=2.0-4.0 \;\&\; M_*\gtrsim10^9M_{\odot}\;.
\end{cases}
\end{equation}
The errors for the low-redshift power law are $\pm0.6$ dex for the coefficient and $\pm0.07$ for the exponent. The errors for the intermediate-redshift power law are $\pm0.3$ dex for the coefficient and $\pm0.03$ for the exponent. As for the high-redshift broken power law, the errors are $\pm0.05$ dex (coefficient), $\pm0.03$ (low-mass exponent), and $\pm0.05$ (high-mass exponent). 

Splitting by phase, the cold energy loading fractions are negligible in all halos compared to the warm and hot energy loading fractions, although the scatter in cold energy loading fractions correlates positively with redshift. Just as for the cold mass and momentum loading fractions, we can approximate the redshift dependence in a simple way (again, excluding the m13 halos):
\begin{equation}
f_{\rm E,cold} = 10^{-3.3} (1+z)^{2.4}
\end{equation}
with errors of $\pm0.1$ dex (coefficient) and $\pm0.3$ (exponent).

The hot energy loading fractions dominate over the warm energy loading fractions by about an order of magnitude for the MW halos, their high redshift dwarf progenitors, and the m13 high-redshift halos. In contrast, a substantial fraction of energy is carried by the warm phase in lower mass halos. We approximate the warm energy loading fractions as a broken power law with the break fixed at $10^{10.5}M_{\odot}$: 
\begin{equation}
f_{\rm E,warm} = 
\begin{cases} 
10^{-0.9} (M_*/10^{10.5}M_{\odot})^{-0.11} & \textrm{for } M_*\lesssim10^{10.5}M_{\odot}\\
10^{-0.9} (M_*/10^{10.5}M_{\odot})^{-1.5} & \textrm{for } M_*\gtrsim10^{10.5}M_{\odot}
\end{cases}
\end{equation}
where the errors are $\pm0.1$ dex (coefficient), $\pm0.05$ (low-mass exponent) and $\pm0.3$ (high-mass exponent). For the hot energy loading fractions, we find
\begin{equation}
f_{\rm E,hot} = 10^{-0.60} (M_*/M_{\odot})^{0.053}
\end{equation}
where the errors are $\pm0.07$ dex (coefficient) and $\pm0.008$ (power law exponent).

\subsection{Multi-phase ISM metal loadings}
\begin{figure*} 
\begin{center}
\includegraphics[width=0.95\hsize]{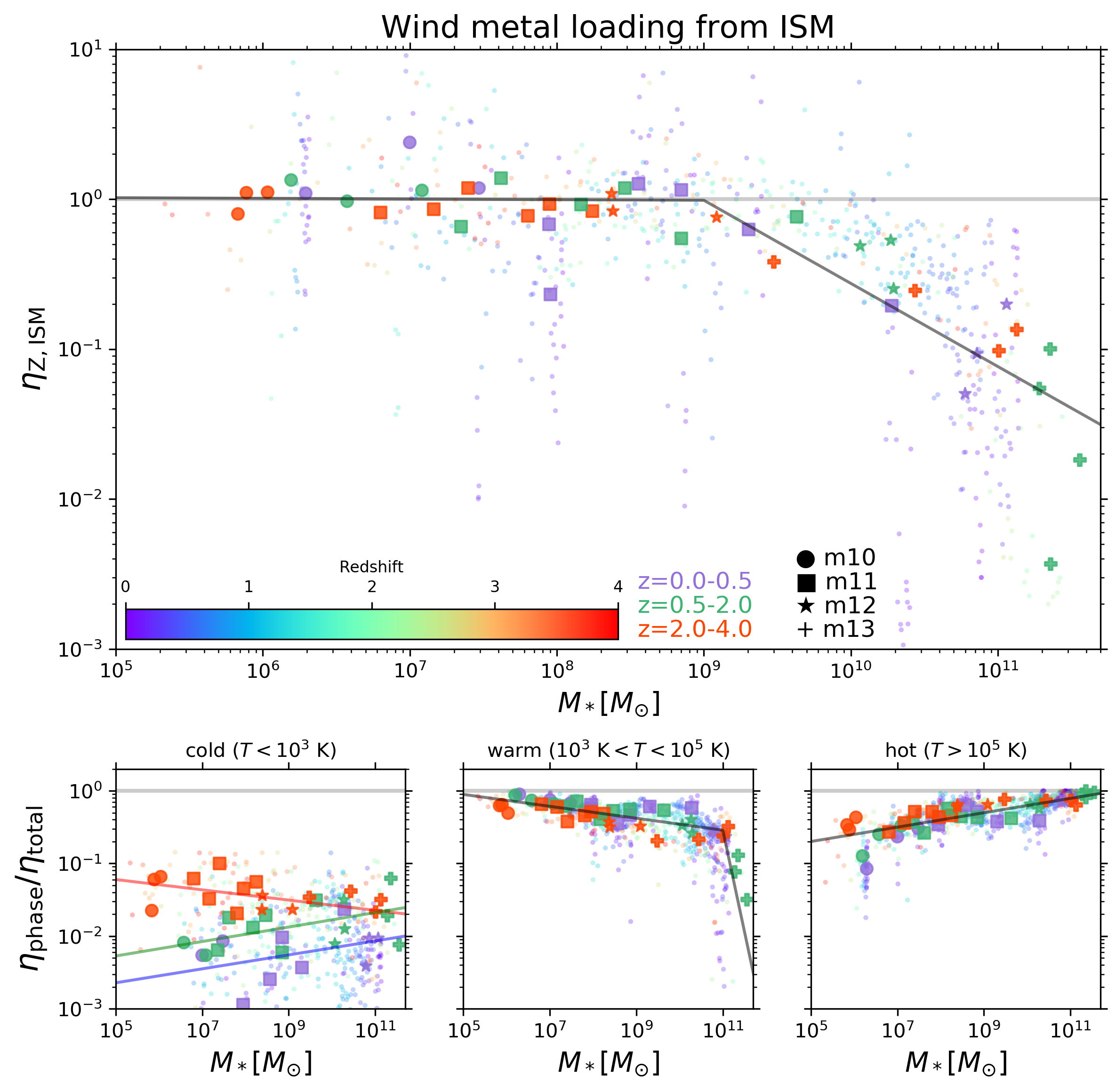}
\end{center}
\caption{Similar to \autoref{fig:etaM_all} but now showing evolution of the ISM metal loading factor. \textit{Top:} Overall metal loadings are of order unity for dwarfs, i.e., nearly all metals produced by SNe are ejected from the ISM of dwarfs. In contrast, the SN metal yield is mostly retained within the ISM of massive halos. \textit{Bottom-left:} The fraction of metals carried by the cold phase is generally negligible for all halos except in high-redshift dwarfs and intermediate-redshift massive halos (for which $f_{\rm cold}\sim10\%$). \textit{Bottom-middle:} The warm phase carries nearly all of the metals in the lowest mass dwarfs, and becomes progressively less important for more massive halos, dropping to $\sim10\%$. \textit{Bottom-right:} In contrast, the hot phase carries nearly all of the metals in massive halos compared to only $\sim10\%$ in the lowest mass halos.}
\label{fig:etaZ_all}
\end{figure*}

In \autoref{fig:etaZ_all}, we plot multi-phase ISM metal loading factors versus stellar mass. Similar to \autoref{fig:previous}, with all phases combined the total ISM metal loadings are of order unity in dwarfs at all redshifts (i.e., including progenitors of MW halos). However, in more massive halos, the ISM metal loadings drop steadily to $\sim0.1$ when averaged over long timescales. There is no strong redshift dependence for the total metal loadings, allowing us to simply parameterize the trends with halo mass using a broken power law (with the break point fixed at $10^9M_{\odot}$):
\begin{equation}
\eta_{\rm Z,ISM} = 
\begin{cases} 
10^{-0.005} (M_*/10^9 M_{\odot})^{-0.04} & \textrm{for } M_*\lesssim10^9M_{\odot}\\
10^{-0.005} (M_*/10^9 M_{\odot})^{-0.44} & \textrm{for } M_*\gtrsim10^9M_{\odot}\;.
\end{cases}
\end{equation}
The errors are $\pm0.05$ dex (coefficient), $\pm0.03$ (low-mass exponent) and $\pm0.05$ (high-mass exponent).

Splitting by phase, metals carried by the cold phase are negligible overall but there is a strong redshift dependence. At high redshift, all halos have roughly constant cold metal loading fractions of $\approx0.05$. At later times, the cold metal loading fractions decrease but there seems to be a positive correlation with stellar mass. Excluding the m13 halos, we parameterize the cold metal loading fractions as 
\begin{equation}
f_{\rm Z,cold} = 
\begin{cases} 
10^{-3.1} (M_*/M_{\odot})^{0.10} & \textrm{for } z=0.0-0.5\\
10^{-2.8} (M_*/M_{\odot})^{0.10} & \textrm{for } z=0.5-2.0\\
10^{-0.9} (M_*/M_{\odot})^{-0.07} & \textrm{for } z=2.0-4.0\;.
\end{cases}
\end{equation}
The errors for the low-redshift power law are $\pm0.8$ dex (coefficient) and $\pm0.08$ (exponent). The errors for both the intermediate- and high-redshift power laws are the same: $\pm0.5$ dex (coefficient) and $\pm0.06$ (exponent).

The hot metal loading fraction is of order unity and the warm metal loading fraction is of order $0.1$ in more massive halos. In contrast, for the lowest mass halos, the warm phase carries nearly all of the metals (the hot metal loading fraction drops to order $0.1$). We fit a broken power law to the warm metal loading fractions assuming a fixed break at $10^{11}M_{\odot}$: 
\begin{equation}
f_{\rm Z,warm} = 
\begin{cases} 
10^{-0.5} (M_*/10^{11}M_{\odot})^{-0.08} & \textrm{for } M_*\lesssim10^{11}M_{\odot}\\
10^{-0.5} (M_*/10^{11}M_{\odot})^{-2.8} & \textrm{for } M_*\gtrsim10^{11}M_{\odot}
\end{cases}
\end{equation}
where the errors are $\pm0.1$ dex (coefficient), $\pm0.04$ (low-mass exponent), and $\pm0.5$ (high-mass exponent). For the hot metal loading fractions, we find: 
\begin{equation}
f_{\rm Z,hot} = 10^{-1.2} (M_*/M_{\odot})^{0.10}
\end{equation}
where the errors are $\pm0.1$ dex (coefficient) and $\pm0.01$ (power law exponent).

\subsection{Trends with SFR and ISM gas mass surface densities}
\begin{figure} 
\begin{center}
\includegraphics[width=0.9\hsize]{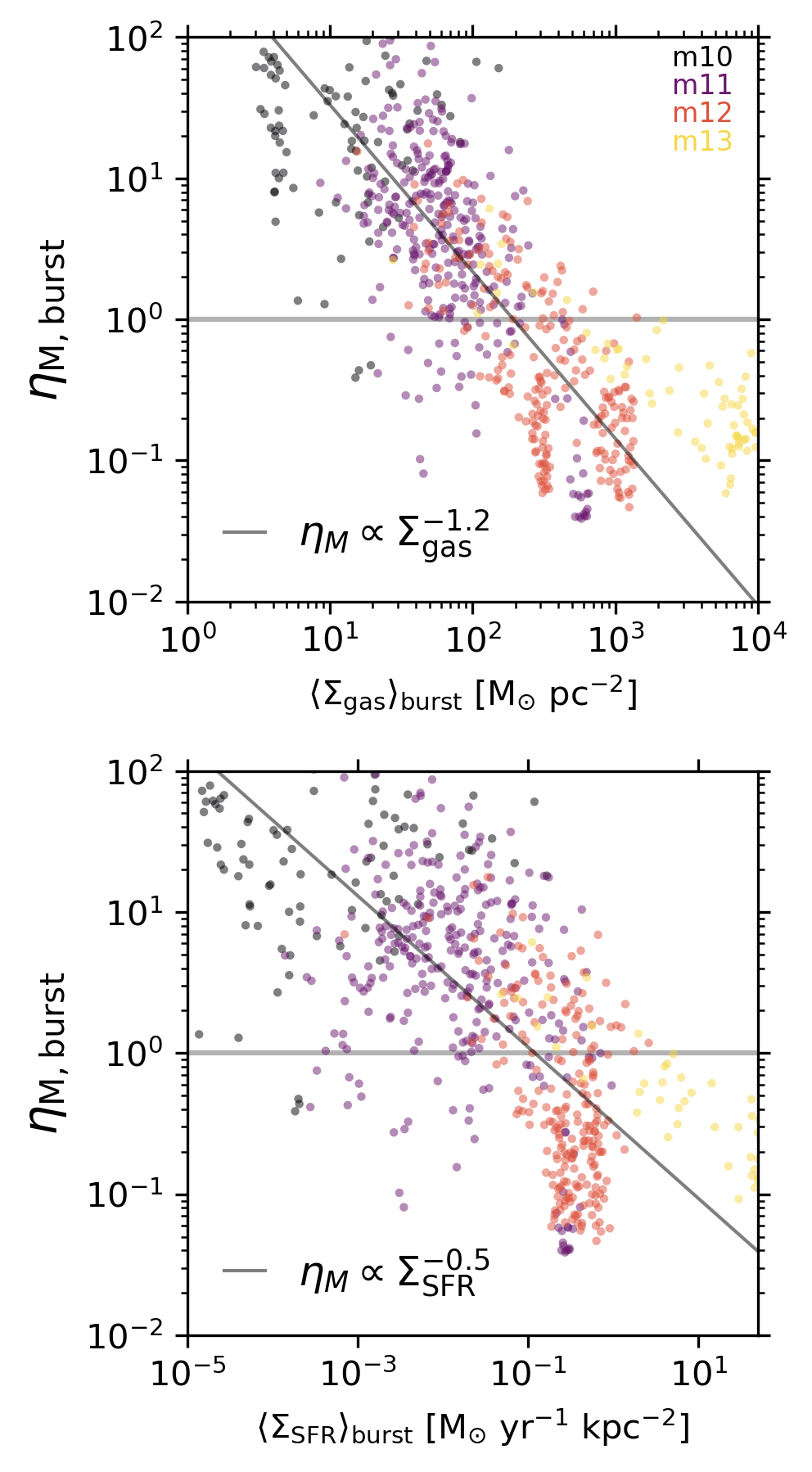}
\end{center}
\caption{Burst-averaged ISM mass loading factor versus $\dot{M}$-weighted average gas mass surface density (top) and SFR surface density (bottom) within individual burst windows. There is a clear negative correlation with gas mass surface density such that $\eta_{\rm M}\propto\Sigma_{\rm gas}^{-1.2}$ (diagonal black line; excluding the m13 halos from the fit). The trend with SFR surface density follows $\eta_{\rm M}\propto\Sigma_{\rm SFR}^{-0.5}$ but the deviation from a simple power law is more apparent.}
\label{fig:resolved}
\end{figure}

In \autoref{fig:resolved}, we plot our burst-averaged burst-integrated mass loading factor as a function of the $\dot{M}$-weighted average $\Sigma_{\rm gas}$ and $\Sigma_{\rm SFR}$ within individual burst windows. $\Sigma_{\rm gas}$ is defined as $M_{\rm gas}/(\pi R_{\rm 3D}^2)$ where $M_{\rm gas}$ is the mass of all gas particles within $0.1R_{\rm vir}$ and $R_{\rm 3D}$ is the 3D stellar half-mass radius (a commonly used definition of galaxy size) computed using star particles within $0.1R_{\rm vir}$. $\Sigma_{\rm SFR}$ is defined similarly except the numerator is the instantaneous SFR as described in \autoref{sec:loadings}. Hence we are assuming, for simplicity, that the ISM gas and star formation within $0.1R_{\rm vir}$ are mostly confined to a flat disk of radius $R_{\rm 3D}$.

Burst-averaged mass loadings drop off as $\Sigma_{\rm gas}$ increases. They also drop off with increasing $\Sigma_{\rm SFR}$ although there is more scatter, especially at low $\Sigma_{\rm SFR}$. The bursts in the m12 halos (red points) show a clear evolution from high mass loadings at low $\Sigma_{\rm gas}$ (i.e., in their dwarf progenitors) to low mass loadings of $\sim0.1$ at high $\Sigma_{\rm gas}$ (i.e., at low redshift).
Most of the bursts in the m13 halos occur at rather large surface densities since these halos were already quite massive by $z=2-4$. For purely illustrative purposes, we parameterize the trends as
\begin{equation}
\eta_{\rm M} = 10^{2.71} \left(\frac{\Sigma_{\rm gas}}{M_{\odot}\, \textrm{pc}^{-2}}\right)^{-1.18}
\end{equation}
and 
\begin{equation}
\eta_{\rm M} = 10^{-0.49} \left(\frac{\Sigma_{\rm SFR}}{M_{\odot}\, \textrm{yr}^{-1}\, \textrm{kpc}^{-2}}\right)^{-0.54}\;.
\end{equation}
The errors for the $\Sigma_{\rm gas}$ scaling are $\pm0.08$ dex (coefficient) and $\pm0.04$ (power law exponent). The errors for the $\Sigma_{\rm SFR}$ scaling are $\pm0.05$ dex (coefficient) and $\pm0.02$ (power law exponent).

\subsection{SF burstiness, dense ISM gas fractions and inner CGM virialization}
The previous global correlations with $M_*$ and $V_{\rm vir}$ are not satisfying in terms of painting a physical picture because they do not address how small-scale ISM conditions may influence the initial properties of winds during breakout. This interpretation-related ambiguity remains even in cases where the global correlations appear statistically strong with minimal scatter. On the other hand, the burst-averaged loading factor trends (or lack thereof) with $\Sigma_{\rm gas}$ and $\Sigma_{\rm SFR}$ are also not sufficiently informative because they lack a proper normalization and hence physical context. Although we cannot establish causality with the FIRE-2 dataset (that would require controlled numerical experiments), we can at least correlate our burst-averaged loading factors against a few relevant ``derived'' physical properties. In this first attempt, we choose to only focus on the following three for simplicity. Do more powerful starbursts (relative to the average SFR over a longer time window) drive winds that are more highly mass-loaded? Are burst-averaged loading factors higher when dense ISM gas fractions are lower since that may enable winds to break out without as much impedance? Does the virialization of the inner halo correlate with the strength of ISM winds? The following is a brief heuristic and empirical exploration of these three questions.

To quantify starburst strength (or ``local burstiness'') for each individual outflow episode, we divide the maximum SFR within the burst window by the 1 Gyr-averaged SFR (i.e., within each burst's overall time chunk). The dense ISM gas fraction is computed as $f_{\rm dense}=M_{\rm gas,dense}/M_{\rm gas}$ where $M_{\rm gas,dense}$ is the mass of all gas particles within $0.1R_{\rm vir}$ that have density $n>1000$ cm$^{-3}$ (this is the SF density threshold in FIRE-2).\footnote{Our $f_{\rm dense}$ statistic is almost certainly too simplistic to capture the full complexity of the multi-phase ISM. A more robust measure of wind breakout conditions would take into account the full temperature--density distribution of the ISM to identify the warmer volume-filling phase fraction \citep[e.g.,][]{li20}. However, it may be challenging to account for the complicated redshift and halo mass dependence of ISM geometry, multi-phase partitioning, and ``contamination'' of hot gas from the inner virialized CGM.} We take the $\dot{M}$-weighted average $f_{\rm dense}$ within each individual burst window. Finally, we take the $t_{\rm cool}/t_{\rm ff}$ ratio at $0.1R_{\rm vir}$ from \citet{stern20}, who analyzed the same simulations. This cooling time to free-fall time ratio is a measure of virialization in the inner CGM (specifically when $t_{\rm cool}/t_{\rm ff}\gtrsim2$, the halo is virialized all the way down to the central galaxy). Following \citet{stern20}, we do not include the low-mass (m10) dwarfs since they have $T_{\rm vir}\sim10^4$ K and the distinction between the dynamically hot and cool phases breaks down. As with $f_{\rm dense}$, we estimate the $\dot{M}$-weighted average $t_{\rm cool}/t_{\rm ff}$ within each individual burst window.

\begin{figure*} 
\begin{center}
\includegraphics[width=0.3\hsize]{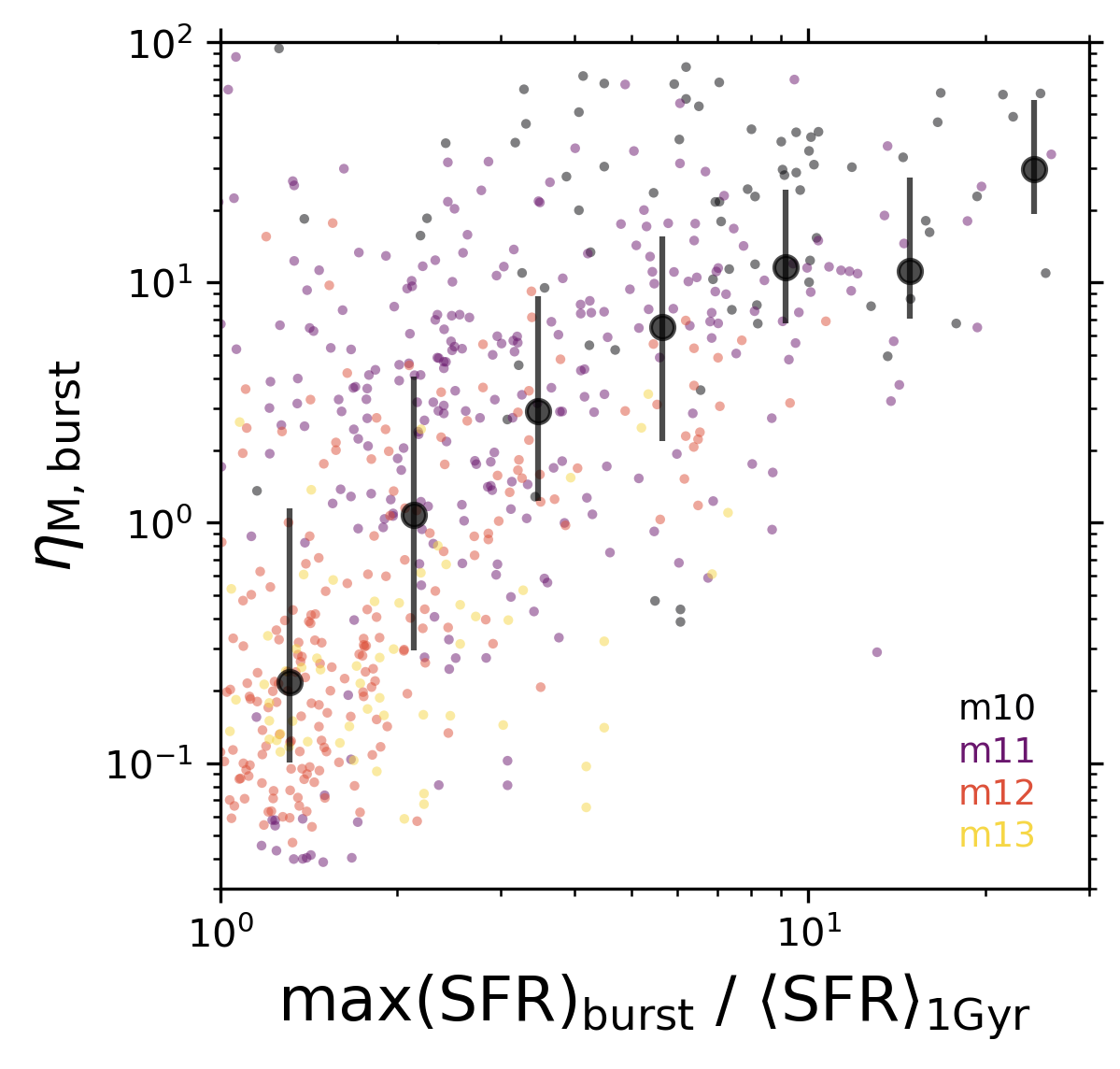}\includegraphics[width=0.31\hsize]{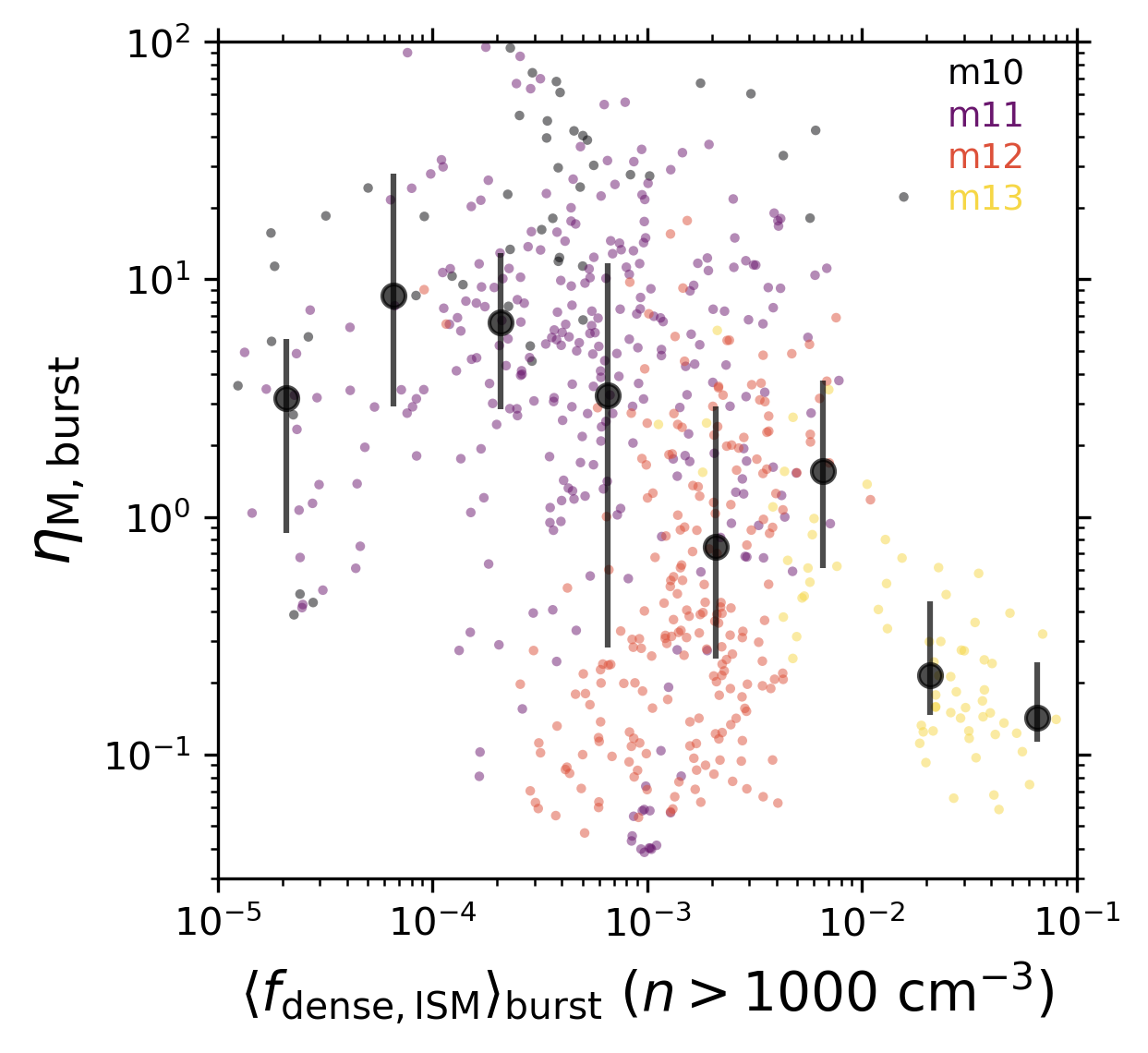}\includegraphics[width=0.3\hsize]{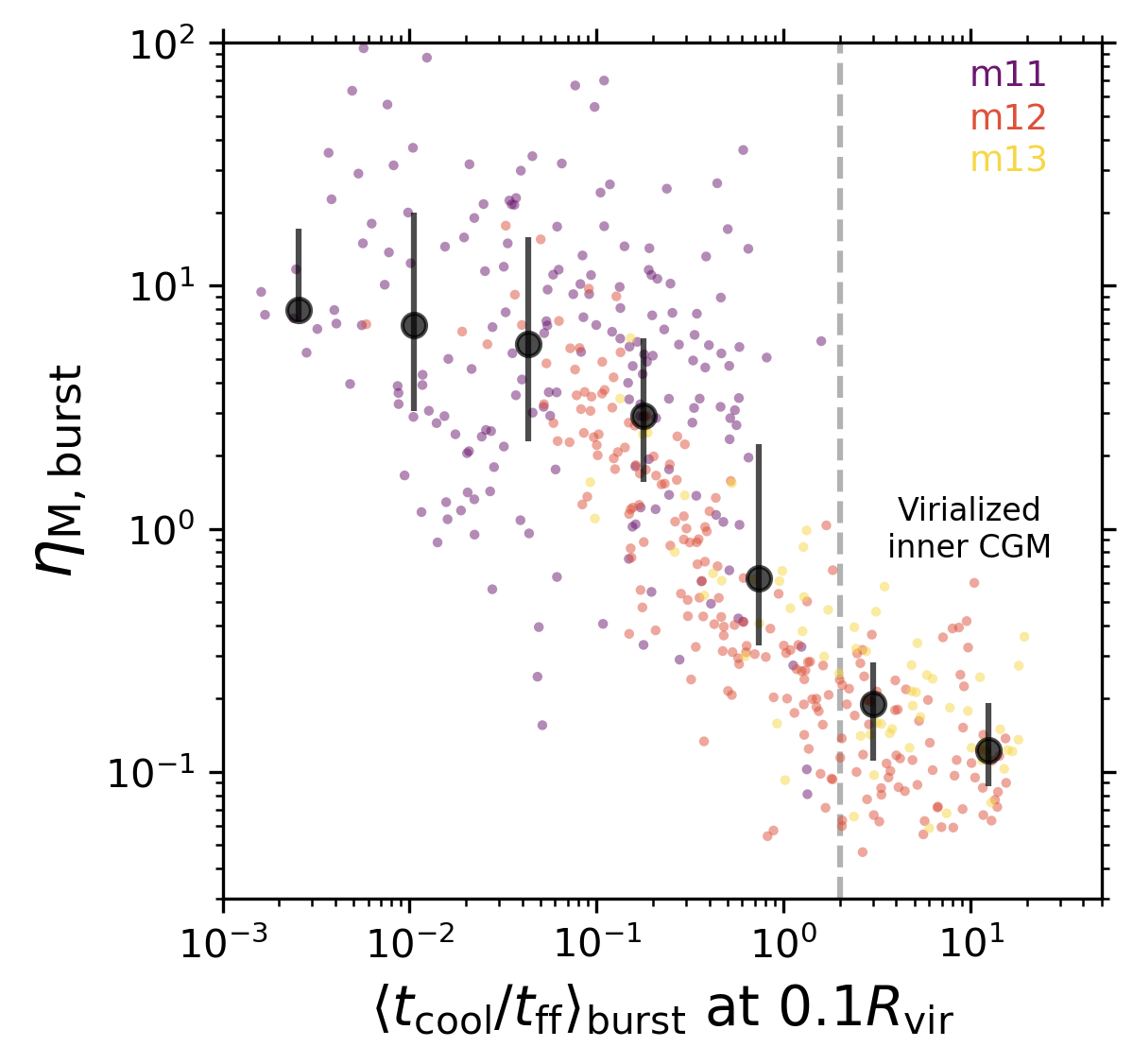}
\end{center}
\caption{Burst-averaged ISM mass loading factors versus three ``derived'' physical properties. From left to right: (1) the maximum SFR in the burst interval divided by the 1 Gyr-averaged SFR, (2) the dense ISM gas fraction, and (3) the cooling time to free-fall time ratio at $0.1R_{\rm vir}$, which is a measure of inner halo virialization. Each of these physical properties are $\dot{M}$-weighted averages within individual burst windows. The large black symbols and errorbars denote binned medians with 25th and 75th percentiles. We see that burst-averaged mass loadings are higher for more powerful starbursts (i.e., when the SFH is locally bursty). There is also a weak trend where burst-averaged $\eta_{\rm M}$ tends to be higher when the dense ISM gas fraction is lower, although there is a lot of scatter. Finally, the burst-averaged $\eta_{\rm M}$ steadily declines as $t_{\rm cool}/t_{\rm ff}$ increases, with the mass loading becoming $\ll1$ when the inner halo is virialized as indicated by $t_{\rm cool}/t_{\rm ff}>2$ (this is the case for massive halos whereas dwarfs have a non-virialized inner halo and higher mass loadings).}
\label{fig:resolved2}
\end{figure*}

\autoref{fig:resolved2} shows our burst-averaged mass loading factors as a function of the aforementioned three physical properties. The burst-averaged mass loadings are clearly larger when starbursts are stronger (i.e., when the peak SFR is more prominent relative to the 1 Gyr-averaged SFR). In contrast, there is a lot of scatter and effectively no trend with $f_{\rm dense}$, especially if we neglect the m13 halos which have large $f_{\rm dense}$ but low $\eta_{\rm M}$ (these halos are so massive that SN-driven winds cannot easily escape). Finally, the burst-averaged mass loading steadily declines as the $t_{\rm cool}/t_{\rm ff}$ ratio gets larger, with $\eta_{\rm M}\ll1$ when the inner halo is virialized (i.e., when $t_{\rm cool}/t_{\rm ff}>2$). This condition is met in the massive halos but not in the dwarfs (including the high-redshift dwarf progenitors of the m12 halos), which instead have high mass loadings and a non-virialized inner halo. These trends will help inform our discussion later.

\section{Halo wind loading factors}\label{sec:results_halo}
We now turn to halo-scale loading factors at $R_{\rm vir}$. The driver of halo-scale outflows is more difficult to disentangle because there can be other input sources for mass, momentum, energy and metals in addition to the ISM outflows \citep[e.g., CGM turbulence stirred by satellite motions and their own outflows;][]{fauchergiguere16,hafen19,hafen20}. As a result, there may be ambiguities in interpreting ``halo loading factors'' which are computed as outflow fluxes in the virial shell normalized by reference fluxes on the ISM scale for type II SN inputs. However, we have verified through animations of the projected particle data that hot outflows generated by the central galaxy do often have enough energy to make it to $R_{\rm vir}$, even in the MW halos at low redshift. Hence, it is informative to compare our broad redshift-averaged measurements of outflows at $R_{\rm vir}$ to those at $0.1R_{\rm vir}$ (the large integration timescale means we are effectively marginalizing over complicated propagation and delay time physics).

\subsection{Bernoulli velocity versus potential depth}
\begin{figure*} 
\begin{center}
\includegraphics[width=0.95\hsize]{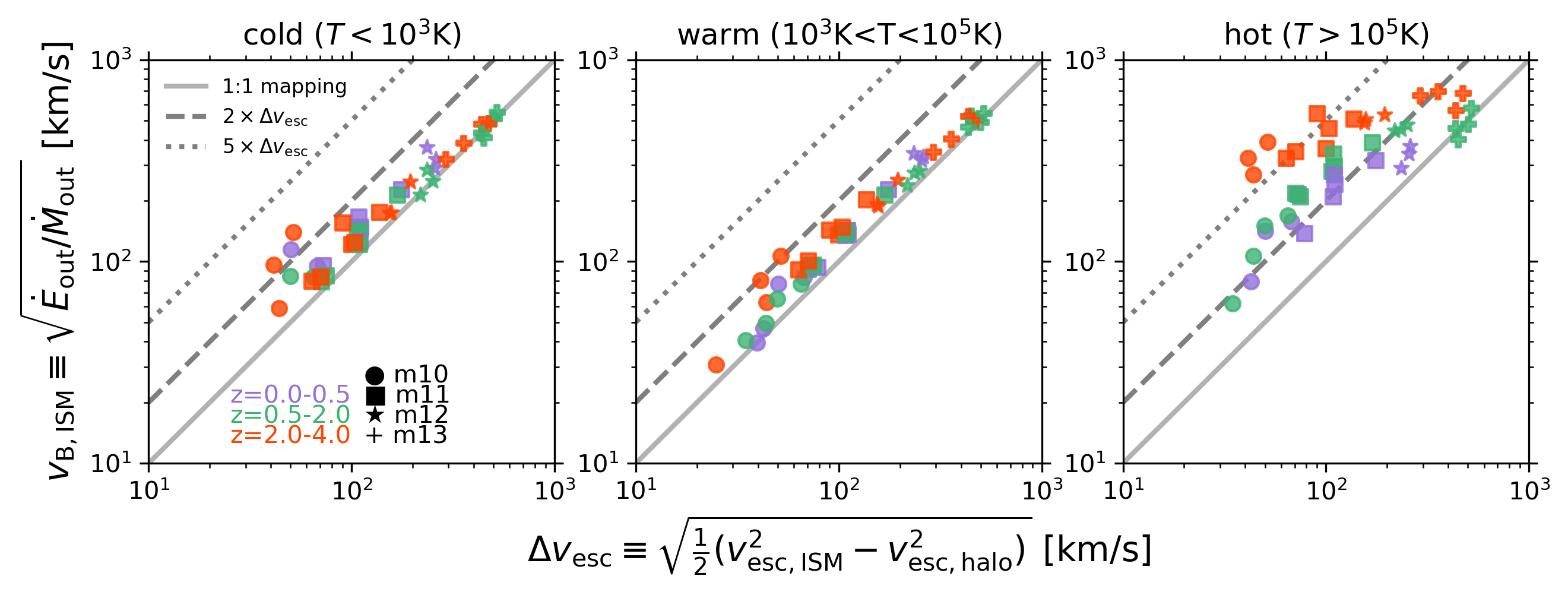}
\end{center}
\caption{Average mass-flux-weighted Bernoulli velocity (i.e., specific kinetic energy plus enthalpy) of multi-phase ISM outflows versus the difference in escape velocity between $0.1R_{\rm vir}$ and $R_{\rm vir}$ (a proxy for the potential difference). This gives a sense of whether outflows can be expected to reach $R_{\rm vir}$ in the absence of interactions (given our ISM wind selection criteria, outflows should make it to $\sim0.5R_{\rm vir}$ at minimum). Cold and warm outflows (left and middle panels) in some dwarfs have up to twice the energy needed to make it to $R_{\rm vir}$, but in the more massive halos the cold/warm outflow energy is comparable to the potential difference. In contrast, hot outflows (right) have Bernoulli velocities that are far in excess of the energy needed to make it to $R_{\rm vir}$. This is obvious for lower mass halos where the hot outflows contain up to $\sim5\times$ more energy than needed to escape the halo (hence these outflows can be expected to travel very large distances, probably becoming unbound). Hot outflows in low-redshift MW halos have just enough energy to reach $R_{\rm vir}$. Hot SN-driven outflows can also escape from m13 halos at high redshift but not necessarily at intermediate redshift.}
\label{fig:vB_vesc}
\end{figure*}

Before presenting the halo loading factors, in \autoref{fig:vB_vesc} we first compare the average mass-flux-weighted Bernoulli velocities ($v_{\rm B}=\sqrt{\dot{E}/\dot{M}}$ following \autoref{eqn:vbern2}) of multi-phase ISM outflows to the difference in escape velocity between $0.1R_{\rm vir}$ and $R_{\rm vir}$ (which quantifies the halo potential depth). As outflows propagate outwards, they gain potential energy at the expense of kinetic and thermal energy; hence in the limiting case of adiabatic outflows, the decrease in Bernoulli velocity should mirror the decrease in escape velocity. Note that the upper limit on the Bernoulli velocity of SN-driven outflows is $\sqrt{\frac{10^{51} \textrm{erg}}{100M_{\odot}}}\approx700$ km s$^{-1}$; comparing this to the potential difference gives a simple estimate of whether SN-driven outflows can escape from halos of a given mass.

We see that cold and warm outflows contain just enough energy to make it to $R_{\rm vir}$ in the dwarfs and even the massive halos. On the other hand, the hot outflows contain much more energy than needed to get to $R_{\rm vir}$; for high-redshift dwarfs, the energy of hot outflows is $\sim5\times$ higher than the escape velocity difference, hence many of these outflows may become unbound from the halo. In the MW halos at low redshift, the hot outflows have just enough energy to make it to $R_{\rm vir}$. This is also true for the m13 halos at high redshift but not at low redshift (where again, outflows may only reach $\sim0.5R_{\rm vir}$ in accordance with our wind selection criteria). We find that roughly half of the specific energy of the hot wind is in kinetic form except in the m13 and low-redshift MW-mass halos where it drops to $\lesssim30\%$ (signifying the prominence of slow but very hot buoyant outflows).

This exercise demonstrates that we should expect to see significant halo wind loading (especially for hot outflows in dwarfs) and that comparing characteristic outflow rates at $R_{\rm vir}$ to those at $0.1R_{\rm vir}$ can help constrain average losses/gains in mass, momentum, energy and metals while winds transit the CGM. In \autoref{fig:ratios}, we compare total mass, momentum, energy and metal loading factors in our ISM shell ($0.1-0.2R_{\rm vir}$) to those in our virial shell ($1.0-1.1R_{\rm vir}$). Both the ISM and halo loading factors in this figure only include outflows that have enough energy to get to at least $2R_{\rm vir}$ if not farther.\footnote{Note that these ``escaping'' ISM outflows are somewhat lower than our fiducial measurements to go from $0.1\to0.5R_{\rm vir}$ since only a subset of ISM outflows will have the greater required initial energy to travel all the way to $2R_{\rm vir}$. However, the overall trends are similar to our results above.} By defining this subset of ``escaping'' ISM and halo outflows, we can constrain what fraction of mass, momentum, energy and metals predicted to escape from the ISM to $2R_{\rm vir}$ may actually do so. We will now describe each of these outflow quantities in turn.

\subsection{Halo mass loading}
We see that even for the low-redshift MW halos, the actual halo mass loading is comparable to, in fact even slightly larger than, the ISM mass loading defined using particles with enough energy to make it to $2R_{\rm vir}$. If we had included slower moving, likely cold and turbulent, ISM outflows -- which never had a chance of getting to $2R_{\rm vir}$ anyway -- then this ratio would be closer to $0.3-0.4$ \citep[Figure 12 of][]{pandya20}. While we do not know whether the identity of the gas leaving the virial shell is the same as the gas that was previously ejected from the ISM (e.g., much of the ISM outflows could have stalled in the CGM while still pushing ambient halo gas outwards), our finding that $\eta_{\rm halo}/\eta_{\rm ISM}\sim1$ in the low-redshift MW halos combined with their relatively large Bernoulli velocities of hot outflows in \autoref{fig:vB_vesc} suggests that outflows can have substantial effects in MW halos (see also our supplementary movies, e.g., \autoref{fig:example}). This agrees with the conclusions drawn from the comparative CGM analysis of diverse simulations by \citet{fielding20b}. 

For dwarfs, the halo mass loading is also larger than the ISM mass loading of winds expected to make it to $2R_{\rm vir}$. However, it is also important to appreciate that a much larger fraction of outflows leaving the ISM of dwarfs have enough energy to reach $2R_{\rm vir}$ as compared to winds in more massive halos (i.e., the ratio $\eta_{\rm M,halo}/\eta_{\rm M,ISM}$ would remain above one for dwarfs even if we relaxed our $0.1\to2R_{\rm vir}$ Bernoulli velocity criterion, unlike for the low-redshift MW halos described above). Since the dwarfs are quite isolated and hence satellite effects are relatively negligible, the halo-scale loading factors being larger than the ISM-scale ones for dwarfs is likely due to entrainment of CGM gas by the winds \citep[see also][]{muratov15,pandya20}. 

In contrast, for the m13 halos, the much higher halo mass loadings than expected are likely due to their rich satellite systems, which can stir up the CGM and have substantial outflows of their own \citep[e.g.,][]{anglesalcazar17,hafen20}. While quantifying these entrainment and satellite effects is beyond the scope of this paper, the time evolution of the radial profile of $\dot{M}_{\rm out}$ and $\dot{M}_{\rm in}$ in our supplementary movies (as in \autoref{fig:example} and \autoref{fig:example2}) can qualitatively reveal these effects. For example, the amplitude and/or width of an outflow spike may increase as it propagates to larger radius, which would be indicative of CGM entrainment.

\subsection{Halo momentum loading}
In the dwarfs, the halo momentum loadings are larger than the ISM momentum loadings, which is expected for energy-conserving outflows (if $\dot{E}\sim\dot{M}v^2$ is roughly constant, then $\dot{p}\sim\dot{M}v$ will increase as the outflow decelerates due to sweeping up mass). Interestingly, the MW halos at low redshift have roughly similar outflow momentum at the halo and ISM boundaries. The m13 halos have anomalously high halo momentum loading factors, which may suggest additional momentum input sources (e.g., their rich satellite systems).

\subsection{Halo energy loading}
The halo energy loadings are comparable to ISM energy loadings in the dwarfs (including the high redshift progenitors of MW halos). This suggests that the relatively higher ISM energy loadings of dwarfs (\autoref{fig:etaE_all}) are conserved to at least $R_{\rm vir}$, which is consistent with the large Bernoulli velocities relative to their potential depth (\autoref{fig:vB_vesc}). The m13 halos also have high halo energy loadings but there is likely significant contamination from their rich satellite systems (which can introduce additional kinetic and thermal energy from stirring turbulence in the CGM, heating from their own energy-rich outflows, etc.). In contrast, for the low redshift MW halos, the halo energy loading factors are only $\sim0.25$ times their ISM energy loadings (i.e., 4 times lower). 

\subsection{Halo metal loading}
In the MW halos at low redshift, the metal loading at $R_{\rm vir}$ is roughly $\sim20\%$ of the ISM-scale metal loading. In contrast, for their dwarf progenitors and all dwarfs more generally, the halo metal loadings are comparable to the ISM metal loadings. This is consistent with the interpretation by \citet{muratov17} for FIRE-1 that metals escape from dwarf halos but are retained within MW halos (perhaps due to substantial interactions in the latter, although we also do not know the level of mixing with pristine gas). The m13 halos also have relatively low halo metal loadings perhaps owing to their deeper potential wells, although again their halo-scale measurements are likely contaminated due to metal-rich outflows from their large satellite systems \citep[see also][]{hafen19}.

\begin{figure*} 
\begin{center}
\includegraphics[width=0.95\hsize]{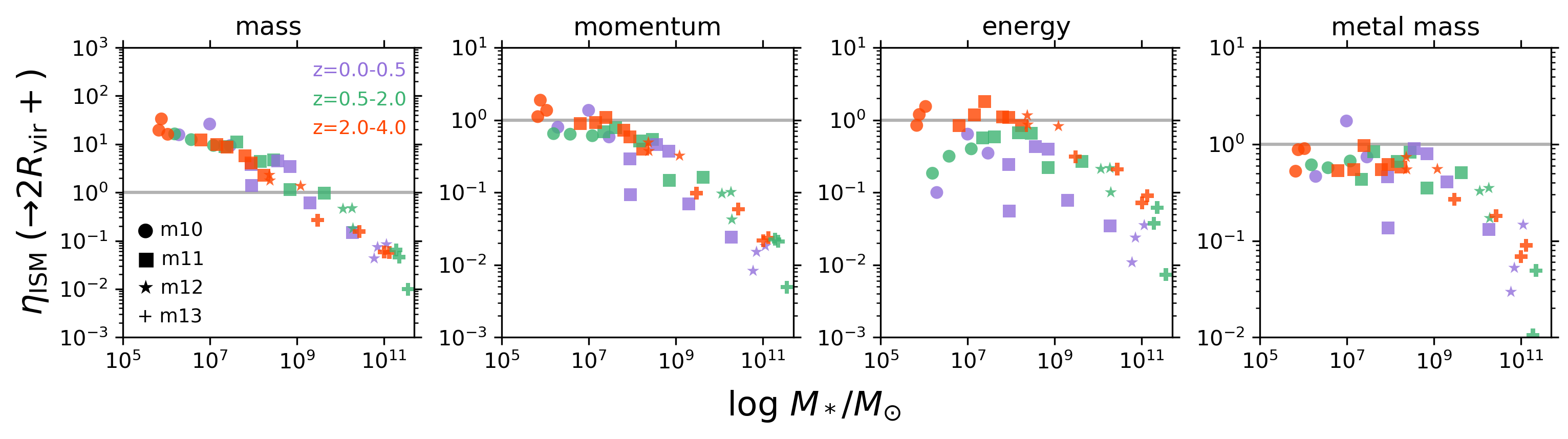}

\includegraphics[width=0.95\hsize]{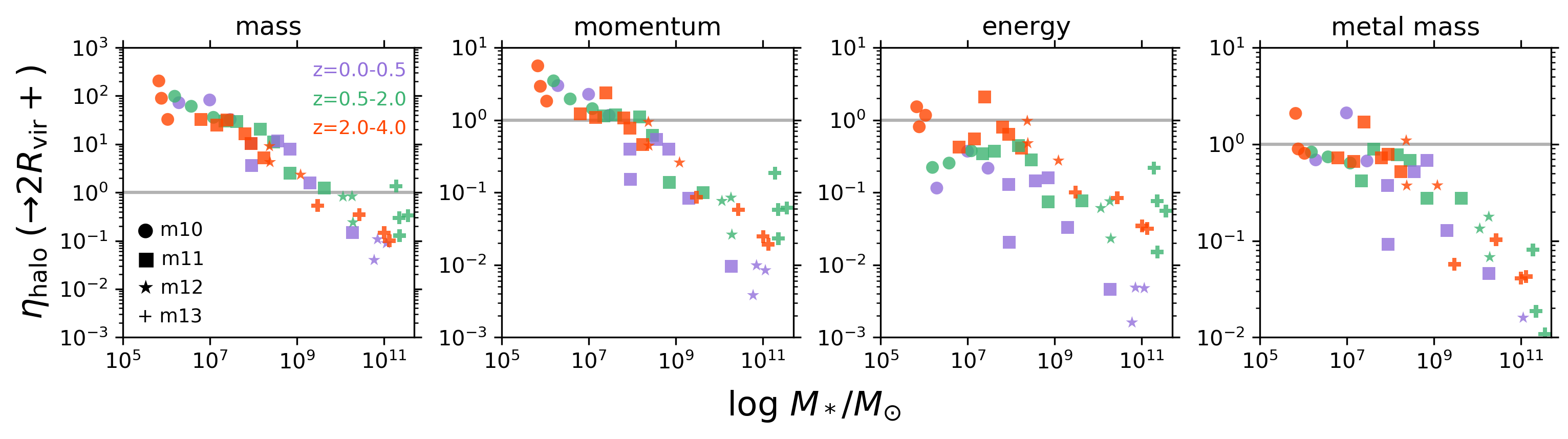}

\includegraphics[width=0.95\hsize]{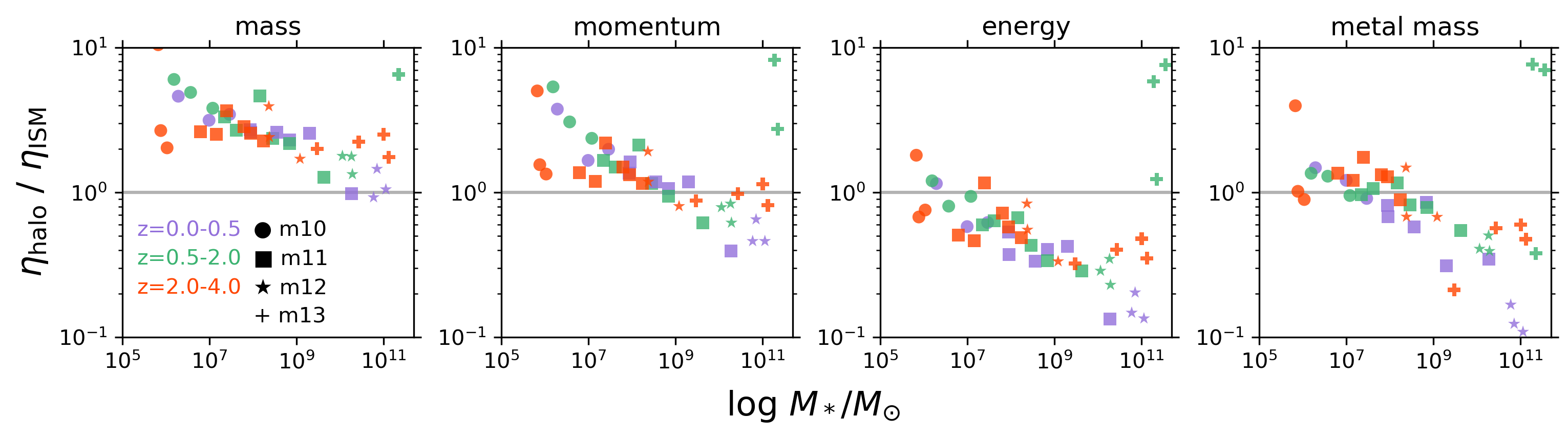}
\end{center}
\caption{Comparing ``escaping" wind loading factors for the ISM and virial shells. The top row is ISM loading factors using only the subset of ISM outflows that have enough energy to reach $2R_{\rm vir}$ instead of just $0.5R_{\rm vir}$. The middle row is halo-scale loading factors using only outflows at $R_{\rm vir}$ that have enough energy to get to at least $2R_{\rm vir}$. The bottom row is the ratio of these halo-scale and ISM-scale loading factors. From left to right we show mass, momentum, energy and metal loading factors. \textit{Mass:} dwarf halo mass loadings are a few times larger than the ISM mass loadings, perhaps indicative of additional swept up material. Low-redshift MW halos have a ratio close to $\sim1$ (recall that our ISM loadings exclude slower outflows, which may be substantial but unlikely to reach $R_{\rm vir}$). m13 halos show a larger ratio at intermediate redshift than at high redshift. \textit{Momentum:} Dwarf outflows often have more momentum at the halo scale than at the ISM scale, in contrast to low-redshift MW halos whose outflows have comparable momentum at $R_{\rm vir}$ and $0.1R_{\rm vir}$. \textit{Energy:} In dwarfs, halo energy loadings are comparable to ISM energy loadings. In contrast, for MW halos at low redshift, the halo-scale energy loadings are $\sim0.1$ smaller than their ISM-scale energy loadings. \textit{Metals:} In dwarf halos, most metals leaving the ISM also leave the halo. In low-redshift MW halos, only $\sim10\%$ of ISM metal outflows leave the halo (surprisingly, intermediate redshift m13 halos have high metal loadings).}
\label{fig:ratios}
\end{figure*}

\section{Discussion}\label{sec:discussion}
Here we summarize the overall story suggested by our results, discuss our findings in the context of previous work, and list some possible systematic uncertainties in our analysis.

The results of our analysis tell a seemingly simple story. We have found that dwarfs have preferentially much higher ISM mass, momentum, energy and metal loadings than MW-mass halos at late times (and even the m13 halos at early times). The cold outflow phase is generally negligible for dwarfs except at high redshift where the cold phase can account for $\sim10\%$ of each of the total loading factors.\footnote{Observed outflows driven by active galactic nuclei (AGN) can have high cold mass loading factors \cite[e.g.,][and references therein]{cicone14,fiore17,fluetsch19}, but the FIRE-2 simulations do not include AGN feedback.} The importance of the warm phase gradually increases toward lower stellar masses (for which the warm phase approaches $\sim100\%$ by mass fraction). The suppression of multi-phase outflows in the lowest mass dwarfs may be a clue that the UV background together with the global thermodynamics of the halo (the virial temperatures of these dwarfs is much lower than our threshold of $10^5$K for the hot phase) either prevents thermal instabilities or rapidly heats up cold outflows due to CGM mixing and/or shocks. In addition, much of the ISM of dwarfs may already be at a warm temperature, so significant cold mass loading may not be expected. In any case, it is remarkable that the overall momentum, energy and metal loadings are of order unity in the lowest mass dwarfs, implying that most of the SN-driven energy, momentum and metals make it quite far out of the ISM; the mass loadings being of order 100 also suggests that the outflows sweep up significant amounts of ambient material. The metal loadings being of order unity suggests that the ISM metallicity of dwarfs is in equilibrium \citep{forbes14} since most of the metals produced as SN ejecta escape via metal-enriched, energy-conserving outflows (hence the ISM metallicity should be roughly constant with time). Note that the FIRE simulations have been shown to agree reasonably well with observed mass--metallicity relations for both gas and stars in the mass ranges that we examine here \citep{ma16,wetzel16,escala18}. The ratio $\eta_{\rm halo}/\eta_{\rm ISM}\gtrsim1$ for the dwarfs, further suggesting that ISM outflows escape to quite large distances ($\gtrsim R_{\rm vir}$) on average with their energy, momentum and metals intact. 

In contrast, for low-redshift MW halos and high-redshift massive (m13) halos, winds are weaker and the hot phase generally carries most of the mass, momentum, energy and metals.\footnote{Had we only used the simpler $v_{\rm rad}>0$ km/s cut, we would select substantially more warm outflows. However, some fraction of these may not travel far beyond $\sim0.1R_{\rm vir}$ and may represent random motions near the ISM edge.} The warm phase is subdominant (though it can carry a substantial fraction of metals in the low-redshift MW halos; see purple stars in \autoref{fig:etaZ_all}) and the cold phase is generally negligible (a few percent by mass fraction). The loading factors for the low-redshift m12 halos are below unity ($\eta_M$ is of order $\sim0.1$ on average, and possibly smaller for individual weak outflow episodes), which means that only a fraction of the SN-driven mass, energy, momentum and metals make it out of the ISM (unlike for the dwarfs). Nevertheless, the ratio $\eta_{\rm M,halo}/\eta_{\rm M,ISM}\sim1$ for the MW halos at low-redshift, suggesting that whenever there is a large breakout of wind from the ISM, there is subsequently also a large outflow from the halo. However, the $\eta_{\rm halo}/\eta_{\rm ISM}$ ratio is far below unity for energy and metals, meaning that a large fraction of wind energy is dissipated while metals are mixed into the CGM due to interactions (or the outflow metallicities are diluted due to sweeping up of metal-poor CGM gas).\footnote{Many of the m13 halos have $\eta_{\rm halo}/\eta_{\rm ISM}$ ratios greater than unity, which is unexpected given their deep potential wells. We think this is likely due to additional input sources of mass, momentum, energy and metals at large radii. Possible sources include outflows and turbulence stirred by their numerous satellites as well as accretion shocks of infalling gas near $R_{\rm vir}$.} Interestingly, the $\eta_{\rm p,halo}/\eta_{\rm p,ISM}$ is closer to 1 for the low-redshift MW halos, and may be driven by the thermal pressure term which would be substantial for their predominantly hot outflows.

\subsection{Comparison to theoretical expectations and other simulations}

\subsubsection{Comparison to simple theoretical arguments}
Traditionally, mass loading factors are correlated against the global halo circular velocity since that is a proxy for the potential depth and because the inferred power law slope may encode whether the winds are ``energy-driven'' ($\eta_{\rm M,ISM} \propto V_{\rm vir}^{-2}$) or ``momentum-driven'' \citep[$\eta_{\rm M,ISM} \propto V_{\rm vir}^{-1}$; e.g.,][]{murray05,hopkins12,muratov15,christensen16}. We find that $\eta_{\rm M,ISM}\propto V_{\rm vir}^{-2}$ at high redshift, with a significant steepening at low redshift. There appears to be no need to appeal to a ``broken'' power law as found for the FIRE-1 halos by \citet{muratov15}. Our power law (particularly at high redshift) is consistent with simple theoretical expectations for energy-conserving winds as laid out in \citet{murray05}. At lower redshifts, our relation becomes even steeper, consistent with a picture in which there are significant losses in the ISM prior to the wind being launched (though winds in the dwarfs still seem to conserve energy after breaking out of the ISM and propagating through the CGM, while the MW halos show a substantial drop in mass loading with redshift and their winds appear to conserve at least some momentum but not energy; \autoref{fig:ratios}).

In addition to correlating the loading factors against the global halo virial velocity and stellar mass, it is important to consider correlations with properties that explicitly characterize the state of the ISM and inner halo \citep[e.g., as suggested by][]{fielding17,li20}. After all, the virial velocity and stellar mass correlations alone do not unambiguously explain what sets the properties of winds upon initial breakout from the ISM. For example, why do winds in high-redshift dwarfs appear to be energy-conserving (i.e., consistent with a $V_{\rm vir}^{-2}$ scaling)? While painting a fully fleshed out physical picture is beyond the scope of this work, we found three important trends that can help guide future work using controlled numerical experiments. First and foremost, burst-averaged mass loading factors are preferentially higher during more powerful starbursts (i.e., when the peak SFR is more prominent compared to the 1 Gyr-averaged SFR). During such locally bursty SF events, we may expect more strongly clustered SNe \citep[][]{fauchergiguere18}. The resulting powerful stellar feedback may clear out the denser phase of the ISM while percolating through the less dense phase, ultimately breaking out of the galaxy prior to losing significant energy via radiative cooling \citep[e.g.,][]{fielding18}.

Correlating burst-averaged mass loading factors with dense ISM gas fractions reveals a lot of scatter and effectively no trend, especially if we ignore the m13 halos at high $f_{\rm dense}$ with low $\eta_{\rm M}$ (SN-driven winds cannot easily escape from these massive halos). The lack of a strong correlation with $f_{\rm dense}$ may reflect the fact that more powerful starbursts are also expected to occur when dense ISM gas fractions are higher, and these in turn may drive more powerful winds despite high $f_{\rm dense}$. On the other hand, our overly simplistic definition of $f_{\rm dense}$ using only particles with $n>1000$ cm$^{-3}$ may not be the best diagnostic of ISM breakout conditions: if the warmer volume-filling ISM phase fraction can be reliably measured, that may lead to a more robust correlation \citep{li20}. On a related note, the ISM may be more turbulent when the overall gas fraction $M_{\rm gas}/(M_{\rm gas}+M_*)$ is higher, which may make it easier to drive strong outflows \citep[this may help explain why winds become weaker in more massive halos at later times, when their overall gas fractions have decreased;][]{hayward17}. Finally, we find that burst-averaged mass loadings are suppressed when the inner halo is virialized (as in the more massive halos) \citep{stern20}. The lack of a virialized inner CGM in dwarfs may allow outflows to propagate relatively unimpeded with minimal energy and momentum losses. Shock heating and entrainment of CGM/IGM gas by these energy-conserving outflows may cause preventative feedback that can suppress future gas accretion and ultimately help reduce the global star formation efficiency of dwarfs \citep{pandya20}. Despite this interesting heuristic exercise, we stress that our analysis of individual outflow episodes groups together events occurring in halos of widely different masses and across $\sim10$ Gyr of cosmic time. It is an enormously challenging task to simultaneously control for all of the possible interplay between global and local conditions using a fully cosmological simulation, but it is encouraging that we at least see some emergent systematic trends with our simple summary statistics.

\subsubsection{Comparison to high-resolution idealized simulations}
It is difficult to say definitively how our FIRE-2 wind scalings compare to those from resolved ISM idealized simulations. A future analysis of the FIRE-2 simulations closer to the ISM while accounting for the complicated geometries of our galaxies can help place these kinds of comparisons on a firmer footing \citep[e.g., following][]{gurvich20}. One seemingly major difference worth commenting on is that \citet{kim20} very clearly predict that cool outflows (with $T<2\times10^4$ K) carry most of the mass whereas hot outflows (with $T>5\times10^5$ K) carry most of the energy in their TIGRESS kpc-scale sub-galactic simulations. In contrast, for our low-redshift MW halos, the hot phase carries both most of the mass and energy (\autoref{fig:etaM_all} and \autoref{fig:etaE_all}). The simplest explanation is that our measurements are made much farther from the galaxy ($0.1R_{\rm vir}$) than in resolved ISM simulations ($\sim1-2$ disk scale heights), and much of the cold and warm outflows may not be expected to make it to $\sim0.5R_{\rm vir}$ anyway. Instead they may recycle as fountain flows much closer to the disk \citep[e.g.,][]{anglesalcazar17,hafen20,gurvich20} or get mixed into the hot phase \citep[e.g.,][]{fielding20,schneider20}. Note also that FIRE includes additional prescriptions for radiative pressure feedback and photoionization that are not modeled in TIGRESS but which may be crucial for heating the ISM and enabling breakout of hot winds. Nevertheless, it is encouraging that the hot mass loadings of our low-redshift MW halos are still only of order $\sim0.1$, which is similar to the TIGRESS predictions \citep{kim20}. 

A related question is why the hot mass loading in our dwarfs is far larger than $\sim0.1$, in fact closer to $\sim10$ (the overall mass loading is $\sim100$ with a $\sim10\%$ hot phase fraction for low $M_*$ galaxies; see \autoref{fig:etaM_all}). Partially, this may be due to shock heating and entrainment of inner CGM gas by energy-conserving outflows in the dwarfs. The warm phase is even more prominent than the hot phase for winds in the FIRE-2 dwarfs. This may be due to the fact that warm outflows may be able to travel farther into the CGM of dwarfs because of their shallower potential well depths. There can also be a disproportionately larger contribution of swept-up warm ISM and inner CGM gas for outflows in dwarfs compared to more massive halos. Note that since the virial temperatures of dwarfs can be lower than our hot phase threshold temperature of $10^5$ K, much of the warm outflows in dwarfs can still be considered ``dynamically hot'' (and vice versa for more massive halos). The idealized, high-resolution global dwarf simulations by \citet{hu19} show that the warm phase (what they call the ionized phase) is indeed very important: it is the dominant phase beyond a few kpc owing to cooling of the hot phase and shock heating of cooler gas.   

As for trends between burst-averaged wind loading factors and ISM and SFR surface densities, the clearest correlation we have found is between $\eta_{\rm M}$ and the $\dot{M}$-weighted average $\Sigma_{\rm gas}$ over individual burst windows. The burst-averaged mass loading tends to drop systematically as $\Sigma_{\rm gas}$ increases, becoming of order $\sim0.1$ in the low-redshift MW-mass halos. This is qualitatively consistent with predictions from TIGRESS where the authors find $\eta_M\propto\Sigma_{\rm gas}^{-1.12}$, albeit much closer to the galaxy \citep[one scale height above/below the disk; Figure 12 in][]{kim20}. The correlation with $\Sigma_{\rm SFR}$ shows more scatter. This may partially be driven by the fact that we are combining bursts from widely different halos and at many different redshifts (up to $z=4$). We also measure our loading factors farther from the ISM than sub-galactic simulations do; in fact, our chosen distance of $0.1R_{\rm vir}$ corresponds to $\sim25-30$ physical kpc for a low-redshift MW-mass halo, which is far beyond the simulation domain of kpc-scale resolved ISM models. Since the properties of a wind may change as it propagates through the inner halo, it is perhaps natural to expect different correlations with more scatter farther from the galaxy (just like we expect halo-scale loadings to be more complicated to interpret).  A fruitful avenue for future work will be to combine measurements of loading factors very close to the galaxy (ideally by defining subpatches of the ISM and properly accounting for the more complicated gravitational potential) with our spherically-averaged loading factors farther out \citep[see also][]{gurvich20}.

\subsubsection{Comparison to other cosmological simulations}
Lastly, it is useful to qualitatively discuss our results in the context of other cosmological simulations (both zooms and large-volume). The FIRE-2 simulations are particularly unique for predicting wind properties in a cosmological context due to their explicit stellar feedback model. In contrast, the modeling of SN-driven winds across different cosmological simulations varies dramatically and often involves ad hoc approaches (e.g., decoupling winds from hydrodynamics, artificially delayed cooling, etc.). 

Compared to the FIRE-1 results of \citet{muratov15,muratov17} that we build on \citep[see also][]{anglesalcazar17}, our overall FIRE-2 mass and metal loadings scale with stellar mass in qualitatively similar ways (despite our more stringent wind selection criteria; \autoref{fig:previous}). However, we have gone further and provided several new insights by explicitly measuring outflow energy and momentum loadings, temperature and velocity distributions, and scalings with quasi-local ISM properties. This allowed us to explicitly demonstrate that winds in dwarfs seem to be energy-conserving whereas the more massive halos show significant outflow energy losses, especially at low redshift. Interestingly, \citet{christensen16} also find that outflows in their GASOLINE zoom-in simulations are consistent with the simple energy-driven scaling ($\eta_{\rm M}\propto V_{\rm vir}^{-2}$; their Figure 11). \citet{tollet19} find a much steeper relation in the NIHAO zoom-in simulations: $\eta_{\rm M}\propto V_{\rm vir}^{-4.6}$, which they attribute to the reduced efficiency of SN feedback in more massive halos. They also cut off their steep scaling for dwarfs with $V_{\rm vir}<45$ km/s since there is a lot of scatter which they claim is due to stochastic SF. Instead, they argue that for these dwarfs, the mass loadings must revert to following at most a predicted $V_{\rm vir}^{-2}$ scaling because there is a ``maximum'' efficiency of SN feedback and because most of their dwarf outflows are cold with radial velocities below the escape velocity. With our more stringent wind selection criterion (which captures the slow component of the hot wind while neglecting cold, turbulent outflows), we find relatively little scatter for the lowest mass dwarfs, which suggests that the steepening of the overall relation with time is indeed driven by higher mass halos. Note that the prominent redshift dependence in \autoref{fig:Vvirscaling} but weaker redshift dependence when plotting against stellar mass (\autoref{fig:etaM_all}) can at least partially be explained by the stellar-to-halo-mass ratio, at fixed halo mass, getting larger at later times (at fixed $M_{\rm vir}$) whereas $V_{\rm vir}$ does not evolve as dramatically. The redshift evolution in the stellar-to-halo-mass ratio is particularly prominent for more massive halos since SF is so inefficient in dwarfs, and so the steepening against virial velocity with time may be driven by the m12 halos.

As for large-volume simulations, energy and momentum loadings are generally not measured. One notable exception is \citet{mitchell20} who measured mass loading factors as well as radial profiles of energy and momentum outflow rates in the EAGLE simulations. They find that $\eta_{\rm M}\propto V_{\rm vir}^{-1.5}$ for lower mass halos where stellar feedback dominates, with the normalization increasing towards higher redshift; this scaling steepens for halo-scale mass loadings to roughly match $\propto V_{\rm vir}^{-2}$ as expected for energy-driven winds. They attribute this steepening to entrainment of CGM gas by winds, as inferred from average radial profiles of mass, momentum and energy outflow rates. Another interesting result of modern large-volume simulations is that they predict multi-phase winds, as recently shown for the IllustrisTNG model by \citet{nelson19}. Despite the very different SN feedback subgrid models of FIRE-2 and IllustrisTNG \citep[decoupled kinetic wind model;][]{springel03}, it is encouraging that the outflow temperature distributions vary in similar systematic fashions with stellar mass, in particular that the cold phase is noticeably absent in the lowest mass dwarfs and more prominent in higher mass halos. This relative agreement likely reflects the fact that the physics of radiative cooling, which is similar amongst all these simulations, is predominantly responsible for setting the general properties of the outflow temperature distributions. However, the inclusion of AGN feedback in many large-volume simulations combined with a more phenomenological treatment of winds (plus different wind analysis methods) makes it difficult to draw direct comparisons with FIRE-2. We do note that the EAGLE \citep[][their equation 6]{schaye15} and IllustrisTNG \citep[][their Figure B2]{pillepich18} large-volume simulations assume that the input SN thermal energy is smaller when the ISM metallicity is larger due to more efficient cooling, which requires that the emergent wind energy loading factors (defined with respect to the nominal input SN energy of $10^{51}$ erg) should be lower in more massive halos and at later times. While we do not explicitly consider scalings with ISM metallicity, their assumption is consistent with our overall findings. A more granular analysis of winds closer to the ISM in FIRE-2 would shed further light on this assumption, including the role of ISM gas metallicity and density in setting the redshift dependence of some of our wind scalings.

In the future, it will be insightful to compare to large-volume simulations that ``plug in'' scalings for mass loadings and wind velocities taken from zoom-in simulations \citep{dave16,dave19,huang20}. Our comprehensive multi-dimensional characterization of FIRE-2 winds in terms of their temperature and velocity distributions, and their energy and momentum loadings, will serve as useful inputs and benchmarks for large-volume models with insufficient resolution to capture stellar feedback. Our FIRE-2 scalings can be implemented in SAMs as is classically done (e.g., ISM mass loading factor versus global halo virial velocity), and it may also be possible to use our relations in radially-resolved SAMs where outflow properties are varied according to local ISM properties \citep[such as the gas mass surface density;][]{forbes19}. Perhaps most importantly, SAMs will benefit from implementing our multi-phase energy and momentum loading factors, which can be used to drive CGM heating and hence suppress ISM accretion, push ambient CGM gas out of the halo via entrainment, and prevent IGM gas from accreting into the halo in the first place \citep[see more discussion of preventative stellar feedback models in][]{pandya20}.

\subsection{Systematic uncertainties}\label{sec:systematics}
Although we have taken the first step to characterize the full thermodynamic properties of outflows in fully cosmological simulations \citep[adapting analysis methods commonly used for high-resolution idealized simulations; e.g.,][]{kim20}, there are several sources of systematic uncertainty that may impact our results and interpretation. Many of these, which we list here, may be fruitful avenues for future work. 

First and foremost, our ISM loading factors are measured with a shell at $0.1-0.2R_{\rm vir}$, which corresponds to nearly $\sim25$ kpc in a MW halo at $z=0$. This was done partly for simplicity (we can use spherical shells, avoid contamination from dense ISM mass flows, ignore the highly non-trivial geometry of galaxies, especially high-redshift dwarfs, etc.), but this is very far from the midplane of the disk and far beyond the domain of highly resolved ISM simulations \citep[e.g., the $\sim1$ kpc scale boxes simulated by][]{kim20}. While it is fundamental to know the properties of outflows this far out near the inner CGM, it is somewhat ambiguous what fraction of our ISM outflows are fresh from the MW versus swept up inner CGM material. A future analysis that considers the properties of outflows directly above and below the galaxy can provide many useful physical insights and sanity checks (this can be done easily at least for MW halos at intermediate and lower redshifts when they have a well-defined disk, but dwarf geometries are more complicated). In parallel, additional metrics for characterizing the inner CGM beyond the $t_{\rm cool}/t_{\rm ff}$ proxy of \citet{stern20} may help us better understand the role of the CGM in modulating outflows and its own susceptibility to being heated/entrained. Combined with a particle tracking approach \citep[e.g.,][]{anglesalcazar17,hafen20}, we would also be able to more confidently constrain the distribution of particle travel times, maximum distances, recycling times, and whether particles expected to conserve energy/momentum actually do so.

Furthermore, the fixed mass resolution of the Lagrangian FIRE-2 simulations may lead to unresolved cooling of hot outflows and propagation of cold outflows. We have found that in MW-mass FIRE-2 halos at low redshift, the hot phase carries not only most of the energy but also most of the mass. In highly resolved ISM simulations closer to the disk, the hot phase carries most of the energy whereas the cool phase carries most of the mass. If we are not resolving the cooling of hot gas, then colder clumps that should form and become entrained in the CGM (leading to smaller outflow rates measured at $R_{\rm vir}$) may not be properly captured. We have also seen that cold outflows are heavily suppressed except in high-redshift dwarfs; this may be physical in the dwarfs, but it could also partially be due to poor resolution in the CGM, especially in the more massive halos at low-redshift. On the other hand, we emphasize that our wind selection criteria exclude outflows that do not have enough starting energy to reach at least $0.5R_{\rm vir}$; most of this excluded material is almost certainly slow, cold outflows. In addition, recall that the mass resolution in the low-mass (m10) dwarfs is $\sim250M_{\odot}$ so the Sedov-Taylor phase is likely well-resolved, whereas in the more massive halos it is unresolved ($\sim2100-33000M_{\odot}$) and that may play a role in them having $\eta_{\rm E}<1$. Note, though, that when SNe are clustered, most of the hot gas may be contained in ``superbubbles'' which are better resolved.

Finally, there are other sources of mass, momentum, energy and metal input beyond type II SNe that are not included in our analytic reference injection rates (but which are included in the FIRE-2 simulations), so our energy and momentum loading factors may be overestimated.\footnote{On a related note, the core FIRE-2 simulations that we use do not include cosmic ray physics, which can significantly affect outflow and CGM properties \citep{hopkins20}. Our simulations also do not include a subgrid model for turbulent diffusion which would otherwise allow metals to no longer strictly follow mass. While this can affect the distribution of outflow metallicities for a given episode, our bulk shell-averaged measurements may be robust \citep[see arguments in][end of their section 5.4]{muratov17}.} It may not be possible to estimate, in a clean way, total injection rates in cosmological simulations due to the number of processes, many of which are approximated using complicated subgrid models. For example, we do not account for type Ia SNe, radiative heating from the stars in the central galaxy, outflows and turbulence stirred by satellites, or gravitational shock heating of infalling gas in more massive halos. All of these effects make it more complicated to interpret the absolute values of $\eta_{\rm E}$ and $\eta_{\rm p}$, especially when they are high. In addition, the m13 halos, which tend to be outliers in some of our relations, are only run down to $z=1$ so we are missing roughly half of their evolution in our intermediate redshift bin (down to $z=0.5$). It is possible that the peculiarly high outflow rates of the m13 halos may decrease significantly at $z<1$ (as happens for the m12 halos at later times).

\section{Summary}\label{sec:summary}
We have characterized the mass, momentum, energy and metal loading factors of multi-phase galactic winds in the FIRE-2 cosmological  ``zoom-in'' simulations. To accomplish this, we implemented a physically motivated Bernoulli velocity wind selection criterion to account for the bulk kinetic, thermal and potential energy of gas particles and exclude slower, turbulent moving outflows from genuine winds. We report outflow measurements at two characteristic radii: close to the ISM boundary ($0.1-0.2R_{\rm vir}$) and the halo boundary ($1.0-1.1R_{\rm vir}$). Given the inherently multi-phase nature of galactic winds, we computed loading factors separately for the cold ($T<10^3$K), warm ($10^3$K$<T<10^5$K) and hot ($T>10^5$K) phases. In order to minimize systematics due to travel time delays, entrainment, etc., our fiducial loading factors were measured as averages over three relatively large redshift bins: low redshift ($z=0.0-0.5$), intermediate redshift ($z=0.5-2.0$) and high-redshift ($z=2.0-4.0$). We also implemented a robust algorithm to derive individual burst-averaged loading factors for the ISM shell to complement our redshift-averaged measurements and explore correlations with physical properties on short timescales. With the large sample size of the core FIRE-2 suite, we analyzed halos in four mass bins: low-mass dwarfs ($M_{\rm vir}\sim10^{10}M_{\odot}$ at $z=0$), intermediate-mass dwarfs ($\sim10^{11}M_{\odot}$ at $z=0$), MW-mass galaxies ($\sim10^{12}M_{\odot}$ at $z=0$), and more massive halos at high redshift ($\sim10^{12.5}-10^{13}M_{\odot}$ by $z=1$). 

Our main takeaways are as follows: 

\begin{enumerate}

\item The ISM mass loading factor is preferentially higher for dwarfs (of order $\sim100$) compared to more massive halos (below unity). Cold mass loading fractions are negligible in all halos except high redshift dwarfs where it approaches order unity. Warm mass loading fractions dominate over cold and hot mass loading fractions in dwarfs, whereas hot outflows carry most of the mass in the more massive halos. Similarly, the ISM momentum, energy and metal loadings are of order unity in the dwarfs (especially at high redshift) and significantly lower in the more massive halos. The warm phase tends to carry most of the momentum, energy and metals in the dwarfs whereas the hot phase dominates for the more massive halos. Correlating total ISM mass loadings with the global halo virial velocity results in a $V_{\rm vir}^{-2}$ dependence at high-redshift (consistent with energy-driven winds), but a steeper scaling at later times.

\item The average Bernoulli velocity of hot outflows is $2-5\times$ the difference in gravitational potential between $0.1R_{\rm vir}$ and $R_{\rm vir}$, especially in high-redshift dwarfs, meaning that we should expect to see substantial outflows at $R_{\rm vir}$. Indeed, mass outflow rates at $R_{\rm vir}$ are several times larger than mass outflow rates at $0.1R_{\rm vir}$ in the dwarfs, indicative of swept up CGM gas. In the low-redshift MW halos, this $\eta_{\rm M,halo}/\eta_{\rm M,ISM}$ ratio is also of order unity when we consider only ``escaping'' ISM outflows. Energy outflow rates at $R_{\rm vir}$ are comparable to those at $0.1R_{\rm vir}$ in dwarfs whereas this $\eta_{\rm E,halo}/\eta_{\rm E,ISM}$ ratio is much lower ($\sim0.25$) in low-redshift MW halos. Halo-scale momentum loading factors exceed ISM momentum loading factors in dwarfs (as expected for energy-conserving outflows) but are comparable in MW-mass halos at later times. Most of the metals that leave the ISM tend to escape from dwarf halos but are retained within low-redshift MW-mass halos.

\item Correlating burst-averaged wind loading factors with $\dot{M}$-weighted physical properties over individual burst windows reveals a few interesting trends. Burst-averaged $\eta_{\rm M}$ shows a clear negative correlation with $\Sigma_{\rm gas}$ but there is substantially more scatter versus $\Sigma_{\rm SFR}$. In contrast, we see a clear positive correlation between the burst-averaged $\eta_{\rm M}$ and a measure of how locally bursty a SF episode is (defined as the peak SFR within a burst interval divided by the 1 Gyr-averaged SFR). We see a lot of scatter and effectively no correlation between $\eta_{\rm M}$ and $f_{\rm dense}$, which may reflect competing trends between how the dense ISM gas fraction affects starburst and wind strengths, and/or that our simple $f_{\rm dense}$ statistic is not an ideal measure of ISM wind breakout conditions. Finally, we see a strong negative correlation between $\eta_{\rm M}$ and $t_{\rm cool}/t_{\rm ff}$ (which is larger than two when the inner halo is virialized): mass loading is preferentially suppressed when the inner halo is virialized (as is the case in massive halos at later times but not in dwarfs or at high redshift).
\end{enumerate}

Our results suggest that the reduced global star formation efficiency of dwarfs may at least partially be driven by their more powerful winds. At the same time, our comprehensive analysis has revealed the multi-phase nature of weaker SN-driven winds in massive halos. Our findings can be used to guide future controlled numerical experiments that aim to clarify the key parameters that determine the properties of galactic winds. In future work, we will use this rich dataset to implement preventative stellar feedback in next-generation SAMs. The traditional approach of relying on mass and metal ejection alone can be improved upon by also considering the energy and momentum injected into the CGM/IGM by SN-driven winds. This may have important physical implications for CGM/IGM heating rates and observable consequences for the redshift evolution of the mass-metallicity relation, the stellar-to-halo-mass relation for dwarfs, and the chemical enrichment of the CGM/IGM.

\section*{Acknowledgements}
We thank Kevin Bundy, Yakov Faerman, Piero Madau, Eve Ostriker, Kung-Yi Su, Cassi Lochhaas and the FIRE and SMAUG teams for helpful discussions. We are grateful to the referee for helping to improve the clarity of this paper. VP was supported by the National Science Foundation Graduate Research Fellowship Program under Grant No. 1339067 and a Flatiron Institute Pre-Doctoral Fellowship. DAA was supported in part by NSF grant AST-2009687. GLB acknowledges financial support from the NSF (grant AST-1615955, OAC-1835509) and computing support from NSF XSEDE. CGK was supported by NASA ATP grant NNX17AG26G. AW received support from NASA through ATP grants 80NSSC18K1097 and 80NSSC20K0513; HST grants GO-14734, AR-15057, AR-15809, and GO-15902 from STScI; the Heising-Simons Foundation; and a Hellman Fellowship. CAFG was supported by NSF through grants AST-1715216 and CAREER award AST-1652522; by NASA through grant 17-ATP17-0067; by STScI through grant HST-AR-16124.001-A; and by a Cottrell Scholar Award and a Scialog Award from the Research Corporation for Science Advancement. DK was supported by NSF grant AST-1715101. ZH was supported by a Gary A. McCue postdoctoral fellowship at UC Irvine. Support for PFH was provided by NSF Research Grants 1911233 \&\ 20009234, NSF CAREER grant 1455342, NASA grants 80NSSC18K0562, HST-AR-15800.001-A. Numerical calculations were run on the Caltech compute cluster ``Wheeler,'' allocations FTA-Hopkins/AST20016 supported by the NSF and TACC, and NASA HEC SMD-16-7592. The data used in this work were, in part, hosted on facilities supported by the Scientific Computing Core at the Flatiron Institute, a division of the Simons Foundation. The simulations were run using XSEDE allocations TG-AST160048 (supported by NSF grant ACI-1548562) and TG-AST120025, and Pleiades via the NASA HEC program through the NAS Division at Ames Research Center.

\vspace*{5mm}
\textit{Software:} yt \citep{yt}, Python 2.7 \citep{python}, IPython \citep{ipython}, NumPy \citep{numpy}, SciPy \citep[][version 1.2.1]{scipy}, Matplotlib \citep{matplotlib}, H5py \citep{h5py}, AstroPy \citep{astropy1,astropy2}

\vspace*{5mm}
\textit{Data Availability Statement:} The data underlying this article were provided by the FIRE Collaboration with permission. Data will be shared on reasonable request to the corresponding author with permission of the FIRE Collaboration.

\bibliographystyle{mnras}
\bibliography{references}

 \appendix
 
\section{Alternative Bernoulli velocity cuts}\label{sec:vBalt}
\begin{figure*} 
\begin{center}
\includegraphics[width=0.45\hsize]{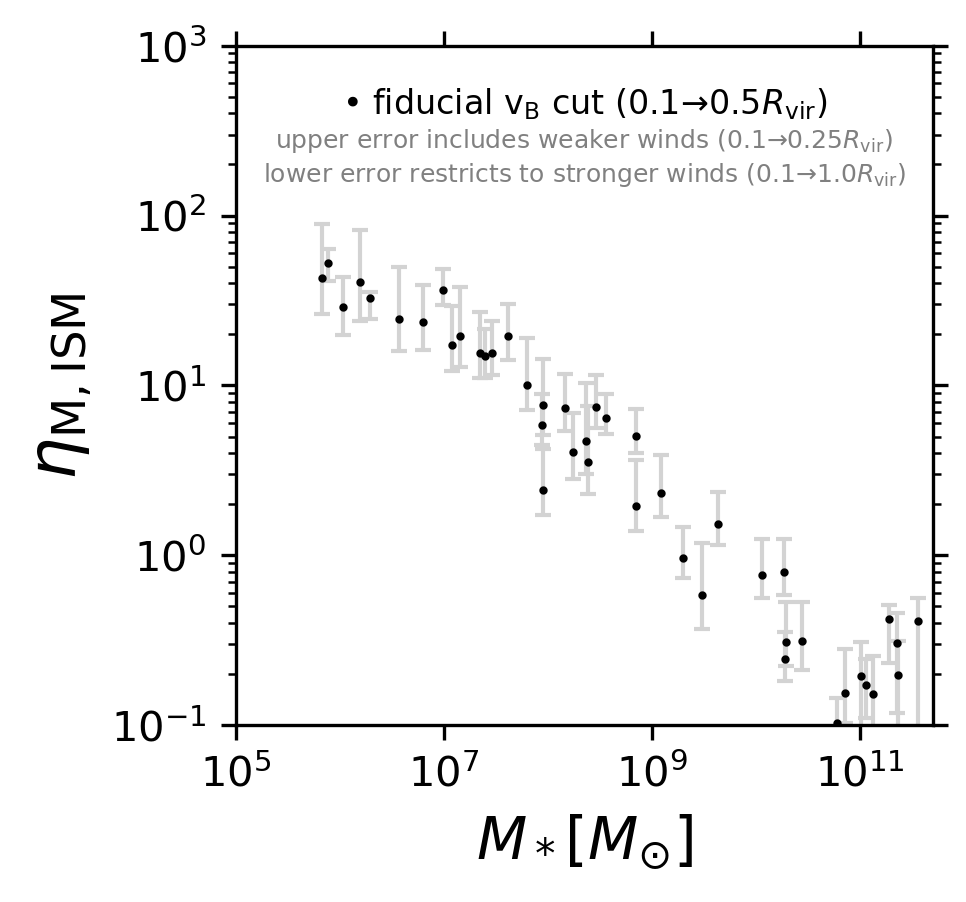}\includegraphics[width=0.45\hsize]{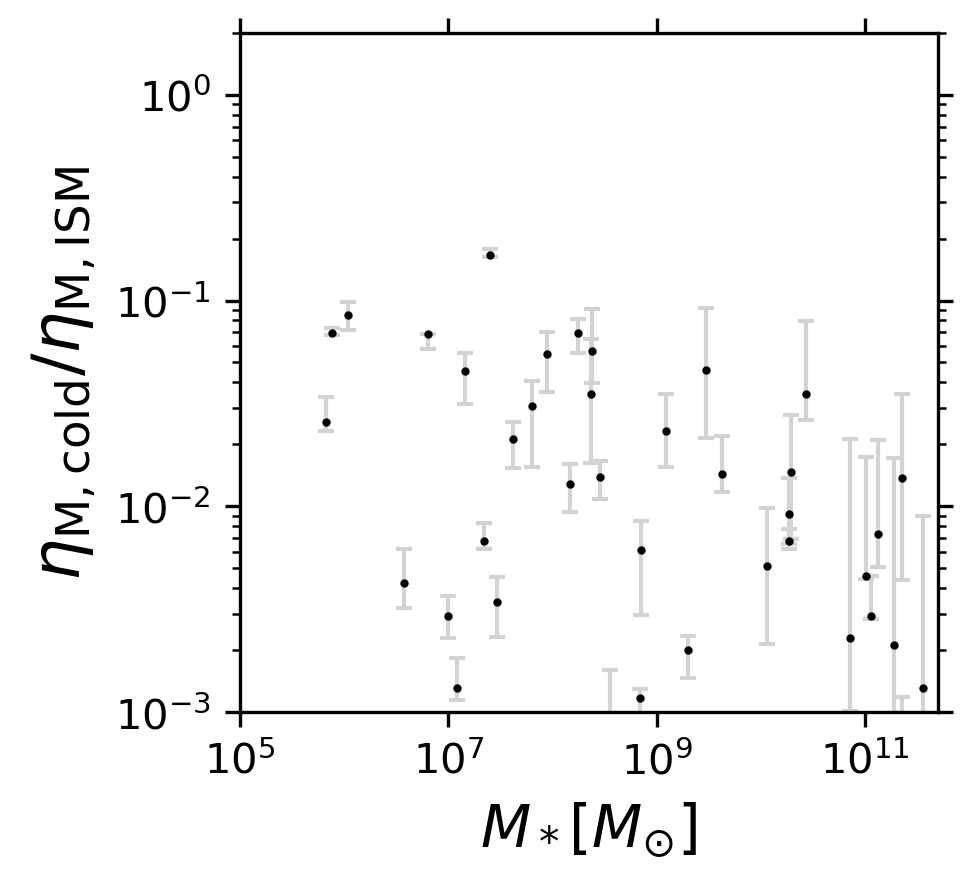}

\includegraphics[width=0.45\hsize]{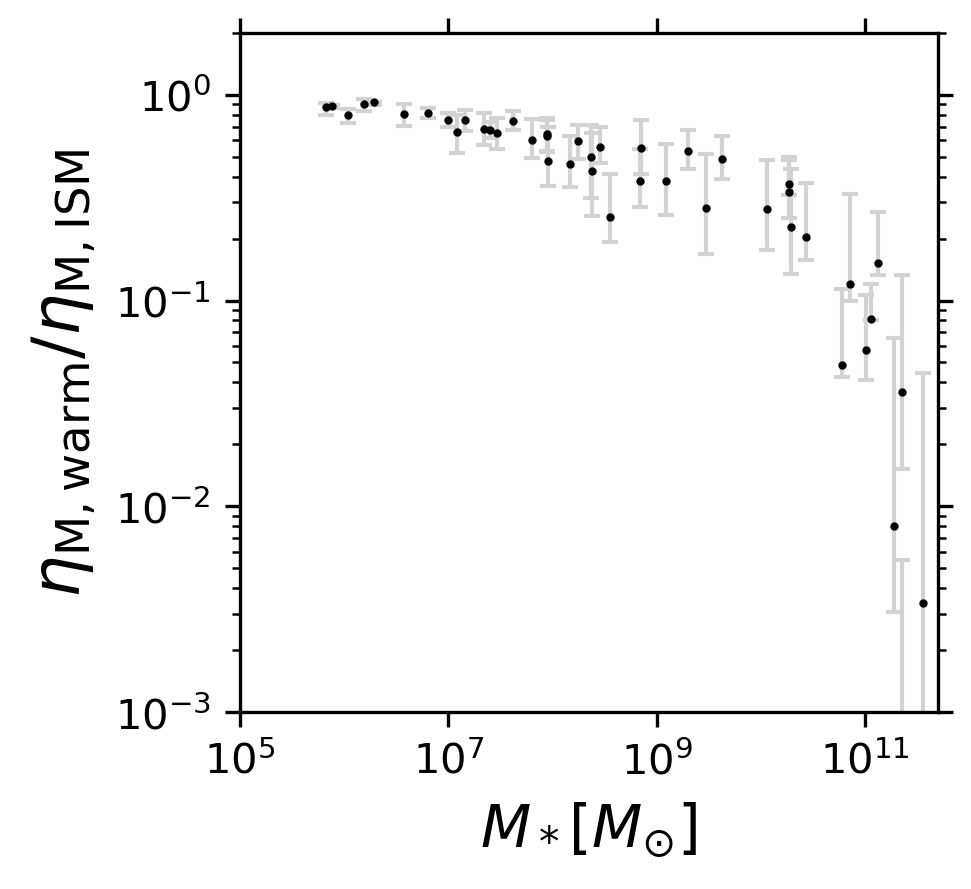}\includegraphics[width=0.45\hsize]{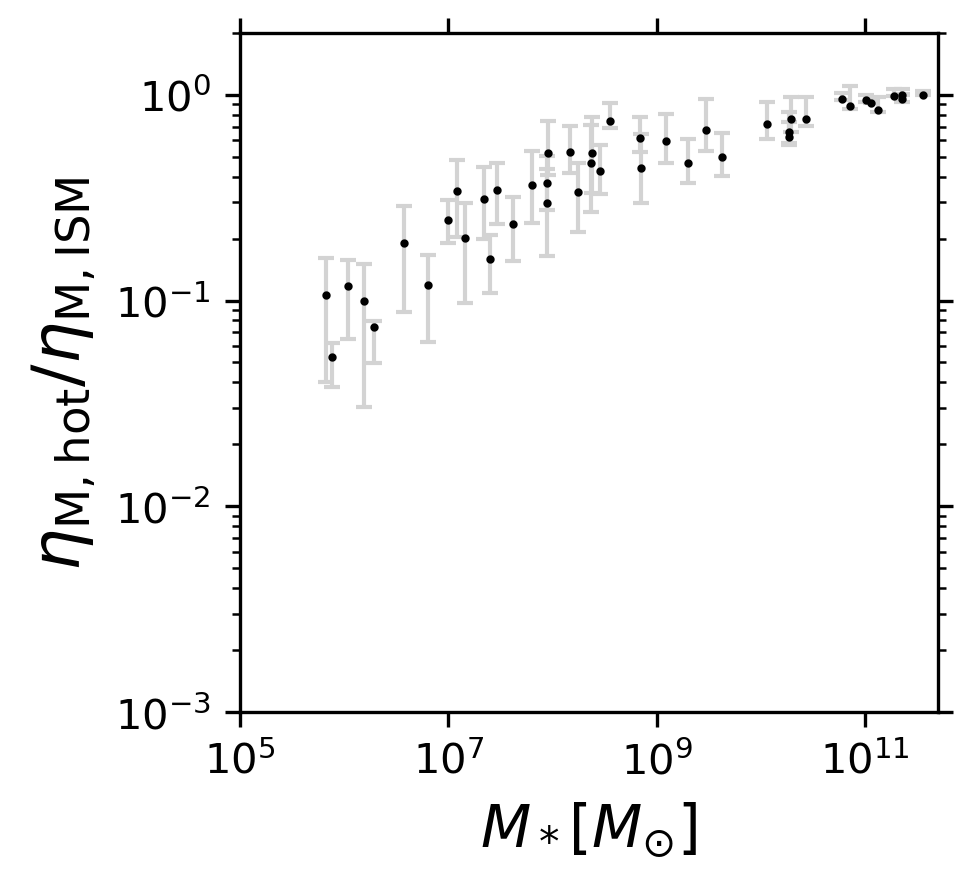}
\end{center}
\caption{Impact of varying the Bernoulli velocity cut for ISM-scale winds. The black points in every panel show our fiducial mass loading factors and phase fractions (as in \autoref{fig:etaM_all}; i.e., selecting outflowing particles that have enough energy to go from $0.1R_{\rm vir}$ to at least $0.5R_{\rm vir}$, if not further). The gray errorbars show how each point would increase or decrease if we included weaker winds (with enough energy to reach only half of our fiducial target distance, $0.25R_{\rm vir}$) or restricted to even stronger winds (with enough energy to get twice as far as our fiducial target distance, $0.5R_{\rm vir}$). Top-left: overall mass loading factor versus stellar mass. Top-right: cold mass loading fraction. Bottom-left: warm mass loading fraction. Bottom-right: hot mass loading fraction.}
\label{fig:vBalt}
\end{figure*}

Here we illustrate how our ISM-scale mass loading factors would change had we adopted a different Bernoulli velocity cut. \autoref{fig:vBalt} shows our fiducial mass loading measurements where we selected outflowing particles that had a high enough $v_{\rm B}$ to go from $0.1R_{\rm vir}$ to at least $0.5R_{\rm vir}$, if not further (i.e., reproducing the data from \autoref{fig:etaM_all}). But now we also show errorbars that indicate how much larger $\eta_{\rm M,ISM}$ would be if we included weaker outflows that only have enough energy to get to $0.25R_{\rm vir}$ (half of our fiducial target distance). In contrast, the lower errorbar indicates how much the mass loading factor would decrease if we restricted the selection to stronger outflows that could get to at least $R_{\rm vir}$ (twice as large as our fiducial target distance). The top-left panel of \autoref{fig:vBalt} shows that changing the target distance by a factor of two for the $v_{\rm B}$ cut correspondingly leads to a factor of a few change in $\eta_{\rm M,ISM}$.

Perhaps more important is to consider how the multi-phase partitioning might change if we included or excluded slower outflows. The other three panels of \autoref{fig:vBalt} show that it is generally the more massive galaxies where the multi-phase partitioning can change by a significant amount, particularly for cold and warm outflows. This makes sense because changing the $v_{\rm B}$ cut will disproportionately affect the cooler outflowing particles which may not have a high enough enthalpy to get to $0.5R_{\rm vir}$ or $R_{\rm vir}$. Nevertheless, even with the errorbars, we still see similar overall trends where, e.g., lower mass galaxies have predominantly warm winds whereas the more massive galaxies have mostly hot winds by mass fraction.

We expect similar results for the dependence of the momentum, energy and metal loading factors on the choice of $v_{\rm B}$ cut.

\section{Tabulated properties and multi-phase loading factors}\label{sec:tables}
In this appendix we provide value-added tables with our loading factor measurements. \autoref{tab:lowz}, \autoref{tab:midz} and \autoref{tab:highz} respectively give the average properties and multi-phase loading factors of the FIRE-2 halos in our low-redshift ($z=0.0-0.5$), intermediate-redshift ($z=0.5-2.0$) and high-redshift ($z=2.0-4.0$) bins. Note that the tables printed in this paper are only a subset of the much longer set of tables that give multi-phase loading factors in both the ISM and virial shells for mass, momentum, energy and metals. The full set of supplementary tables is available for download online. We note that we have provided the average global SFR which can be used to convert the dimensionless loading factors back into raw mass, momentum, energy and metal outflow rates in physical units (following \autoref{sec:loadings}). The mass-weighted average Bernoulli velocity (excluding the gravitational term) can be approximated as $\sqrt{\dot{E}/\dot{M}}$ and the mass-weighted average radial velocity (including the thermal momentum component) can be approximated as $\dot{p}/\dot{M}$.

Additionally, \autoref{tab:instant} provides a catalog of properties for individual outflow episodes in all halos at $z<4$. This catalog includes individual integrated burst stellar masses, wind masses and mass loading factors (both combined and split into cold, warm and hot phases). Similarly, integrated multi-phase momentum, energy and metal loadings are also provided for each individual burst. Burst interval averaged SFR and gas surface densities, dense ISM gas fractions, global stellar mass and halo virial velocity, etc. are also provided as discussed in the main text.

\begin{table*}
\tiny
\begin{tabular}{cccccccccccHHHHHHHHHHHHHHHHHHHHHHHHHHHH}
Halo & log $M_{\rm vir}$ & log $M_*$ & $V_{\rm vir}$ & log SFR & $V_{\rm esc,ISM}$ & $V_{\rm esc,halo}$ & $\eta_{\rm M,ISM}$ & $f_{\rm M,ISM,cold}$ & $f_{\rm M,ISM,warm}$ & $f_{\rm M,ISM,hot}$ & $\eta_{\rm p,ISM}$ & $f_{\rm p,ISM,cold}$ & $f_{\rm p,ISM,warm}$ & $f_{\rm p,ISM,hot}$ & $\eta_{\rm E,ISM}$ & $f_{\rm E,ISM,cold}$ & $f_{\rm E,ISM,warm}$ & $f_{\rm E,ISM,hot}$ & $\eta_{\rm Z,ISM}$ & $f_{\rm Z,ISM,cold}$ & $f_{\rm Z,ISM,warm}$ & $f_{\rm Z,ISM,hot}$ & $\eta_{\rm M,halo}$ & $f_{\rm M,halo,cold}$ & $f_{\rm M,halo,warm}$ & $f_{\rm M,halo,hot}$ & $\eta_{\rm p,halo}$ & $f_{\rm p,halo,cold}$ & $f_{\rm p,halo,warm}$ & $f_{\rm p,halo,hot}$ & $\eta_{\rm E,halo}$ & $f_{\rm E,halo,cold}$ & $f_{\rm E,halo,warm}$ & $f_{\rm E,halo,hot}$ & $\eta_{\rm Z,halo}$ & $f_{\rm Z,halo,cold}$ & $f_{\rm Z,halo,warm}$ & $f_{\rm Z,halo,hot}$ \\
\hline\\
m10q & 9.8546 & 6.2894 & 26.4429 & -4.3402 & 65.9863 & 26.7786 & 32.8769 & 0.0000 & 0.9164 & 0.0746 & 1.7546 & 0.0000 & 0.8129 & 0.0848 & 0.1622 & 0.0000 & 0.8034 & 0.1897 & 1.1000 & 0.0000 & 0.9070 & 0.0856 & 72.2407 & 0.0000 & 0.9601 & 0.0006 & 2.9885 & 0.0000 & 0.7783 & 0.0018 & 0.1150 & 0.0000 & 0.9562 & 0.0087 & 0.6875 & 0.0000 & 0.9692 & 0.0006 \\
m10v & 9.8418 & 4.5565 & 26.2860 & -4.3757 & 62.1762 & 27.1094 & 45.9859 & 0.0000 & 0.9687 & 0.0000 & 2.8840 & 0.0000 & 0.7698 & 0.0000 & 0.1433 & 0.0000 & 0.9719 & 0.0000 & 0.0129 & 0.0000 & 0.9833 & 0.0000 & 152.7317 & 0.0000 & 0.9203 & 0.0000 & 11.2564 & 0.0000 & 0.6963 & 0.0000 & 0.2645 & 0.0000 & 0.9204 & 0.0000 & 0.0167 & 0.0000 & 0.9179 & 0.0000 \\
m10y & 10.1182 & 6.9964 & 32.5870 & -3.2242 & 78.8187 & 33.9041 & 36.6056 & 0.0029 & 0.7505 & 0.2461 & 1.6676 & 0.0041 & 0.6541 & 0.3356 & 0.6900 & 0.0041 & 0.4738 & 0.5220 & 2.4067 & 0.0055 & 0.7570 & 0.2373 & 82.2951 & 0.0000 & 0.9736 & 0.0235 & 2.2670 & 0.0000 & 0.9343 & 0.0289 & 0.3711 & 0.0000 & 0.9414 & 0.0571 & 2.1072 & 0.0000 & 0.9778 & 0.0203 \\
m10z & 10.5395 & 7.4673 & 44.3759 & -2.3722 & 105.8409 & 46.0409 & 15.5688 & 0.0034 & 0.6518 & 0.3447 & 0.8104 & 0.0033 & 0.5331 & 0.4606 & 0.4051 & 0.0024 & 0.3457 & 0.6519 & 1.1901 & 0.0086 & 0.6821 & 0.3093 & 32.3414 & 0.0000 & 0.8190 & 0.1795 & 1.1601 & 0.0000 & 0.7876 & 0.1847 & 0.2163 & 0.0000 & 0.7041 & 0.2949 & 0.6704 & 0.0000 & 0.8480 & 0.1514 \\
m11a & 10.5537 & 7.9441 & 46.0388 & -1.5776 & 113.1283 & 48.0383 & 5.8186 & 0.0003 & 0.6273 & 0.3724 & 0.3669 & 0.0003 & 0.4769 & 0.5217 & 0.2650 & 0.0001 & 0.2492 & 0.7506 & 0.6827 & 0.0011 & 0.6515 & 0.3474 & 10.3391 & 0.0000 & 0.5591 & 0.4402 & 0.3963 & 0.0000 & 0.4606 & 0.5288 & 0.1298 & 0.0000 & 0.2981 & 0.7017 & 0.3727 & 0.0000 & 0.7552 & 0.2447 \\
m11b & 10.5742 & 7.9563 & 46.7756 & -1.8808 & 120.7866 & 47.6938 & 2.4105 & 0.0000 & 0.4769 & 0.5229 & 0.1374 & 0.0000 & 0.3837 & 0.6105 & 0.0677 & 0.0000 & 0.2938 & 0.7061 & 0.2333 & 0.0000 & 0.5652 & 0.4347 & 3.6542 & 0.0000 & 0.9049 & 0.0912 & 0.1506 & 0.0000 & 0.8541 & 0.0905 & 0.0205 & 0.0000 & 0.8398 & 0.1575 & 0.0918 & 0.0000 & 0.9242 & 0.0735 \\
m11c & 11.1002 & 8.8442 & 70.1262 & -1.0490 & 172.2316 & 72.7489 & 5.0571 & 0.0012 & 0.3799 & 0.6189 & 0.4646 & 0.0010 & 0.2924 & 0.7055 & 0.4354 & 0.0006 & 0.1644 & 0.8349 & 1.1606 & 0.0097 & 0.4685 & 0.5217 & 7.8742 & 0.0000 & 0.2456 & 0.7540 & 0.3947 & 0.0000 & 0.1870 & 0.8028 & 0.1601 & 0.0000 & 0.1338 & 0.8660 & 0.6816 & 0.0000 & 0.3757 & 0.6242 \\
m11f & 11.5972 & 10.2756 & 106.2536 & 0.7062 & 271.1128 & 109.0921 & 0.2455 & 0.0068 & 0.3687 & 0.6244 & 0.0341 & 0.0059 & 0.3148 & 0.6765 & 0.0401 & 0.0043 & 0.2305 & 0.7652 & 0.1958 & 0.0238 & 0.5852 & 0.3910 & 0.1468 & 0.0000 & 0.1466 & 0.8532 & 0.0095 & 0.0000 & 0.1331 & 0.8591 & 0.0046 & 0.0000 & 0.1313 & 0.8686 & 0.0455 & 0.0000 & 0.5381 & 0.4619 \\
m11q & 11.1147 & 8.5553 & 70.5149 & -1.4055 & 169.3732 & 73.2543 & 6.4407 & 0.0004 & 0.2534 & 0.7461 & 0.5759 & 0.0004 & 0.1961 & 0.8002 & 0.4778 & 0.0003 & 0.1235 & 0.8762 & 1.2677 & 0.0026 & 0.3553 & 0.6421 & 11.5998 & 0.0000 & 0.1876 & 0.8118 & 0.5392 & 0.0000 & 0.1560 & 0.8256 & 0.1438 & 0.0000 & 0.1230 & 0.8766 & 0.5164 & 0.0000 & 0.3699 & 0.6298 \\
m11v & 11.1405 & 9.3015 & 73.6367 & -0.4062 & 174.0530 & 79.0590 & 0.9672 & 0.0020 & 0.5307 & 0.4672 & 0.0908 & 0.0014 & 0.4087 & 0.5886 & 0.0862 & 0.0007 & 0.2397 & 0.7595 & 0.6289 & 0.0037 & 0.6155 & 0.3808 & 1.5588 & 0.0000 & 0.0998 & 0.8999 & 0.0827 & 0.0000 & 0.0655 & 0.9245 & 0.0330 & 0.0000 & 0.0477 & 0.9522 & 0.1271 & 0.0000 & 0.2502 & 0.7497 \\
m12f & 12.0816 & 10.8553 & 156.5903 & 1.1966 & 395.9206 & 160.9728 & 0.1542 & 0.0023 & 0.1202 & 0.8774 & 0.0286 & 0.0017 & 0.1023 & 0.8913 & 0.0353 & 0.0017 & 0.1053 & 0.8930 & 0.0936 & 0.0094 & 0.2701 & 0.7205 & 0.1073 & 0.0000 & 0.0096 & 0.9902 & 0.0099 & 0.0000 & 0.0092 & 0.9801 & 0.0049 & 0.0000 & 0.0116 & 0.9883 & 0.0065 & 0.0000 & 0.1641 & 0.8358 \\
m12m & 12.0184 & 11.0586 & 152.7088 & 1.3188 & 399.2244 & 156.7018 & 0.1728 & 0.0029 & 0.0816 & 0.9154 & 0.0334 & 0.0026 & 0.0731 & 0.9206 & 0.0470 & 0.0022 & 0.0651 & 0.9326 & 0.2001 & 0.0094 & 0.2362 & 0.7544 & 0.0885 & 0.0000 & 0.0144 & 0.9854 & 0.0084 & 0.0000 & 0.0124 & 0.9785 & 0.0048 & 0.0000 & 0.0134 & 0.9865 & 0.0160 & 0.0000 & 0.2877 & 0.7123 \\
m12i & 11.9294 & 10.7769 & 139.7645 & 1.0767 & 360.6920 & 143.3503 & 0.1034 & 0.0003 & 0.0486 & 0.9510 & 0.0201 & 0.0002 & 0.0398 & 0.9498 & 0.0177 & 0.0005 & 0.0664 & 0.9330 & 0.0506 & 0.0039 & 0.2864 & 0.7097 & 0.0398 & 0.0000 & 0.0185 & 0.9811 & 0.0038 & 0.0000 & 0.0211 & 0.9601 & 0.0016 & 0.0000 & 0.0407 & 0.9590 & 0.0050 & 0.0000 & 0.3786 & 0.6213 \\
\end{tabular}
\caption{Average properties and loading factors of the FIRE-2 halos in our low redshift bin ($z=0.0-0.5$). We provide some basic global properties: halo virial mass ($M_{\odot}$), stellar mass ($M_{\odot}$), virial velocity (km s$^{-1}$), SFR ($M_{\odot}$ yr$^{-1}$), escape velocity from the ISM at $0.1R_{\rm vir}$ (km s$^{-1}$) and escape velocity from the halo at $R_{\rm vir}$ (km s$^{-1}$). For the loading factors, we provide the average total loading factor (dimensionless) and the corresponding cold, warm and hot phase fractions ($\eta_{\rm phase}/\eta$). As the full table is much longer, in the text here we only show a limited set of columns for the ISM-scale mass loading factors; there are additional columns giving the loading factors and their corresponding phase fractions for momentum, energy and metal outflows (for both the ISM and virial shells). The m13 halos are not shown since they are only run down to $z=1$ (they appear in the subsequent two tables). \texttt{This supplementary table is available for download online in the journal.}}
\label{tab:lowz}
\end{table*}

\begin{table*}
\tiny
\begin{tabular}{cccccccccccHHHHHHHHHHHHHHHHHHHHHHHHHHHH}
Halo & log $M_{\rm vir}$ & log $M_*$ & $V_{\rm vir}$ & log SFR & $V_{\rm esc,ISM}$ & $V_{\rm esc,halo}$ & $\eta_{\rm M,ISM}$ & $f_{\rm M,ISM,cold}$ & $f_{\rm M,ISM,warm}$ & $f_{\rm M,ISM,hot}$ & $\eta_{\rm p,ISM}$ & $f_{\rm p,ISM,cold}$ & $f_{\rm p,ISM,warm}$ & $f_{\rm p,ISM,hot}$ & $\eta_{\rm E,ISM}$ & $f_{\rm E,ISM,cold}$ & $f_{\rm E,ISM,warm}$ & $f_{\rm E,ISM,hot}$ & $\eta_{\rm Z,ISM}$ & $f_{\rm Z,ISM,cold}$ & $f_{\rm Z,ISM,warm}$ & $f_{\rm Z,ISM,hot}$ & $\eta_{\rm M,halo}$ & $f_{\rm M,halo,cold}$ & $f_{\rm M,halo,warm}$ & $f_{\rm M,halo,hot}$ & $\eta_{\rm p,halo}$ & $f_{\rm p,halo,cold}$ & $f_{\rm p,halo,warm}$ & $f_{\rm p,halo,hot}$ & $\eta_{\rm E,halo}$ & $f_{\rm E,halo,cold}$ & $f_{\rm E,halo,warm}$ & $f_{\rm E,halo,hot}$ & $\eta_{\rm Z,halo}$ & $f_{\rm Z,halo,cold}$ & $f_{\rm Z,halo,warm}$ & $f_{\rm Z,halo,hot}$ \\
\hline\\
m10q & 9.6679 & 6.1910 & 28.3840 & -3.9136 & 68.7518 & 29.8466 & 40.4091 & 0.0000 & 0.8977 & 0.0999 & 1.3719 & 0.0000 & 0.8011 & 0.1602 & 0.2686 & 0.0000 & 0.6593 & 0.3394 & 1.3436 & 0.0000 & 0.8700 & 0.1285 & 98.2785 & 0.0000 & 0.9841 & 0.0005 & 3.4761 & 0.0000 & 0.8672 & 0.0010 & 0.2217 & 0.0000 & 0.9846 & 0.0024 & 0.8289 & 0.0000 & 0.9880 & 0.0006 \\
m10v & 9.6041 & 3.6395 & 26.8659 & -inf & 57.1233 & 29.1378 & inf & nan & nan & nan & inf & nan & nan & nan & inf & nan & nan & nan & inf & nan & nan & nan & inf & nan & nan & nan & inf & nan & nan & nan & inf & nan & nan & nan & inf & nan & nan & nan \\
m10y & 9.9107 & 6.5727 & 34.4039 & -2.9452 & 79.6701 & 37.1780 & 24.6485 & 0.0042 & 0.8047 & 0.1907 & 0.9709 & 0.0049 & 0.6635 & 0.3253 & 0.3838 & 0.0039 & 0.4420 & 0.5540 & 0.9755 & 0.0082 & 0.7387 & 0.2530 & 60.8850 & 0.0000 & 0.9731 & 0.0231 & 1.9546 & 0.0000 & 0.9188 & 0.0295 & 0.2544 & 0.0000 & 0.9308 & 0.0666 & 0.7384 & 0.0000 & 0.9465 & 0.0512 \\
m10z & 10.2583 & 7.0819 & 44.9341 & -2.4946 & 103.5742 & 47.8982 & 17.2078 & 0.0013 & 0.6583 & 0.3403 & 0.8754 & 0.0012 & 0.5090 & 0.4869 & 0.4657 & 0.0007 & 0.2901 & 0.7092 & 1.1508 & 0.0055 & 0.6646 & 0.3298 & 35.9543 & 0.0000 & 0.7176 & 0.2811 & 1.4359 & 0.0000 & 0.6619 & 0.3167 & 0.3760 & 0.0000 & 0.5464 & 0.4531 & 0.6378 & 0.0000 & 0.7931 & 0.2063 \\
m11a & 10.3552 & 7.3467 & 48.9326 & -2.0557 & 112.5164 & 51.5667 & 15.4777 & 0.0068 & 0.6832 & 0.3100 & 0.9227 & 0.0048 & 0.5190 & 0.4748 & 0.6371 & 0.0021 & 0.2791 & 0.7188 & 0.6571 & 0.0065 & 0.6172 & 0.3763 & 29.8515 & 0.0000 & 0.5667 & 0.4326 & 1.1407 & 0.0000 & 0.4732 & 0.5131 & 0.3412 & 0.0000 & 0.3310 & 0.6687 & 0.4185 & 0.0000 & 0.6617 & 0.3380 \\
m11b & 10.3977 & 7.6180 & 49.8646 & -2.1166 & 118.1167 & 52.8372 & 19.6582 & 0.0213 & 0.7432 & 0.2355 & 1.1172 & 0.0166 & 0.6182 & 0.3643 & 0.6760 & 0.0089 & 0.3953 & 0.5958 & 1.3827 & 0.0182 & 0.7162 & 0.2655 & 29.4548 & 0.0000 & 0.6677 & 0.3313 & 1.1796 & 0.0000 & 0.5647 & 0.4186 & 0.3729 & 0.0000 & 0.3836 & 0.6160 & 0.8900 & 0.0000 & 0.7479 & 0.2517 \\
m11c & 10.8999 & 8.4603 & 74.6748 & -0.9751 & 173.6736 & 78.3749 & 7.4433 & 0.0139 & 0.5591 & 0.4270 & 0.6934 & 0.0110 & 0.4333 & 0.5553 & 0.7258 & 0.0059 & 0.2330 & 0.7611 & 1.1906 & 0.0194 & 0.5329 & 0.4477 & 11.0761 & 0.0000 & 0.1745 & 0.8253 & 0.6157 & 0.0000 & 0.1239 & 0.8697 & 0.2837 & 0.0000 & 0.0914 & 0.9085 & 0.6790 & 0.0000 & 0.3444 & 0.6556 \\
m11f & 11.3949 & 9.6310 & 111.3398 & 0.4287 & 264.9979 & 116.9103 & 1.5264 & 0.0144 & 0.4861 & 0.4995 & 0.2114 & 0.0114 & 0.3841 & 0.6041 & 0.2990 & 0.0067 & 0.2273 & 0.7660 & 0.7684 & 0.0312 & 0.5453 & 0.4235 & 1.2194 & 0.0000 & 0.1391 & 0.8608 & 0.0997 & 0.0000 & 0.1139 & 0.8833 & 0.0765 & 0.0000 & 0.0771 & 0.9229 & 0.2768 & 0.0000 & 0.3438 & 0.6562 \\
m11q & 10.9105 & 8.1635 & 73.8962 & -1.3263 & 169.5099 & 78.1021 & 7.3854 & 0.0128 & 0.4615 & 0.5257 & 0.6839 & 0.0096 & 0.3423 & 0.6466 & 0.7379 & 0.0051 & 0.1748 & 0.8201 & 0.9280 & 0.0132 & 0.4074 & 0.5794 & 20.4283 & 0.0000 & 0.1178 & 0.8819 & 1.0955 & 0.0000 & 0.0915 & 0.9001 & 0.4460 & 0.0000 & 0.0796 & 0.9203 & 0.7718 & 0.0000 & 0.2452 & 0.7547 \\
m11v & 10.8051 & 8.8465 & 70.3905 & -0.5639 & 170.5885 & 74.2094 & 1.9429 & 0.0061 & 0.5524 & 0.4414 & 0.1920 & 0.0039 & 0.3781 & 0.6175 & 0.2403 & 0.0015 & 0.1733 & 0.8253 & 0.5512 & 0.0060 & 0.5601 & 0.4339 & 2.5113 & 0.0000 & 0.2671 & 0.7327 & 0.1387 & 0.0000 & 0.1820 & 0.8118 & 0.0743 & 0.0000 & 0.1075 & 0.8924 & 0.2745 & 0.0000 & 0.4100 & 0.5900 \\
m12f & 11.8129 & 10.2684 & 155.6512 & 0.9783 & 370.1400 & 163.9351 & 0.7978 & 0.0092 & 0.3351 & 0.6557 & 0.1412 & 0.0072 & 0.2700 & 0.7221 & 0.2529 & 0.0046 & 0.1581 & 0.8373 & 0.5342 & 0.0318 & 0.4040 & 0.5642 & 0.8327 & 0.0004 & 0.0624 & 0.9372 & 0.0847 & 0.0003 & 0.0555 & 0.9415 & 0.0754 & 0.0002 & 0.0438 & 0.9559 & 0.1772 & 0.0006 & 0.2946 & 0.7048 \\
m12m & 11.8957 & 10.2896 & 159.2550 & 1.2817 & 393.2716 & 170.7018 & 0.3074 & 0.0147 & 0.2272 & 0.7580 & 0.0584 & 0.0097 & 0.1629 & 0.8266 & 0.1170 & 0.0048 & 0.0931 & 0.9021 & 0.2544 & 0.0126 & 0.2630 & 0.7244 & 0.2406 & 0.0000 & 0.0541 & 0.9458 & 0.0262 & 0.0000 & 0.0461 & 0.9491 & 0.0231 & 0.0000 & 0.0431 & 0.9568 & 0.0676 & 0.0000 & 0.2492 & 0.7508 \\
m12i & 11.7536 & 10.0605 & 147.9378 & 0.9407 & 344.9041 & 155.8538 & 0.7647 & 0.0052 & 0.2783 & 0.7165 & 0.1300 & 0.0033 & 0.2026 & 0.7937 & 0.2404 & 0.0015 & 0.1007 & 0.8978 & 0.4891 & 0.0079 & 0.3374 & 0.6548 & 0.8192 & 0.0000 & 0.0292 & 0.9707 & 0.0762 & 0.0000 & 0.0253 & 0.9715 & 0.0608 & 0.0000 & 0.0197 & 0.9803 & 0.1335 & 0.0000 & 0.1834 & 0.8166 \\
A1 & 12.4241 & 11.3586 & 276.6363 & 1.8137 & 695.7145 & 287.7350 & 0.1981 & 0.0000 & 0.0000 & 0.9998 & 0.1072 & 0.0000 & 0.0000 & 0.9849 & 0.0639 & 0.0000 & 0.0001 & 0.9998 & 0.0037 & 0.0000 & 0.0004 & 0.9994 & 0.1275 & 0.0000 & 0.0002 & 0.9997 & 0.0231 & 0.0000 & 0.0001 & 0.9911 & 0.0151 & 0.0000 & 0.0003 & 0.9996 & 0.0036 & 0.0000 & 0.0043 & 0.9956 \\
A2 & 12.6153 & 11.5553 & 324.3633 & 1.9258 & 786.7868 & 340.2850 & 0.4118 & 0.0013 & 0.0034 & 0.9952 & 0.1946 & 0.0007 & 0.0017 & 0.9901 & 0.1885 & 0.0015 & 0.0035 & 0.9949 & 0.0182 & 0.0077 & 0.0319 & 0.9604 & 0.3343 & 0.0000 & 0.0004 & 0.9996 & 0.0615 & 0.0000 & 0.0003 & 0.9946 & 0.0556 & 0.0000 & 0.0006 & 0.9994 & 0.0108 & 0.0003 & 0.0068 & 0.9929 \\
A4 & 12.4939 & 11.2806 & 288.3921 & 1.8442 & 687.2344 & 306.7316 & 0.4200 & 0.0021 & 0.0081 & 0.9898 & 0.1417 & 0.0013 & 0.0053 & 0.9885 & 0.1738 & 0.0019 & 0.0083 & 0.9897 & 0.0548 & 0.0193 & 0.0772 & 0.9036 & 1.3513 & 0.0000 & 0.0003 & 0.9997 & 0.1852 & 0.0000 & 0.0003 & 0.9982 & 0.2181 & 0.0000 & 0.0004 & 0.9996 & 0.0808 & 0.0000 & 0.0093 & 0.9906 \\
A8 & 12.7956 & 11.3537 & 366.9799 & 2.2786 & 831.0209 & 388.9996 & 0.3029 & 0.0137 & 0.0360 & 0.9503 & 0.1291 & 0.0083 & 0.0220 & 0.9646 & 0.1983 & 0.0125 & 0.0321 & 0.9553 & 0.1002 & 0.0628 & 0.1288 & 0.8084 & 0.2978 & 0.0000 & 0.0007 & 0.9993 & 0.0578 & 0.0000 & 0.0006 & 0.9966 & 0.0764 & 0.0000 & 0.0006 & 0.9994 & 0.0186 & 0.0005 & 0.0250 & 0.9746 \\
\end{tabular}
\caption{Identical to \autoref{tab:lowz} but now for our intermediate redshift bin ($z=0.5-2.0$). \texttt{This supplementary table is available for download online in the journal.}}
\label{tab:midz}
\end{table*}

\begin{table*}
\tiny
\begin{tabular}{cccccccccccHHHHHHHHHHHHHHHHHHHHHHHHHHHH}
Halo & log $M_{\rm vir}$ & log $M_*$ & $V_{\rm vir}$ & log SFR & $V_{\rm esc,ISM}$ & $V_{\rm esc,halo}$ & $\eta_{\rm M,ISM}$ & $f_{\rm M,ISM,cold}$ & $f_{\rm M,ISM,warm}$ & $f_{\rm M,ISM,hot}$ & $\eta_{\rm p,ISM}$ & $f_{\rm p,ISM,cold}$ & $f_{\rm p,ISM,warm}$ & $f_{\rm p,ISM,hot}$ & $\eta_{\rm E,ISM}$ & $f_{\rm E,ISM,cold}$ & $f_{\rm E,ISM,warm}$ & $f_{\rm E,ISM,hot}$ & $\eta_{\rm Z,ISM}$ & $f_{\rm Z,ISM,cold}$ & $f_{\rm Z,ISM,warm}$ & $f_{\rm Z,ISM,hot}$ & $\eta_{\rm M,halo}$ & $f_{\rm M,halo,cold}$ & $f_{\rm M,halo,warm}$ & $f_{\rm M,halo,hot}$ & $\eta_{\rm p,halo}$ & $f_{\rm p,halo,cold}$ & $f_{\rm p,halo,warm}$ & $f_{\rm p,halo,hot}$ & $\eta_{\rm E,halo}$ & $f_{\rm E,halo,cold}$ & $f_{\rm E,halo,warm}$ & $f_{\rm E,halo,hot}$ & $\eta_{\rm Z,halo}$ & $f_{\rm Z,halo,cold}$ & $f_{\rm Z,halo,warm}$ & $f_{\rm Z,halo,hot}$ \\
\hline\\
m10q & 9.4611 & 5.8281 & 32.5904 & -3.1300 & 71.2789 & 34.8426 & 42.9513 & 0.0258 & 0.8679 & 0.1061 & 1.6363 & 0.0192 & 0.6404 & 0.3377 & 0.9544 & 0.0079 & 0.3067 & 0.6853 & 0.7983 & 0.0228 & 0.6327 & 0.3443 & 203.8417 & 0.0000 & 0.8481 & 0.1510 & 5.5965 & 0.0000 & 0.7223 & 0.2658 & 1.5242 & 0.0000 & 0.5087 & 0.4911 & 2.0830 & 0.0000 & 0.8160 & 0.1835 \\
m10v & 8.7289 & 3.6450 & 18.4540 & -inf & 40.5621 & 20.3317 & inf & nan & nan & nan & inf & nan & nan & nan & inf & nan & nan & nan & inf & nan & nan & nan & inf & nan & nan & nan & inf & nan & nan & nan & inf & nan & nan & nan & inf & nan & nan & nan \\
m10y & 9.3987 & 5.8869 & 31.3700 & -2.7648 & 67.8895 & 34.8569 & 52.7396 & 0.0694 & 0.8775 & 0.0531 & 2.2888 & 0.0750 & 0.7708 & 0.1535 & 1.2594 & 0.0531 & 0.4769 & 0.4700 & 1.1102 & 0.0605 & 0.6459 & 0.2936 & 89.2438 & 0.0000 & 0.7638 & 0.2349 & 2.9221 & 0.0000 & 0.6741 & 0.3097 & 0.8065 & 0.0000 & 0.5529 & 0.4467 & 0.8921 & 0.0000 & 0.8540 & 0.1456 \\
m10z & 9.6626 & 6.0321 & 38.4115 & -2.0593 & 84.1841 & 42.0798 & 28.8917 & 0.0855 & 0.7972 & 0.1173 & 1.7555 & 0.0887 & 0.6269 & 0.2841 & 1.6434 & 0.0584 & 0.3155 & 0.6261 & 1.1175 & 0.0667 & 0.4999 & 0.4334 & 32.5719 & 0.0000 & 0.5344 & 0.4651 & 1.8244 & 0.0000 & 0.4058 & 0.5897 & 1.1641 & 0.0000 & 0.2441 & 0.7558 & 0.8038 & 0.0001 & 0.4669 & 0.5329 \\
m11a & 9.8915 & 6.7998 & 45.7891 & -2.0118 & 102.0125 & 48.6025 & 23.7544 & 0.0686 & 0.8121 & 0.1192 & 1.2835 & 0.0517 & 0.6573 & 0.2907 & 0.9378 & 0.0220 & 0.3382 & 0.6398 & 0.8182 & 0.0623 & 0.6639 & 0.2738 & 32.1922 & 0.0000 & 0.7264 & 0.2733 & 1.2252 & 0.0000 & 0.6160 & 0.3788 & 0.4256 & 0.0000 & 0.4871 & 0.5128 & 0.7203 & 0.0000 & 0.8141 & 0.1858 \\
m11b & 10.0150 & 7.1604 & 50.6737 & -1.8223 & 113.9029 & 55.1251 & 19.5089 & 0.0454 & 0.7537 & 0.2008 & 1.2644 & 0.0289 & 0.5443 & 0.4265 & 1.2738 & 0.0097 & 0.2345 & 0.7558 & 0.8609 & 0.0331 & 0.6044 & 0.3625 & 24.9300 & 0.0000 & 0.6788 & 0.3210 & 1.0934 & 0.0002 & 0.5080 & 0.4884 & 0.5449 & 0.0004 & 0.2954 & 0.7041 & 0.6615 & 0.0001 & 0.6658 & 0.3340 \\
m11c & 10.4254 & 7.8025 & 69.1876 & -1.2756 & 159.2269 & 74.5694 & 10.0930 & 0.0309 & 0.6033 & 0.3658 & 0.9402 & 0.0189 & 0.4154 & 0.5653 & 1.1981 & 0.0077 & 0.1855 & 0.8068 & 0.7772 & 0.0206 & 0.4586 & 0.5208 & 16.3973 & 0.0000 & 0.3941 & 0.6059 & 1.0719 & 0.0000 & 0.2784 & 0.7206 & 0.7968 & 0.0000 & 0.1480 & 0.8520 & 0.7208 & 0.0000 & 0.4294 & 0.5706 \\
m11f & 10.8506 & 8.2414 & 96.9847 & -0.0968 & 218.4961 & 102.2143 & 4.0647 & 0.0699 & 0.5948 & 0.3353 & 0.5277 & 0.0450 & 0.4247 & 0.5299 & 0.9197 & 0.0190 & 0.2158 & 0.7653 & 0.8380 & 0.0565 & 0.4856 & 0.4579 & 5.1715 & 0.0000 & 0.3020 & 0.6980 & 0.4565 & 0.0000 & 0.2277 & 0.7708 & 0.4068 & 0.0000 & 0.1402 & 0.8598 & 0.5182 & 0.0000 & 0.4439 & 0.5561 \\
m11q & 10.3770 & 7.3952 & 66.5144 & -1.1906 & 146.0839 & 71.5715 & 14.9112 & 0.1663 & 0.6745 & 0.1592 & 1.3732 & 0.1348 & 0.4975 & 0.3674 & 1.9230 & 0.0618 & 0.2141 & 0.7241 & 1.1939 & 0.1002 & 0.3798 & 0.5200 & 32.0541 & 0.0005 & 0.4907 & 0.5088 & 2.3641 & 0.0007 & 0.3626 & 0.6360 & 2.0954 & 0.0007 & 0.1961 & 0.8032 & 1.6884 & 0.0005 & 0.3567 & 0.6428 \\
m11v & 10.5013 & 7.9483 & 74.4011 & -0.8721 & 166.8365 & 80.8988 & 7.6534 & 0.0549 & 0.6469 & 0.2983 & 0.7773 & 0.0327 & 0.4261 & 0.5407 & 1.1813 & 0.0109 & 0.1844 & 0.8047 & 0.9321 & 0.0454 & 0.5159 & 0.4387 & 10.3547 & 0.0000 & 0.4029 & 0.5971 & 0.7710 & 0.0000 & 0.2922 & 0.7069 & 0.6320 & 0.0000 & 0.1674 & 0.8325 & 0.7892 & 0.0000 & 0.4462 & 0.5538 \\
m12f & 11.3000 & 9.0861 & 138.5344 & 0.6098 & 311.4647 & 145.4994 & 2.3211 & 0.0231 & 0.3822 & 0.5947 & 0.4221 & 0.0154 & 0.2598 & 0.7243 & 0.9083 & 0.0072 & 0.1256 & 0.8672 & 0.7600 & 0.0232 & 0.3282 & 0.6486 & 2.3285 & 0.0000 & 0.1029 & 0.8970 & 0.2586 & 0.0000 & 0.0822 & 0.9169 & 0.2750 & 0.0000 & 0.0518 & 0.9482 & 0.3740 & 0.0001 & 0.2878 & 0.7121 \\
m12m & 11.0576 & 8.3822 & 115.4186 & 0.0730 & 255.1137 & 132.6741 & 3.5408 & 0.0568 & 0.4249 & 0.5183 & 0.5264 & 0.0323 & 0.2598 & 0.7072 & 0.9839 & 0.0122 & 0.1127 & 0.8751 & 0.8360 & 0.0366 & 0.3176 & 0.6459 & 4.2473 & 0.0000 & 0.1305 & 0.8695 & 0.4424 & 0.0000 & 0.0845 & 0.9122 & 0.4782 & 0.0000 & 0.0471 & 0.9528 & 0.3723 & 0.0001 & 0.1920 & 0.8079 \\
m12i & 11.1531 & 8.3699 & 121.9358 & -0.0764 & 258.3670 & 133.6648 & 4.7308 & 0.0350 & 0.4974 & 0.4676 & 0.6859 & 0.0202 & 0.3112 & 0.6680 & 1.2962 & 0.0077 & 0.1282 & 0.8641 & 1.0926 & 0.0233 & 0.3744 & 0.6024 & 9.0717 & 0.0000 & 0.1524 & 0.8475 & 0.9400 & 0.0000 & 0.1160 & 0.8826 & 0.9713 & 0.0000 & 0.0767 & 0.9232 & 1.0908 & 0.0006 & 0.2502 & 0.7493 \\
A1 & 12.1771 & 11.0075 & 278.5722 & 2.0537 & 682.5476 & 294.1653 & 0.1946 & 0.0046 & 0.0574 & 0.9379 & 0.0630 & 0.0036 & 0.0485 & 0.9440 & 0.1201 & 0.0034 & 0.0509 & 0.9457 & 0.0974 & 0.0217 & 0.2406 & 0.7377 & 0.1465 & 0.0002 & 0.0214 & 0.9784 & 0.0247 & 0.0002 & 0.0227 & 0.9751 & 0.0345 & 0.0002 & 0.0205 & 0.9793 & 0.0409 & 0.0024 & 0.1870 & 0.8106 \\
A2 & 12.2491 & 11.1264 & 299.9680 & 2.3505 & 736.2077 & 311.7270 & 0.1531 & 0.0074 & 0.1518 & 0.8408 & 0.0540 & 0.0053 & 0.1141 & 0.8778 & 0.1307 & 0.0040 & 0.0892 & 0.9068 & 0.1356 & 0.0321 & 0.3235 & 0.6444 & 0.1002 & 0.0000 & 0.0331 & 0.9669 & 0.0192 & 0.0000 & 0.0307 & 0.9672 & 0.0316 & 0.0000 & 0.0240 & 0.9760 & 0.0426 & 0.0003 & 0.2811 & 0.7186 \\
A4 & 12.0143 & 10.4359 & 244.9935 & 1.9061 & 567.1822 & 260.1953 & 0.3140 & 0.0352 & 0.2043 & 0.7606 & 0.0941 & 0.0235 & 0.1422 & 0.8329 & 0.2551 & 0.0129 & 0.0821 & 0.9050 & 0.2467 & 0.0420 & 0.2142 & 0.7438 & 0.3458 & 0.0004 & 0.0283 & 0.9713 & 0.0576 & 0.0006 & 0.0247 & 0.9728 & 0.0840 & 0.0007 & 0.0176 & 0.9817 & 0.1025 & 0.0077 & 0.1349 & 0.8574 \\
A8 & 11.8331 & 9.4767 & 212.0698 & 1.5120 & 475.8170 & 237.4089 & 0.5838 & 0.0459 & 0.2813 & 0.6728 & 0.1566 & 0.0270 & 0.1802 & 0.7915 & 0.3867 & 0.0143 & 0.1020 & 0.8838 & 0.3821 & 0.0343 & 0.2054 & 0.7602 & 0.5345 & 0.0001 & 0.0234 & 0.9765 & 0.0857 & 0.0000 & 0.0150 & 0.9813 & 0.1007 & 0.0001 & 0.0110 & 0.9889 & 0.0571 & 0.0001 & 0.0580 & 0.9419 \\
\end{tabular}
\caption{Identical to \autoref{tab:lowz} but now for our high redshift bin ($z=2.0-4.0$). \texttt{This supplementary table is available for download online in the journal.}}
\label{tab:highz}
\end{table*}

\begin{table*}
\tiny
\begin{tabular}{ccccccccccccHHHHHHHHHHHHHHHHHHHHHHHHHHHHHHHHHHHHHHHHH}
Halo & Redshift & dt$_{\rm lag}$ & log $M_{\rm *,burst}$ & log $M_{\rm wind}$ & log $\eta_{\rm M}$ & log $M_{\rm wind,cold}$ & log $M_{\rm wind,warm}$ & log $M_{\rm wind,hot}$ & log $\eta_{\rm M,cold}$ & log $\eta_{\rm M,warm}$ & log $\eta_{\rm M,hot}$ & log\_pref & log\_pwind & log\_etap & log\_pwind\_cold & log\_pwind\_warm & log\_pwind\_hot & log\_etap\_cold & log\_etap\_warm & log\_etap\_hot & log\_eref & log\_ewind & log\_etaE & log\_ewind\_cold & log\_ewind\_warm & log\_ewind\_hot & log\_etaE\_cold & log\_etaE\_warm & log\_etaE\_hot & log\_zref & log\_zwind & log\_etaZ & log\_zwind\_cold & log\_zwind\_warm & log\_zwind\_hot & log\_etaZ\_cold & log\_etaZ\_warm & log\_etaZ\_hot & peak\_prominence & peak\_height & peak\_baseline\_Myr & log\_fdense\_weighted & log\_sigmaSFR\_weighted & log\_sigmaGas\_weighted & log\_global\_mstar & SFR\_max\_Gyr\_normed & log\_tcool\_tff & global\_Vvir & global\_Vvir\_ism & log\_global\_mvir & global\_rvir & SFR\_max \\
\hline \\
m10q & 3.8854 & 0.0000 & 4.7019 & 6.7134 & 2.0114 & 5.0387 & 6.6871 & 5.2872 & 0.3368 & 1.9852 & 0.5853 & 8.1024 & 8.5237 & 0.4214 & 6.7894 & 8.4274 & 7.7776 & -1.3130 & 0.3250 & -0.3248 & 53.7019 & 53.6829 & -0.0190 & 51.7770 & 53.3795 & 53.3734 & -1.9249 & -0.3224 & -0.3285 & 3.0029 & 2.8984 & -0.1046 & 1.1896 & 2.7883 & 2.2086 & -1.8134 & -0.2146 & -0.7944 & 0.1097 & 0.1182 & 127.4735 & -5.8013 & -3.5129 & 1.4382 & 5.3844 & 2.3541 & nan & 32.9762 & 22.4603 & 9.2915 & 8.5036 & 0.0028 \\
m10q & 3.5071 & 0.0000 & 4.4318 & 7.0711 & 2.6392 & 5.0237 & 6.9779 & 6.3361 & 0.5919 & 2.5461 & 1.9043 & 7.8323 & 9.1044 & 1.2721 & 6.7841 & 8.8263 & 8.7734 & -1.0482 & 0.9940 & 0.9411 & 53.4318 & 54.5382 & 1.1064 & 51.9881 & 54.0700 & 54.3557 & -1.4438 & 0.6381 & 0.9239 & 2.7329 & 3.6123 & 0.8795 & 1.4289 & 3.3916 & 3.2054 & -1.3040 & 0.6588 & 0.4725 & 0.2642 & 0.2690 & 150.9894 & -inf & -3.5388 & 1.0837 & 5.5731 & 3.0608 & nan & 33.1544 & 22.9220 & 9.3355 & 9.3852 & 0.0036 \\
m10q & 3.0765 & 0.0000 & 3.2909 & 6.8650 & 3.5742 & 5.4829 & 6.8036 & 5.8213 & 2.1920 & 3.5127 & 2.5304 & 6.6913 & 8.7759 & 2.0846 & 7.3389 & 8.6156 & 8.2082 & 0.6476 & 1.9242 & 1.5169 & 52.2909 & 53.9587 & 1.6678 & 52.2677 & 53.6247 & 53.6714 & -0.0232 & 1.3338 & 1.3805 & 1.5919 & 3.3464 & 1.7545 & 1.9377 & 3.1948 & 2.7541 & 0.3458 & 1.6029 & 1.1622 & 0.1891 & 0.1922 & 139.0900 & -inf & -6.6940 & 0.7531 & 5.6805 & 0.0772 & nan & 34.5621 & 25.3727 & 9.4801 & 11.6787 & 0.0001 \\
m10q & 2.6924 & 0.0000 & 6.0332 & 7.4680 & 1.4348 & 5.9496 & 7.4100 & 6.4444 & -0.0836 & 1.3768 & 0.4112 & 9.4337 & 9.4748 & 0.0411 & 7.8258 & 9.2966 & 8.9714 & -1.6079 & -0.1371 & -0.4622 & 55.0332 & 54.8803 & -0.1529 & 52.8001 & 54.3090 & 54.7397 & -2.2331 & -0.7242 & -0.2935 & 4.3342 & 4.1185 & -0.2157 & 2.5293 & 3.9142 & 3.6619 & -1.8049 & -0.4200 & -0.6723 & 0.7180 & 0.7184 & 106.0190 & -2.9877 & -1.6573 & 1.2354 & 6.0491 & 34.4819 & nan & 33.3745 & 26.7728 & 9.5066 & 12.9912 & 0.0406 \\
m10q & 2.0145 & 67.2997 & 4.6990 & 5.4566 & 0.7575 & -inf & 5.4462 & 3.5149 & -inf & 0.7472 & -1.1841 & 8.0995 & 7.5229 & -0.5766 & -inf & 7.4518 & 5.9862 & -inf & -0.6477 & -2.1133 & 53.6990 & 51.9894 & -1.7096 & -inf & 51.9662 & 50.6040 & -inf & -1.7329 & -3.0950 & 3.0001 & 1.6698 & -1.3302 & -inf & 1.6540 & 0.0538 & -inf & -1.3461 & -2.9463 & 0.0034 & 0.0037 & 160.8799 & -4.5771 & -3.2027 & 0.9245 & 6.0577 & 30.6242 & nan & 30.9185 & 28.5277 & 9.5470 & 16.4710 & 0.0017 \\
m10q & 1.9858 & 67.2997 & 3.6776 & 5.1241 & 1.4465 & -inf & 5.1025 & 3.6053 & -inf & 1.4249 & -0.0723 & 7.0781 & 7.2752 & 0.1972 & -inf & 7.1694 & 6.0699 & -inf & 0.0913 & -1.0082 & 52.6776 & 51.8256 & -0.8520 & -inf & 51.5933 & 51.4311 & -inf & -1.0843 & -1.2465 & 1.9787 & 1.3505 & -0.6281 & -inf & 1.2920 & 0.3925 & -inf & -0.6867 & -1.5861 & 0.0004 & 0.0028 & 114.8874 & -inf & -3.7705 & 0.8875 & 6.0627 & 9.1144 & nan & 30.8776 & 28.7582 & 9.5489 & 16.5927 & 0.0005 \\
\end{tabular}
\caption{A subset of columns and rows from our full individual outflow episode catalog for all halos at $z<4$. Integrated starburst and wind masses are in $M_{\odot}$, momentum in $M_{\odot}$ km/s, energy in erg, metal mass in $M_{\odot}$. Gas mass surface densities are in $M_{\odot}$/yr/pc$^2$ and SFR surface densities in $M_{\odot}$/yr/kpc$^2$. Global stellar mass and virial mass are in $M_{\odot}$, virial radius in proper kpc, and virial velocity at $R_{\rm vir}$ and $0.1R_{\rm vir}$ in km/s. The time lag and burst baseline are in Myr (note that because of our algorithm, all bursts in the same 1 Gyr time slice have the same time lag). Unless noted otherwise, all other quantities are dimensionless (see also main text). \texttt{This supplementary table is available for download online in the journal.}}
\label{tab:instant}
\end{table*}

\section{Radial velocities}\label{sec:vrad}
Although we focused on the temperature distribution of loading factors, the radial velocity distributions are also fundamental for characterizing the thermodynamics of outflows. In addition, outflow velocities are more easily constrained by observations than total masses, momenta, energies, Bernoulli velocities, etc. 

\begin{figure*} 
\begin{center}
\includegraphics[width=0.8\hsize]{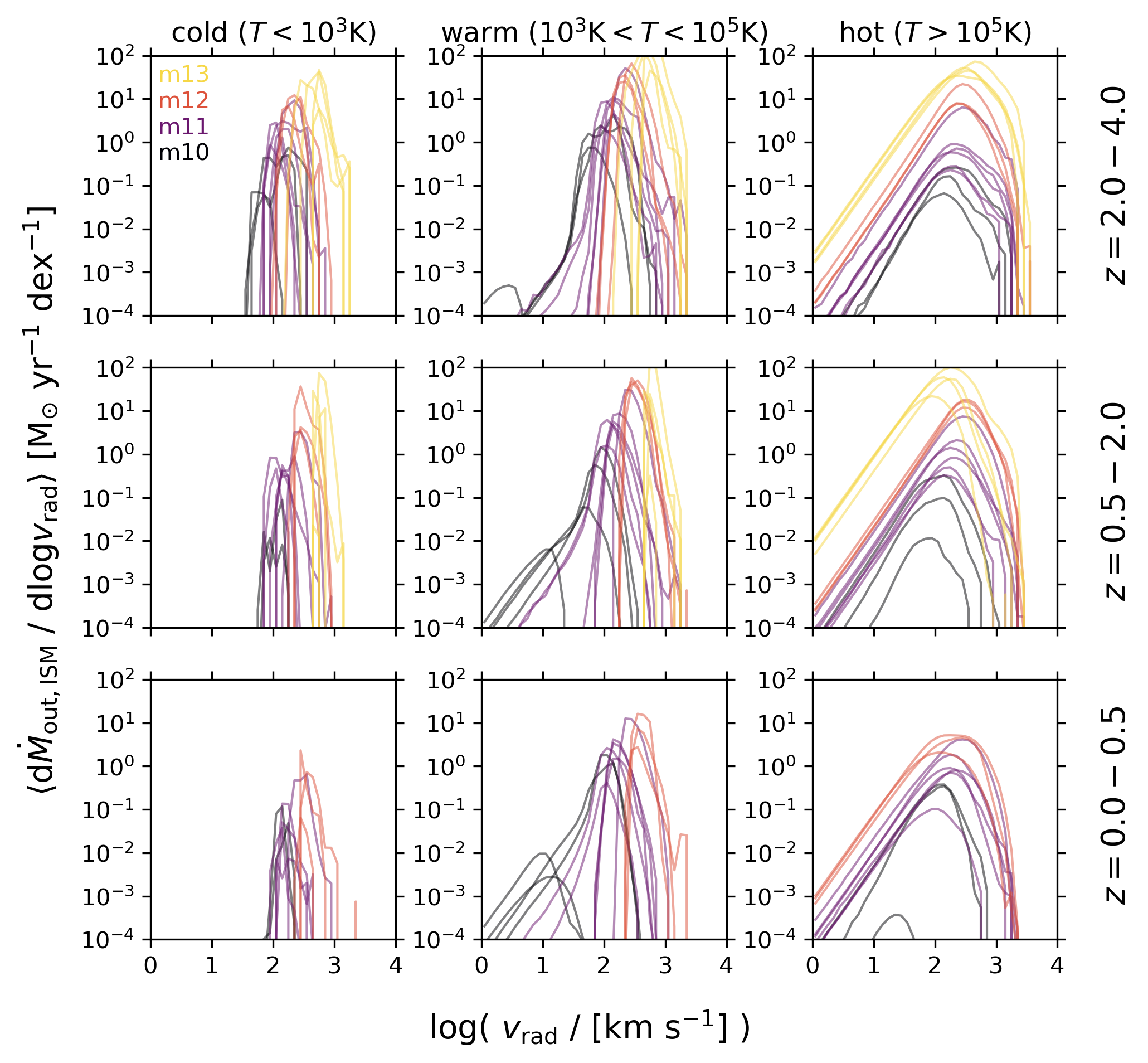}
\end{center}
\caption{Similar to \autoref{fig:mdot_temp_hist} but now showing the distributions of ISM mass outflow rate in bins of radial velocity for wind particles classified as cold (left column), warm (middle column) or hot (right column). The different rows show the average distributions over our fiducial large redshift bins, where snapshots with higher total $\dot{M}_{\rm out,ISM}$ are given higher weight in the average.}
\label{fig:vrad_pdf}
\end{figure*}

\autoref{fig:vrad_pdf} shows the distributions of ISM mass outflow rate in bins of radial velocity for wind particles classified as cold, warm or hot. The distributions are averaged over our fiducial redshift ranges weighting by the overall $\dot{M}_{\rm out,ISM}$ in each contributing snapshot. As can also be seen in the velocity panel of our movies (\autoref{fig:example} and \autoref{fig:example2}), the full radial velocity distributions generally extend to very low values for hot and even warm outflows. These slow moving particles have enough energy from their temperature, compared to the halo potential, to still be classified as wind particles. However cold outflows generally do not extend to very low velocities since their thermal energy is negligible and cannot get them classified as winds according to our Bernoulli velocity criterion. All three phases show a sharp cutoff above a few thousand km/s since these are likely the fastest that stellar-driven winds can propagate (AGN feedback may lead to even faster winds but that is not implemented here).

\begin{figure*} 
\begin{center}
\includegraphics[width=0.8\hsize]{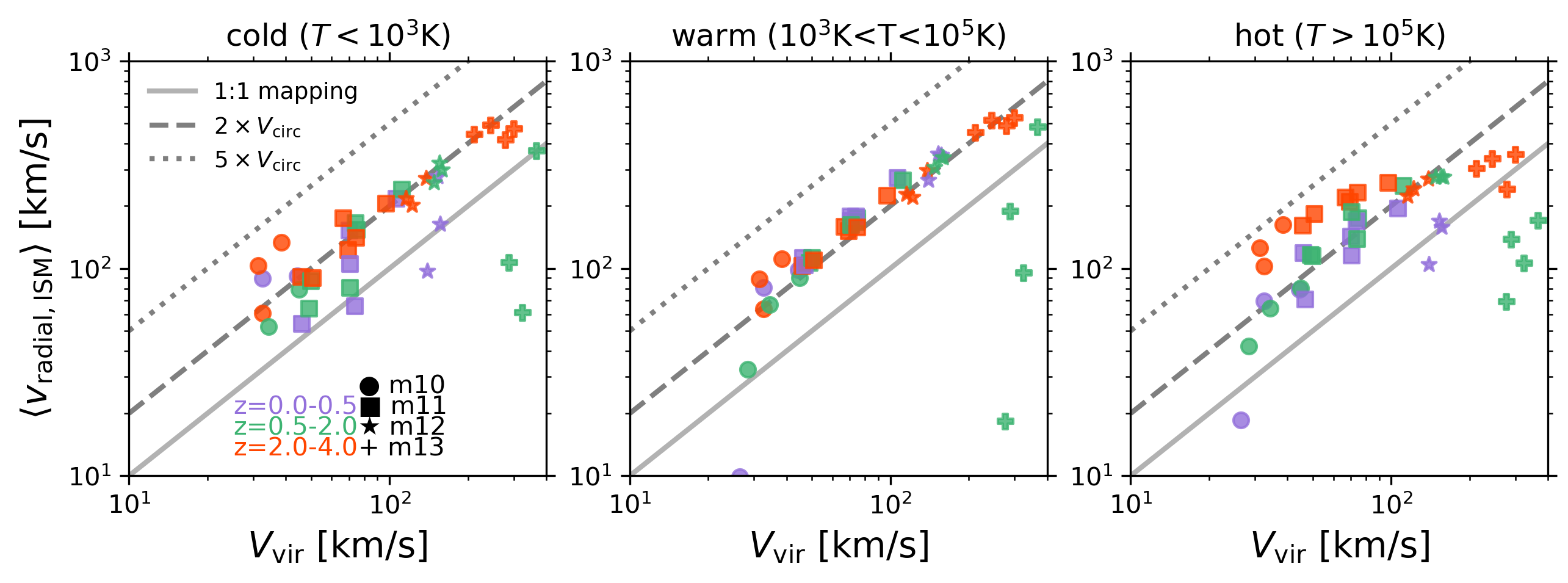}
\end{center}
\caption{Analogous to \autoref{fig:vB_vesc} but now, following common practice, we plot the mass-flux-weighted average radial velocity versus halo virial velocity. From left to right: we plot this for the cold, warm and hot ISM outflows. The solid gray line is the one-to-one mapping between radial velocity halo circular velocity; the dashed and dotted lines are twice and five times the circular velocity, respectively. We see that generally cold and warm outflows cluster around $\approx2V_{\rm vir}$, with slightly lower radial velocities in the m13 halos. The hot outflows tend to be faster, approaching $\approx5\times V_{\rm vir}$ on average for some dwarfs. Interestingly the radial velocity of hot outflows in some halos can be less than $V_{\rm vir}$, which either suggests deceleration due to interactions or that the slower component of the hot wind dominates (as illustrated in the bottom-right panel of \autoref{fig:example} and \autoref{fig:example2}).}
\label{fig:vrad_vcirc}
\end{figure*}

\autoref{fig:vrad_vcirc} collapses the full multi-phase radial velocity distributions into mass-flux-weighted average radial velocities in each phase as a function of halo circular velocity. This is a simpler, more traditional plot compared to \autoref{fig:vB_vesc}, where we compared the average mass-flux-weighted Bernoulli velocity to the difference in escape velocity between $0.1R_{\rm vir}$ and $1.0R_{\rm vir}$. Consistent with previous work, we find a positive correlation between average outflow radial velocity and halo circular velocity, with the cold and warm outflows generally clustering around $v_{\rm rad}\approx2\times V_{\rm vir}$. The relatively large radial velocities of cold and warm outflows is likely driven by the fact that most of their Bernoulli velocity must come from the kinetic term, which alone needs to be sufficiently large for traveling to $\gtrsim0.5R_{\rm vir}$. The hot outflows can have substantially larger radial velocities, especially in some high-redshift dwarfs where $v_r\approx5\times V_{\rm vir}$. Interestingly, in some halos, the average hot outflow radial velocities can be lower than $V_{\rm vir}$; this may reflect deceleration of outflows due to high density gas within the ISM and inner CGM. However, these slower moving outflows still have enough energy to travel deep into the CGM owing to their hot temperatures and hence sufficiently high Bernoulli velocities. 

It is beyond the scope of this work to present a detailed two-dimensional analysis of loading factors simultaneously in temperature and radial velocity bins (but see the bottom-right panel of \autoref{fig:example} and \autoref{fig:example2}). Such an analysis would provide useful constraints for launching of galactic winds in SAMs and lower resolution cosmological simulations \citep[see discussion in][]{kim20b}. It is also beyond the scope of our work to investigate full radial velocity profiles and compare to observations. However, our analysis can be adapted in the future to study outflows closer to the ISM and make predictions for observables based on the trajectories and intrinsic evolution of wind particles \citep[following, e.g.,][]{anglesalcazar17,hafen20}.

\end{document}